\documentclass[conference,letterpaper]{IEEEtran}

\usepackage[
top    = 0.54in,
bottom = 0.84in,
left   = 0.54in,
right  = 0.54in]{geometry}

\usepackage{tcolorbox}
\usepackage[normalem]{ulem}
\usepackage[colorlinks=false]{hyperref}

\usepackage[utf8]{inputenc} 
\usepackage[T1]{fontenc}
\usepackage{url}
\usepackage{ifthen}
\usepackage{cite}
\usepackage[cmex10]{amsmath} % Use the [cmex10] option to ensure complicance                           % with IEEE Xplore (see bare_conf.tex)
\usepackage{amssymb}
\usepackage{graphicx}
\usepackage{bm}
\usepackage{bbm}
\usepackage{color}
\usepackage{subfig}
\usepackage{mathtools}
\usepackage{algorithm}
\usepackage{algorithmic}
\usepackage{graphicx}
\usepackage{float}
\newtheorem{example}{Example}[section]
\newtheorem{lemma}{Lemma}[section]
\newtheorem{theorem}{Theorem}[section]
\newtheorem{proposition}{Proposition}[section]
\newtheorem{corollary}{Corollary}[section]
\newtheorem{definition}{Definition}[section]
\newtheorem{remark}{Remark}[section]

\newtcolorbox{mybox}[3][]
{
  colframe = #2!25,
  colback  = #2!10,
  coltitle = #2!20!black,  
  title    = {#3},
  #1,
}

\newcommand{\lp}{\left(}
\newcommand{\rp}{\right)}

\newcommand{\lnorm}{\left\lVert}
\newcommand{\rnorm}{\right\rVert}

\newcommand{\argmin}{\mathop{\mathrm{arg\,min}}}

\newcommand{\MinS}{{\rm MinS}}
\newcommand{\MinJ}{{\rm MinJ}}
\newcommand{\F}{{\rm F}}

\newcommand{\plnR}{R}
\newcommand{\one}{{\cs_{\plnR_{\dc_{i}}^{1}}}}

\newcommand{\oneutwo}{{\cs_{\plnR_{\dc_{i}}^{1}\bigcup \plnR_{\dc_{i}}^{2}}}}
\newcommand{\oneatwo}{{\cs_{\plnR_{\dc_{i}}^{1}\bigcap \plnR_{\dc_{i}}^{2}}}}
\newcommand{\twosone}{{\cs_{\plnR_{\dc_{i}}^{2}\setminus \plnR_{\dc_{i}}^{1}}}}

\newcommand{\fm}{$P(G, \leq t)$}

\newcommand{\HH}{{H}}
\newcommand{\KK}{{K}}

\newcommand{\Eline}{E_{\rm line}}
\newcommand{\Earc}{E_{\rm arc}}

\newcommand{\itv}[1]{\overline{#1}}
\newcommand{\pln}{d}
\newcommand{\cs}{s}
\newcommand{\dc}{x}
\newcommand{\upd}{\mathcal{D}}
\newcommand{\nb}{b}

\newcommand{\e}[2]{\dc_{#1}\dc_{#2}}

\begin{document}
%\title{On the Noisy Coin Weighing Problem} 
%\title{On the Noisy Combinatorial Quantitative Group Testing Problem}
%\title{On the Combinatorial Quantitative Group Testing Problem with Noisy Query and Partial Recovery}
\title{Reducing Data Fragmentation in Data Deduplication Systems via Partial Repetition and Coding}

\author{
\IEEEauthorblockN{Yun-Han Li, Jin Sima, Ilan Shomorony and Olgica Milenkovic
}
\IEEEauthorblockA{
%\IEEEauthorrefmark{1}
Department of Electrical and Computer Engineering, University of Illinois Urbana-Champaign, USA\\ 
Email: \url{yunhanl2,jsima,ilans,milenkov@illinois.edu}
}
\IEEEauthorblockA{
%\IEEEauthorrefmark{1}
}
}
\maketitle

\begin{abstract}
    Data deduplication, one of the key features of modern Big Data storage devices, is the process of removing replicas of data chunks stored by different users. Despite the importance of deduplication, several drawbacks of the method, such as storage robustness and file fragmentation, have not been previously analyzed from a theoretical point of view. Storage robustness pertains to ensuring that deduplicated data can be used to reconstruct the original files without service disruptions and data loss. Fragmentation pertains to the problems of placing deduplicated data chunks of different user files in a proximity-preserving linear order, since neighboring chunks of the same file may be stored in sectors far apart on the server. This work proposes a new theoretical model for data fragmentation and introduces novel graph- and coding-theoretic approaches for reducing fragmentation via limited duplication (repetition coding) and coded deduplication (e.g., linear coding). In addition to alleviating issues with fragmentation, limited duplication and coded deduplication can also serve the dual purpose of increasing the robusteness of the system design. 
    The contributions of our work are three-fold. First, we describe a new model for file structures in the form of self-avoiding (simple) paths in specialized graphs. Second, we introduce several new metrics for measuring the fragmentation level in deduplication systems on graph-structured files, including the \emph{stretch metric} that captures the worst-case ``spread'' of adjacent data chunks within a file  when deduplicated and placed on the server; and, the \emph{jump metric} that captures the worst-case number of times during the reconstruction process of a file that one has to change the readout location on the server. For the stretch metric, we establish a connection between the level of fragmentation and the \emph{bandwidth} of the file-graph. In particular, we derive lower and upper bounds on degree of fragmentation and describe instances of the problem where repetition and coding reduce fragmentation. The key ideas behind our approach are graph folding and information-theoretic arguments coupled with graph algorithms such as matching. For the jump metric, we provide a new algorithm for computing the jump number of hierarchical data structures captured by trees. Third, we describe how controlled repetition and coded redundancy added after deduplication can ensure valuable trade-offs between the storage volume and the degree of fragmentation.     
\end{abstract}

\section{Introduction}\label{sec:intro}
%\begin{tcolorbox}
Data deduplication is a data management technique employed in modern large-scale data storage systems to increase storage efficiency by removing content replicas of different users~\cite{comprehensive,kaur2018data,armknecht2015transparent,elrouayheb2016synchronization}. For example, cloud storage providers such as Microsoft and Dropbox, as well as intelligent data infrastructure companies such as NetApp, actively support deduplication schemes in their storage systems.
    
A typical deduplication algorithm first splits the input files into smaller data subunits, referred to as ``chunks''~\cite{chunking,chunking_random_substitute,chunking_edit}, then removes duplicate data chunks and proceeds to place the chunks on servers. For duplicate chunk identification, one typically uses hash values~\cite{chunking,chunking_random_substitute,chunking_edit}. The chunk placement is based either on some predetermined protocol or the order of arrival of new chunks. After deduplication, each file is stored in `compressed' form, involving a set of pointers explaining how to assemble it (e.g., a set of hash values pointing to the actual locations of the data chunks and labels indicating their order). During the file restoration phase, the server uses the compressed description to recreate files by retrieving chunks guided by the pointers. The trade-off between the chunk size and the length of the assembly instruction under various chunking strategies is an important practical system compression feature, addressed in~\cite{chunking}. Since the volume of the assembly instruction data is usually negligible compared to the actual user data, this consideration is irrelevant for protocol-level analyses of deduplication.
    
Although deduplication algorithms provide significant storage savings, they introduce a number of readout issues, including \emph{data fragmentation}~\cite{idedup} and \emph{data instability}~\cite{reliability,reliability2}. Data fragmentation arises due to storing information chunks of a single file at multiple potentially distal locations, which slows down data retrieval and potentially creates problems with accessing large volumes of secondary, unwanted interspersing data. Deduplication instability, on the other hand, refers to the vulnerability to data loss (i.e., lack of robustness) of the file system when only one replica of each chunk is maintained. These problems are partly mitigated in practice by using caches and specialized chunk placements~\cite{kesavan2020countering}, as well as `incomplete deduplication' protocols, where not all redundant chunks are removed but only a portion of them, dictated by their popularity~\cite{comprehensive,kaur2018data}. 

Despite the widespread use of deduplication and the many practical implementations of the method, little is known about the \emph{theoretical}  performance limits of such systems with regards to data fragmentation. Prior work~\cite{elrouayheb2016synchronization} provided a mathematical analysis of the influence of deduplication on distributed storage codes, while~\cite{chunking} initiated an information-theoretic study of deduplication as a form of compression. Some other works~\cite{chunking_edit,chunking_random_substitute} focused on examining the influence of errors on chunking and deduplication. None of these contributions considered fragmentation problems or issues associated with partial deduplication and file retrieval robustness.

Here, we initiate the first analytical study of the trade-off between storage redundancy and file fragmentation. The key features supporting our analysis are: a) A new model for files that share chunks in terms of self-avoiding (simple) paths/walks in what we call a \emph{chunk or file graph}. In particular, we model the data as being either generated by a sparse Hamiltonian-path graph in which nodes represent chunks and nodes on connected paths correspond to files to be stored; or, as hierarchical structures whose chunks are organized on a tree. The Hamiltonian path requirement is used to both model a natural linear ordering of concatenated files and to simplify challenging analyses. Similarly, tree models are used to capture chunk organizations in omnipresent hierarchical data;  b) Several new metrics for measuring the level of fragmentation, including a worst-case metric which examines the maximum stretch of allotted locations of adjacent data chunks (i.e., the \emph{stretch metric}) and a metric that captures the worst-case number of times the location on a server needs to be changed to retrieve a file (i.e., the \emph{jump metric}). Average-case metrics, although formally defined in this work will be examined elsewhere. 

Our main contributions are bounds on the worst-case fragmentation metrics and algorithms for chunk placements that produce provably good fragmentation results. Most importantly, we examine these key features of deduplication systems for the case of \emph{completely deduplicated files} (for which the solution involves optimization over permutations, i.e., over the symmetric group), \emph{partially deduplicated files with data chunk repetitions} (repetition-coded deduplication) and \emph{coded deduplication}, in which the stored chunks may represent linear (or, in general nonlinear) combinations of systematic user chunks. 

The outlined analysis involves information-theoretic arguments but also strongly relies on results in graph theory pertaining to the \emph{graph bandwith} and related concepts~\cite{lai1999survey}, as well as matchings in graphs. In the former context, we introduce \emph{graph folding} which allows one to transfer the domain of the problem from graphs to functions. Additional ideas used in this context include graph factorization and concatenation of different permutations of the chunks, and results already established in the context of the related problem of graph bandwidth analysis~\cite{blache1997approximation}. We also describe a number of examples pertaining to deduplication systems in which one is either allowed to retain multiple copies of the same chunk or use coded chunks. In these settings, it is \emph{a priori clear} that redundant chunks can help reduce fragmentation, since concatenating all files (e.g., not performing deduplication) results in no fragmentation. What is at this point completely unknown is how much redundancy is needed to reduce fragmentation to practically acceptable levels. Furthermore, repeated and/or coded chunks both increase the robustness of the system, which is desirable, but at the cost of increasing the storage overhead of the system. 

The work closest to our new study is the work that established ultimate performance limits of information retrieval with and without redundancy~\cite{coffey2000fundamental}. There, the questions were posed in a \emph{probabilistic} context that does not include deduplication nor fragmentation models. Furthermore, information is modelled as arising from a Markovian grid or torus, by traversing a path whose length increases in an unbounded manner. The main objective of the work was to explore the expansion of the average walk lengths of 1-D retrieval versus 2-D retrieval approaches. The work also included an analysis of coding approaches that can aid in reducing the gap introduced by lowering the dimension of the data space. A more detailed description of this work, which is of independent interest, is provided 
in Section~\ref{sec:related}.
% at the end of the manuscript.  

The paper is organized as follows. Section~\ref{sec:practice} provides a review of deduplication approaches and introduces terms needed for our analysis and the problem formulations that follow. Section~\ref{sec:formulation} describes the problem formulation(s), based on the description of the file and chunk store model and the metrics used to measure the fragmentation level. Results pertaining to fragmentation bounds and limits for uncoded, repetition- and linearly encoded chunk stores under the stretch metrics are discussed in Section~\ref{sec:stretch}. This section also contains the description of our graph folding approach and pertinent bounds based on it. Similar results for the jump fragmentation metric are presented in Section~\ref{sec:jump}. A collection of open problems is discussed in Section~\ref{sec:future}.

\section{Review of Deduplication Approaches}\label{sec:practice}
%Data deduplication is a data reduction  technique employed in modern large-scale data storage systems that has the goal to increase storage efficiency by removing replicas of data stored by different users~\cite{comprehensive,kaur2018data,armknecht2015transparent,elrouayheb2016synchronization}.
As already described, deduplication systems first perform decompositions of files into chunks, then identify repeated and unique chunks using hash values, and then proceed to place the chunks into a chunk store. 

Hashing algorithms (such as SHA-1~\cite{penard2008secure} and MD5~\cite{kaliski1995message}) are used to produce hash values assigned to individual data chunks. For example, the MD5 hashing method is a \emph{message-digest one-way function} that takes as its input a message of arbitrary length and outputs a fixed-length hash as the message's signature; although originally designed for cryptographic protocols and message authentication in particular, it has found other applications, including deduplication. Data chunks are termed \emph{unique} if they have not been stored before, and \emph{duplicates} if they are already present in the storage system. Both the unique and duplicate chunks are represented via hash value, and their presence in the cloud detected through hash rather than direct comparisons of their binary encodings.

Deduplication algorithms generally fall into two main categories, \emph{inline} and \emph{post-processing} deduplication. 

\begin{figure}[H]
 \vspace{-3mm}
    \centering
    \includegraphics[scale=0.27]{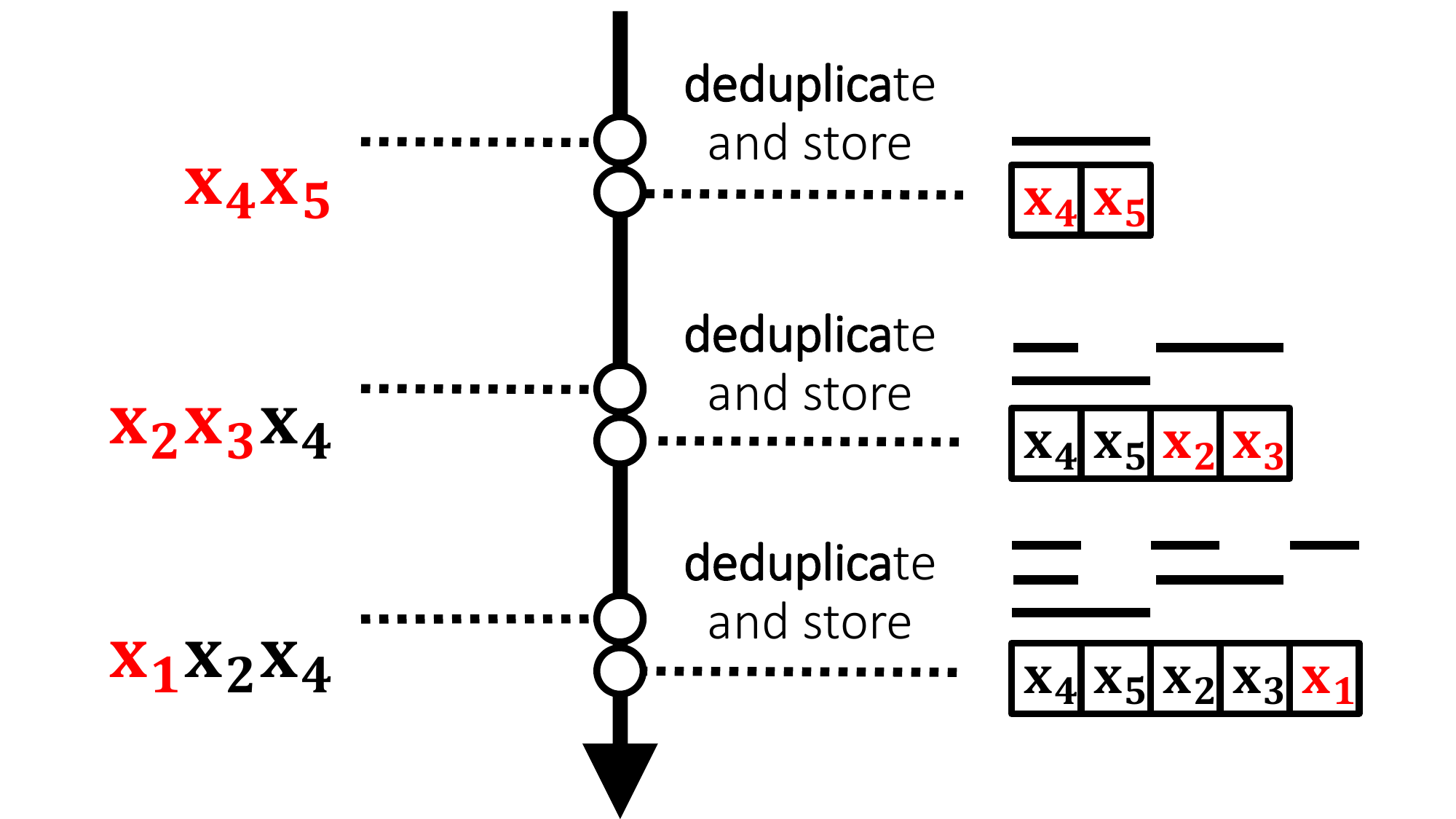}
    \caption{Inline deduplication. Files arrive sequentially and are processed before being stored. Unique chunks from each file are marked in red, while the number of bars over a chunk symbol indicates the number of times it has been observed up to the given point in time.} \label{fig:inline}
\end{figure}

Inline deduplication first deduplicates chunks in files arriving at the write path and then stores deduplicated data on the server. Inline deduplication requires few Input/Output operations but tends to highly scatter data chunks on the server and lead to severe fragmentation (see~\cite{li2014efficient} for a more detailed discussion). To illustrate the process of inline deduplication, we denote the $j$-th %\iscomment{later, the notation $j$th is used, stick with one} 
data chunk by $\dc_{j}$ and provide an example of how to deduplicate three files in Figure~\ref{fig:inline}. When encountering the first file comprising two data chunks,  $(\dc_{4},\dc_{5})$, the server has no previously stored chunks, so both $\dc_{4}$ and $\dc_{5}$ are declared unique chunks and stored in consecutive linear order. Hence, the whole file (marked in red) is stored directly. When encountering the second file $(\dc_{2},\dc_{3},\dc_{4})$, only the chunks $(\dc_{2},\dc_{3})$ are unique (marked in red) and stored on the server. The process continues until the final chunk store content subsequently fragments as $(\dc_{4},\dc_{5},\dc_{2},\dc_{3},\dc_{1}).$ Since complete deduplication is performed, each chunk is stored only once, and the three files are represented by pointers to the store that include chunks of the files and their orders, e.g., $(1,2), (3,4,1), (5,3,1)$. Note that the pointers describe both the chunks as well as their orders within each individual file.

\begin{figure}[H]
 \vspace{-5mm}
    \centering
    \includegraphics[scale=0.27]{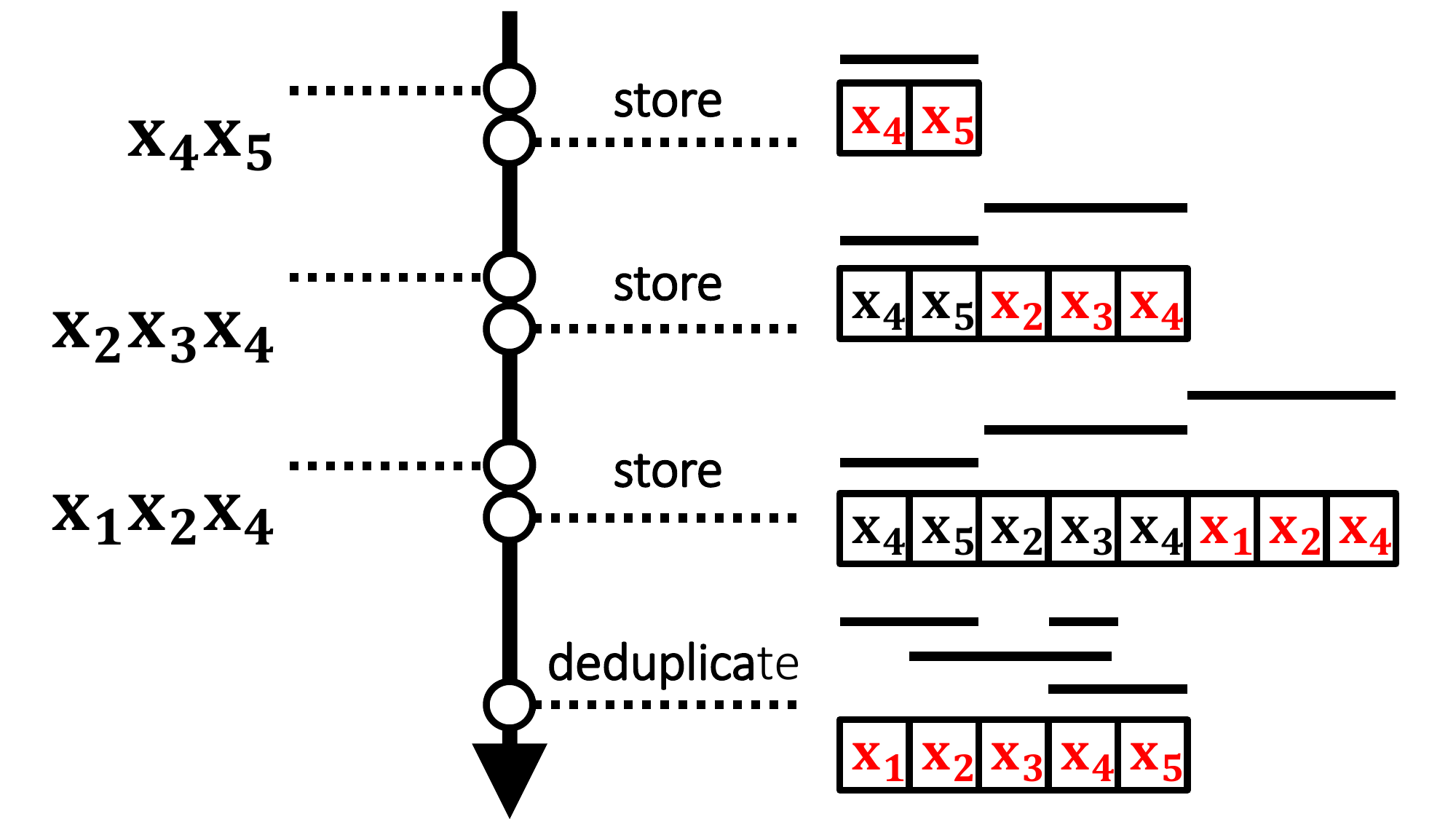}
    \caption{Post-processing deduplication. Chunks are stored as soon as presented to the system, and after all chunks are placed in the chunk store, duplicate chunks are removed and the unique chunks reordered based on some set of optimization criteria.} \label{fig:offline}
\end{figure}

As opposed to inline deduplication, during post-processing deduplication one first stores all data chunks on the server, then deduplicates the content through an optimization process. An example of postprocessing deduplication is shown in Figure~\ref{fig:offline}, where as before one has to deduplicate the content of three files that arrived in some order. Instead of performing deduplication online, the post-processing deduplication algorithm considers the three files as a batch that is to be compacted jointly. In this particular example, the final chunk store content fragments as $(\dc_{1},\dc_{2},\dc_{3},\dc_{4},\dc_{5}),$ and the three files are replaced by pointers as previously described. The file content is less fragmented in the post-processing chunk store than in the inline chunk store, which is easy to see for this simple case.
   
Although both methods of deduplication offer significant storage savings, it is clear from the above example that they introduce \emph{fragmentation}~\cite{li2014efficient,idedup,kesavan2020countering} and create problems with chunk \emph{robustness}~\cite{reliability,reliability2}. Data fragmentation refers to the already mentioned problem of storing information chunks of a file at multiple distal locations, which slows down data retrieval and potentially creates problems with accessing large volumes of secondary, unwanted interspersing data. For example, from the figure illustrating post-processing deduplication, one can observe that in order to restore the content of the third file, $(\dc_{1},\dc_{2},\dc_{4})$ one needs to read the first four chunks from the chunk store and then parse out the data chunk $\dc_{3}$.

Robustness, on the other hand, refers to the degree of tolerance of data loss in the file system when only one replica of each unique chunk is maintained. For example, from the figure illustrating post-processing deduplication, one can see that the data chunk $\dc_{4}$ is present in the files of all three users, but stored only once which may lead to a multi-client information loss in case the data chunk is compromised. Robustness problems are mitigated in practice by using specialized chunk placements and ``incomplete deduplication,'' where not all redundant chunks are removed but only a portion of them, as dictated by the popularity of the data~\cite{comprehensive}. In the context of the given example, this would suggest storing, say, one more copy of the chunk $\dc_{4}$.

The above discussion raises several natural questions. First, how does one have to place data chunks in a store to minimize some predefined fragmentation metric across all files? Second, can repeating a limited number of data chunks reduce the fragmentation metric for arbitrary or special collections of files? %A priori, it is unclear if redundancy actually reduces or increases fragmentation, since adding redundant chunks increases the number of chunks placed on the server that have to be potentially traversed during file reconstruction. 
Third, whenever redundancy helps, can \emph{coding across chunks} outperform simple \emph{repetitions}? Fourth, can carefully selected redundant chunks both reduce fragmentation and increase robustness simultaneously?

 \begin{figure}[H]
    \centering
    \includegraphics[scale=0.27]{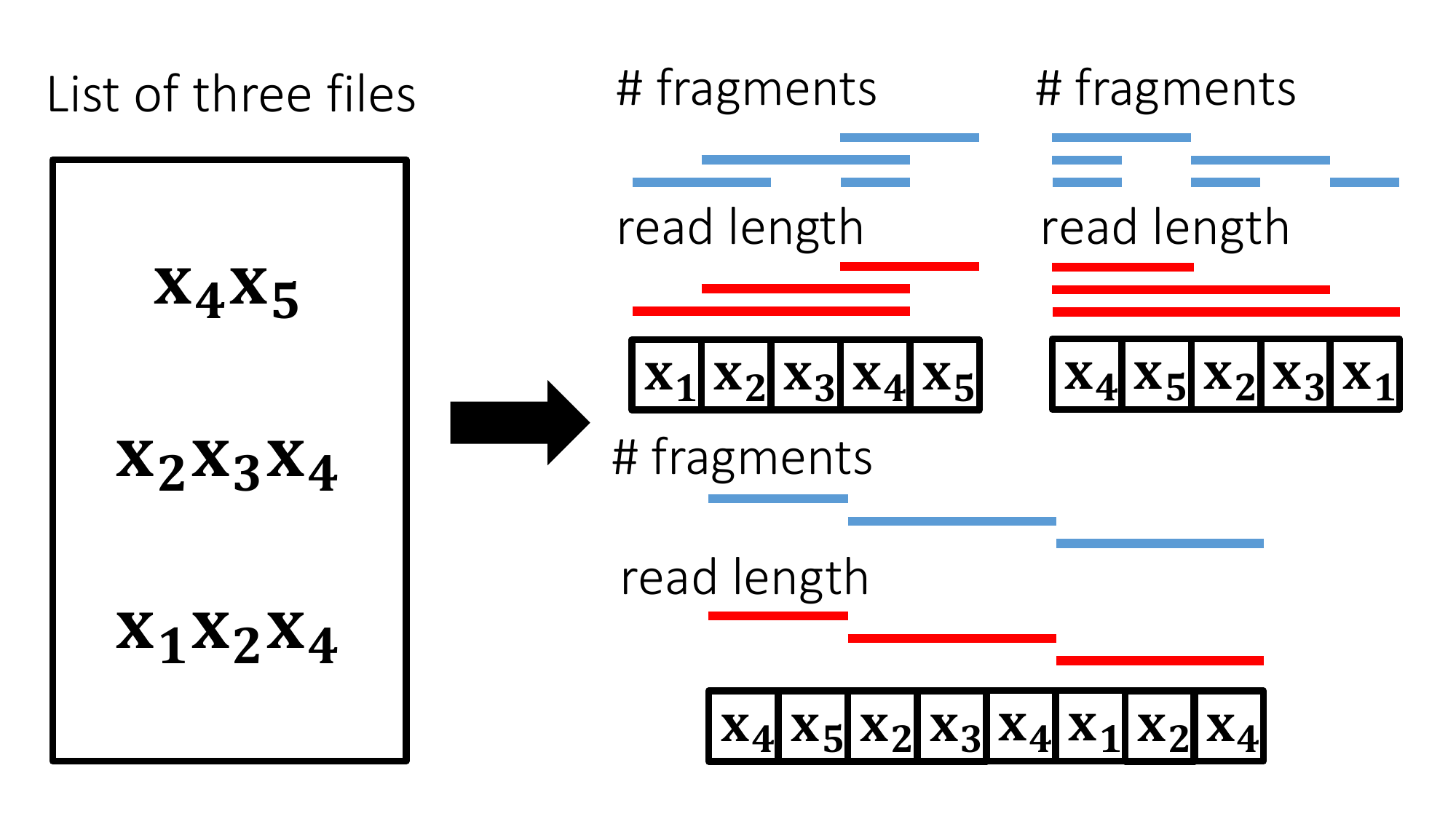}
    \caption{An illustrative example of the trade-off between storage redundancy and fragmentation in deduplication systems. Blue and red lines correspond to the minimum number of fragments that need to be read to recover a file without accessing any chunks not in the file, and the minimum fragment length that covers all chunks of a file, respectively.}
    \label{fig:motivation}
\end{figure}
\begin{figure}[H]
 \vspace{-5mm}
    \centering
    \includegraphics[scale=0.27]{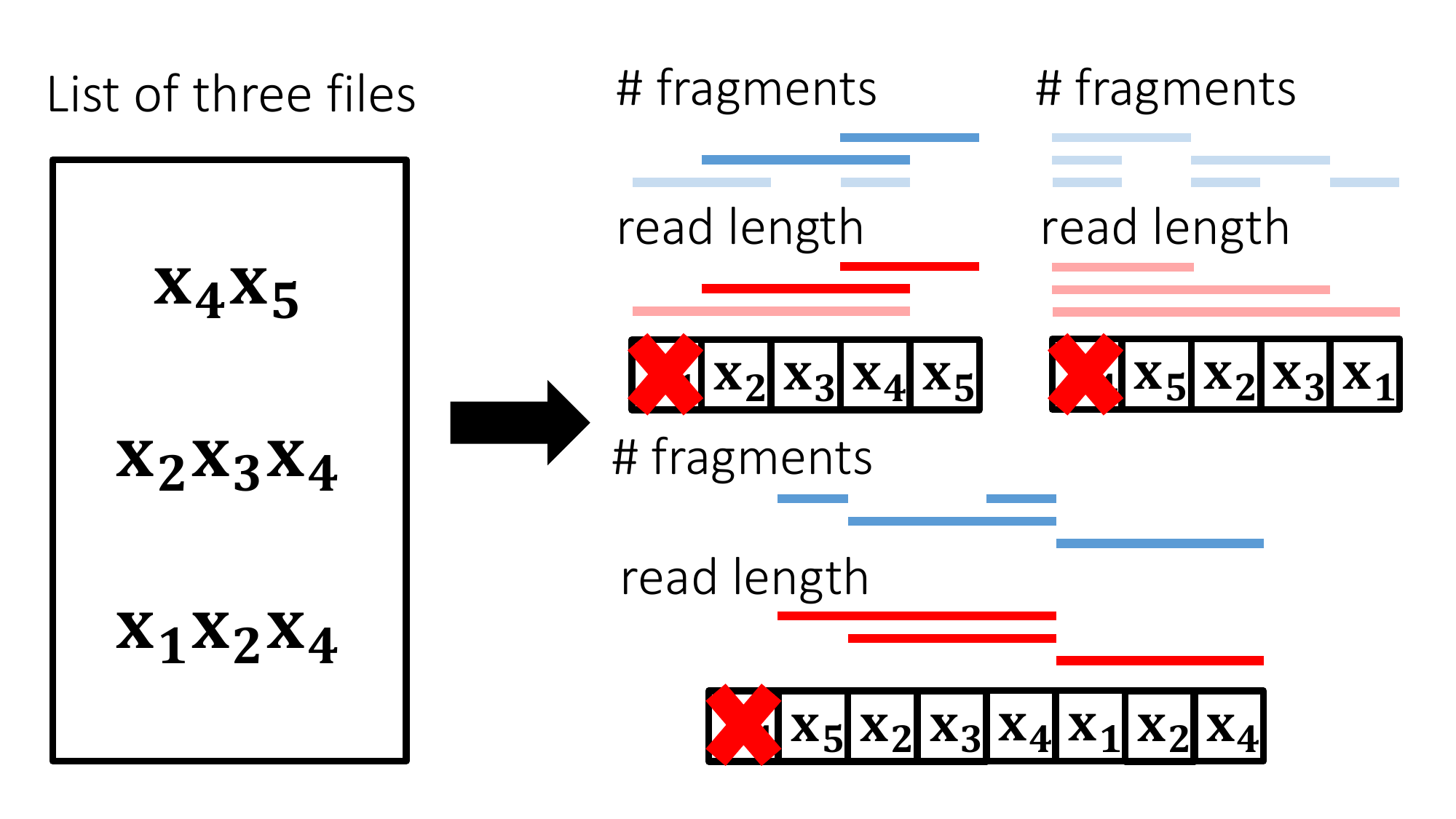}
    \caption{An illustrative example of the trade-offs among storage redundancy, fragmentation, and robustness in deduplication systems.} \label{fig:motivation2}
\end{figure}

To gain more insight into these questions, consider the examples in Figures~\ref{fig:motivation} and~\ref{fig:motivation2}. Figure~\ref{fig:motivation} provides an illustration of the redundancy-fragmentation trade-off. Suppose that we have to store three files with data chunks $(\dc_{4},\dc_{5})$, $(\dc_{2},\dc_{3},\dc_{4})$, and $(\dc_{1},\dc_{2},\dc_{4})$. If the three files are directly stored without deduplication, as shown in the figure, one needs to set aside space for $8$ data chunks in total (bottom, right). To reduce the number of redundant data chunks stored, one might store each data chunk once (top, left), and as a result, the stored data chunk string is a permutation of $\dc_{1}$ to $\dc_{5}$. In this case, retrieving a file requires either a) accessing multiple segments of consecutive data chunks stored on the server to prevent reading extra data chunks that do not belong to the file, which is costly; or, b) reading a single or a small number of long data chunk fragments, including data chunks not originally in the file. As shown in Figure~\ref{fig:motivation}, different permutations of data chunks stored on the server result in different levels of fragmentation, the definition of which will be formalized later. For example, when $(\dc_{1},\dc_{2},\dc_{3},\dc_{4},\dc_{5})$ is the chunk sore string, the files $(\dc_{2},\dc_{3},\dc_{4})$ and $(\dc_{4},\dc_{5})$ can be read from $1$ sequential fragment without including any extra chunks. However, when retrieving $(\dc_{1},\dc_{2},\dc_{4})$, either it needs $2$ fragments to prevent reading extra chunks $\dc_{4}$, or $1$ longer fragment for including it. 

\begin{figure}[H]
    \centering
    \includegraphics[scale=0.27]{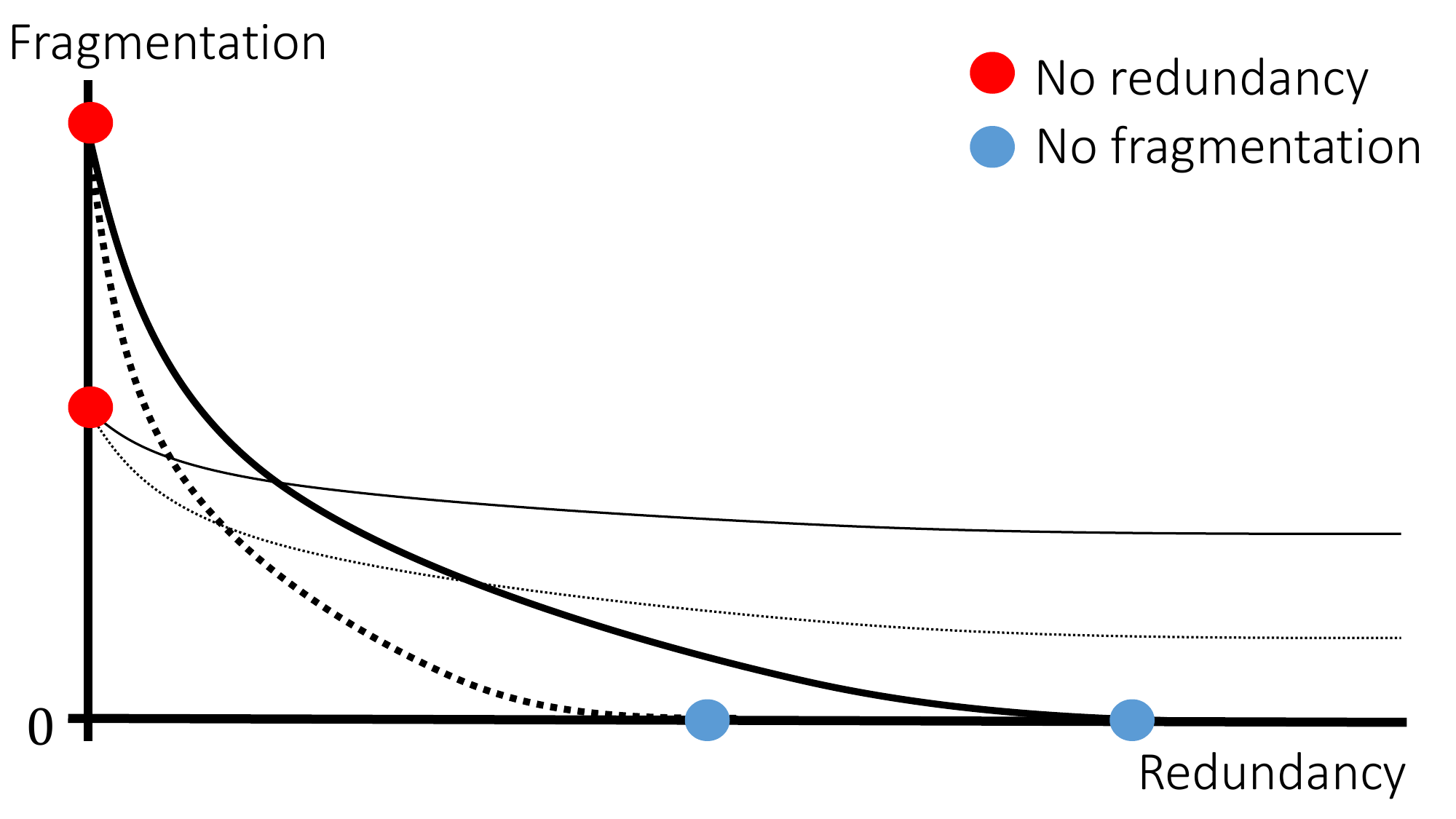}
    \caption{An illustrative example of potential trade-off curves for storage redundancy versus fragmentation. The curves are file-graph specific, and shown for two different graphs using solid lines. The dashed lines correspond to the trade-off curves for the same file-graphs, but are meant to illustrate the potential gain achievable by allowing coding versus simple repetition redundancy. Note that these curves only serve as an illustration, and are not exact mathematical characterizations.} \label{fig:curve}
\end{figure}

Figure~\ref{fig:motivation2} is an illustration of the redundancy, fragmentation, and robustness trade-off paradigm. Now all three chunk stores lost their first chunk, which hurts their performance differently. Chunk store $(\dc_{1},\dc_{2},\dc_{3},\dc_{4},\dc_{5})$ lost chunk $\dc_{1}$ and therefore lost one file $(\dc_{1},\dc_{2},\dc_{4})$. Chunk store $(\dc_{4},\dc_{5},\dc_{2},\dc_{3},\dc_{1})$ lost chunk $\dc_{4}$ and therefore lost all three files. Chunk store $(\dc_{4},\dc_{5},\dc_{2},\dc_{3},\dc_{4},\dc_{1},\dc_{2},\dc_{4})$ lost the first chunk $\dc_{4}$ and therefore does not lose any file. However for the file $(\dc_{4},\dc_{5})$, the number of fragments without reading redundant chunks increases from $1$ to $2$, and the read length for one consecutive fragment increases from $2$ to $4$. 

The goal of our work is to initiate the study of redundancy and fragmentation trade-offs for post-processing deduplication. The discussion of the interplay between redundancy, fragmentation, and robustness trade-offs for post-processing
settings will be addressed in follow-up work. 
More specifically, we are interested in characterizing the trade-off curves for redundancy and fragmentation, for which potential examples are shown in~Figure~\ref{fig:curve}. The two solid lines are hypothetical trade-offs for two distinct file-graphs and simple repetition redundancy. Since coding in the form of redundant (non)linear combinations of chunks may lead to better results, two additional dashed lines are plotted to illustrate the potential gains. On each curve, two special points are highlighted in red, corresponding to the no-redundancy case, and blue, corresponding to the no fragmentation case.

%\begin{figure}[H]
%    \centering
%    \includegraphics[scale=0.27]{p3x.pdf}
%    \caption{An illustrative example of potential trade-off curves for storage redundancy versus fragmentation. The curves are file-graph specific \textcolor{red}{what is dashed, what is solid?}} \label{fig:curve}
%\end{figure}

\section{Problem Formulation}\label{sec:formulation}
%In our model, each file is a string comprising distinct data chunks labeled from the set $\{{x_{1},\ldots, x_{n}\}},$ where $n$ is the total number of unique data chunks across all files. As an example, $(x_{1},x_{2},x_{1},x_{2})$ is not an allowed file format. This modeling assumption is reasonable for most file types (such as movies and books), since users often remove their locally redundant data chunks (e.g., they self-deduplicate). 
%To store the files, the server only stores the data chunks labeled from $x_{1}$ to $x_{n}$ (each data chunk may be stored multiple times at different locations) in a sequential manner. To retrieve each file, one or multiple segments of consecutive data chunks stored on the server are accessed such that all data chunks associated with the file are read. 
%Note that the order in which data chunks are stored on the server is not necessarily the same as that of the data chunk string of a file (e.g., one can read a segment $(x_{1},x_{3},x_{2})$ on the server to recover the file $(x_{1},x_{2},x_{3})$).  
 
In what follows, we provide formal definitions of the file model, storage model, and metrics for measuring the fragmentation level. % We need to describe the file generation mechanism, the server model, and the criteria for assessing the fragmentation level of the files stored on the server. 
%\begin{definition}
    %\begin{enumerate}
    
\begin{figure}[H]
     \vspace{-0.1in}
        \centering
        \includegraphics[trim={0 3cm 0 4cm},clip,scale=0.26]{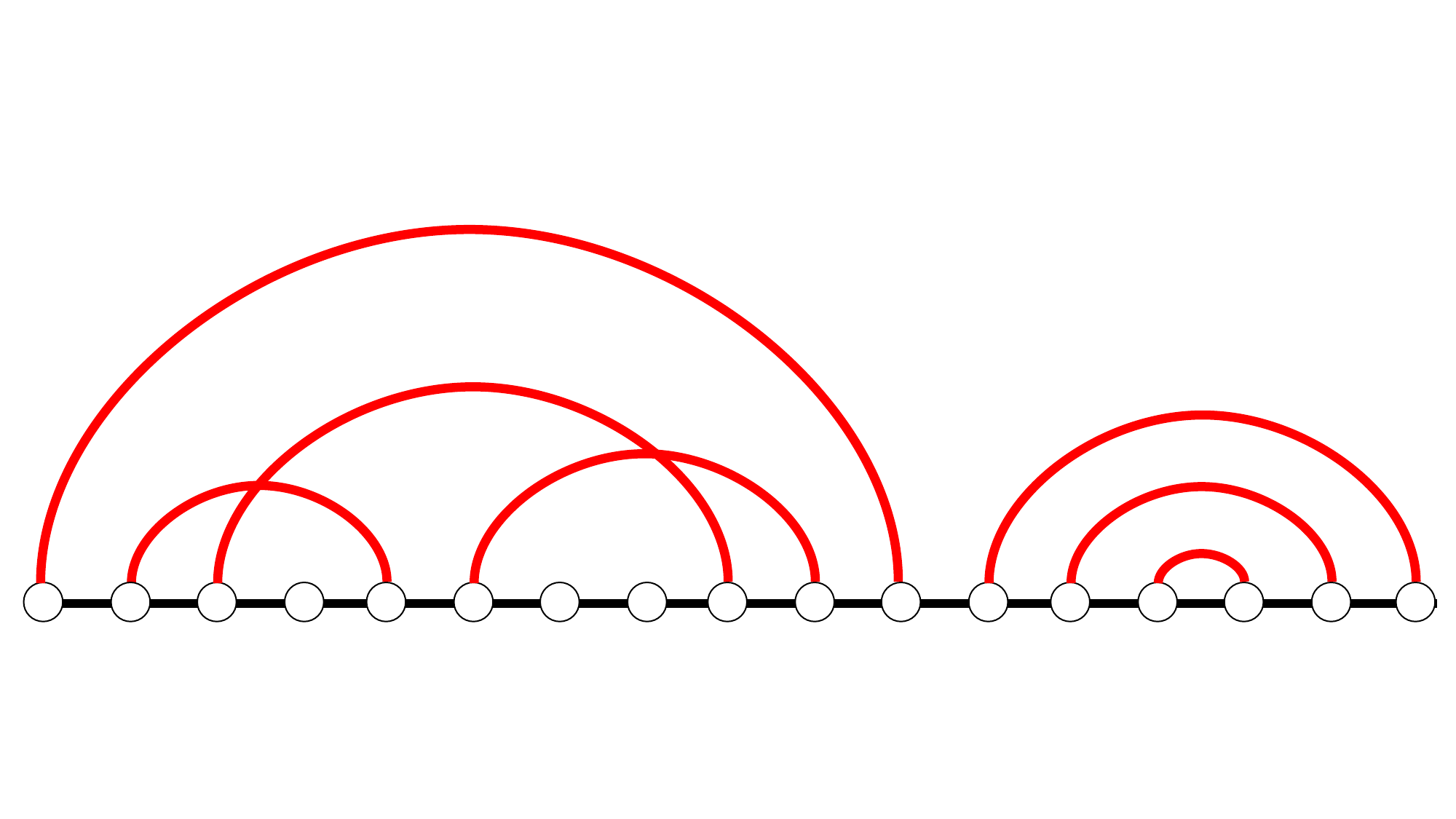}
        \vspace{-0.1in}
    \caption{Example of a sparse Hamiltonian graph, with arcs shown in red. Sparsity refers to the number of arcs scaling as $o(n)$.}
    \label{fig:sham}
\end{figure}

\begin{figure}[H]
     \vspace{-0.1in}
        \centering
        \includegraphics[trim={0 3cm 0 1cm},clip,scale=0.26]{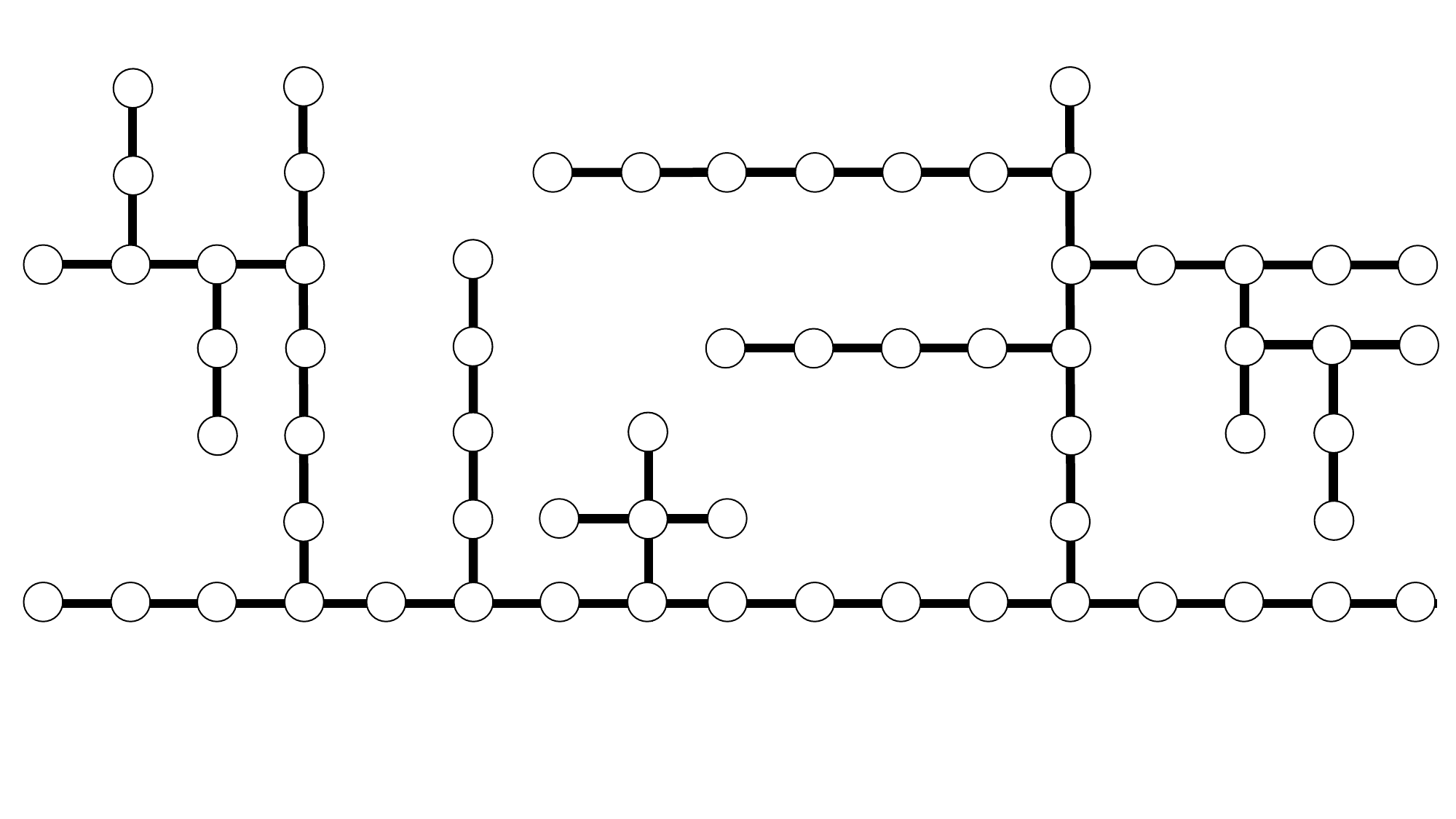}
        \vspace{-0.1in}
    \caption{Example of a tree file structure.}
    \label{fig:tree}
\end{figure}

\underline{File model.} The unique data chunks that constitute the file system are represented as vertices in an undirected graph
$G=(V,E)$, so that $V=\dc_{[n]}=\{\dc_{1},\ldots,\dc_{n}\}$ and $\dc_{i} \neq \dc_{j},$ whenever $i \neq j$. For simplicity, we use $\dc_{i}$ to represent both the node and the corresponding data chunk. Furthermore, we assume that all data chunks are of equal size of $d$ bits. Each file is a simple (nonbacktracking, self-avoiding) path $p=({\dc_{i_1}},\ldots,\dc_{i_{\ell(p)}}),$ where $\ell(p)$ stands for the number of data chunks of the file. Clearly, $(\dc_{i_{j}}, \dc_{i_{j+1}}) \in E,$ for all $j\in[\ell(p)-1]$ (where $[s]=\{{1,2,\ldots,s\}}$). 
The set of all files is denoted by \fm, comprising all simple paths $p$ of length $\ell(p)\leq t$, with $t\leq n$. Two classes of graphs of particular interest are \emph{sparse Hamiltonian graphs} and trees. A sparse Hamiltonian graph is a line graph with a set of $o(n)$ additional edges, to which we refer to with a slight abuse of terminology as \emph{arcs} (Figure~\ref{fig:sham}). Sparse Hamiltonian graphs capture the fact that some files (such as videos) are read by skipping parts in the middle or represent edited versions of a master file with certain chunks removed. 
Trees, on the other hand, capture hierarchical data structures where initial portions of file may agree but then diverge at some chunk index  (Figure~\ref{fig:tree}).

%A sparse Hamiltonian graph consists of a Hamilton path that passes through every node in $V$ once, i.e., a permutation of nodes in $V$, $(v_{i_1},\ldots,v_{i_n})$, such that $v_{i_j}$ and $v_{i_{j+1}}$ are connected for $j\in[n-1]$, and a set of additional edges, which we refer to as \emph{arcs}, that connect the nodes in $V$. The sparse Hamiltonian graphs capture the fact that some files (such as videos) are read by skipping parts in the middle. Trees capture the hierarchical data.

%A caterpillar graph is a tree that consists of is a path (not Hamiltonian) $p_0=(v_{i_1},\ldots,v_{i_u})$ and additional paths $p_1,\ldots,p_w$, $w\le u$, such that each path $p_j$, $j\in[w]$ has a starting point $v\in\{v_{i_1},\ldots,v_{i_w}\}$ and that no two paths $p_{j_1}$ and $p_{j_2}$, $j_1,j_2\in[w]$ share the same node.

\underline{Storage model.} The server is assumed to store chunks in linear order. When there is no redundancy, the order in which the server stores the data chunks can be described by a permutation of nodes in $V$. The ultimate goal is to minimize a properly defined fragmentation metric either over all chunk store sequences $\cs$ that are permutations of the $n$ unique chunks (for the case of no redundancy); or, sequences $\cs$ that contain all $n$ unique chunks, in addition to ${m-n}$ repeated chunks (for chunk stores that contain repetitions only); or, sequences $\cs$ that include linear combinations of data chunks in $V$ with possible repetitions and allow exact recovery of the files from linear combinations of their entries (for the case of coded redundancy). %Note that every coded chunk store plan has to have the additional property that it allows for exact reconstruction of all the user files based on linear combinations of chunks in $s$. 
Here, a linear combination of chunks equals the bitwise XOR of their content. To reduce the notational burden, we use the same notation $\cs$ for nonredundant, redundant and coded chunk stores of length $m \geq n,$ whenever clear from the context. %Each node $v_i$  at location $j$ in $s$ \iscomment{A bit confusing with the $S_V$ notation below} represents a copy of data chunk with label $i$ stored as the $j$th data chunk on the server.  
%Since we consider linear codes, we use a subset $\{v_{i_1},\ldots,v_{i_u}\}\subseteq V$ to denote a coded data chunk obtained by bitwise XOR of all the systematic data chunks $x_{i_1},\ldots,x_{i_u}$ contained in data chunks with labels $i_1,\ldots,i_u$. Let $S_V$ be the set of all subsets of $V$. As a result, we define a \emph{coded chunk store plan} for $m$ data chunks as a string $s\in S^m_V$ of sets of length $m$, where each set represents a bitwise XOR of all data chunks in the set. 
%A \emph{coded chunk store plan} for $m$ data chunks as a string $S=s_{1},s_{2},\hdots,s_{m}$ is specified by the generator matrix $G\in\{0,1\}^{n\times m}$ with $s_{j}=\sum_{i=1}^{n}G(i,j)x_{i}$. Where sum is in the bitwise fashion. 

\underline{Fragmentation metrics.} We introduce several metrics that capture the level of fragmentation, depending on how files are retrieved from the server and which aspects of file storage or retrieval are of interest. First, we define the \emph{stretch metric}, relevant for systems in which repositioning the read head frequently is undesirable and costly. In this setting, one should be able to retrieve each file from one substring of $\cs$ that contains sufficient information to reconstruct all data chunks in the file. More precisely, the stretch number of a file $p$ and chunk store $\cs$ is denoted by $\MinS(\cs,p)$, and it equals the ratio of the length of the shortest substring of $\cs$ that allows for recovering all chunks in $p$ and the total file length $\ell(p)$. For example, if $\cs= (\dc_{5},\dc_{1},\dc_{2},\dc_{6},\dc_{3},\dc_{4},\dc_{4},\dc_{1})$, and $p=(\dc_{1},\dc_{3},\dc_{4})$, then $\text{MinS}(\cs,p)=\frac{4}{3}$. For another choice of the store which involves both repeated and coded chunks, $\cs=(\dc_{5},\dc_{1},\dc_{2},\dc_{6},\dc_{3},\dc_{3}\oplus \dc_{4},\dc_{4},\dc_{1})$, we have $\text{MinS}(\cs,p)=1$.

%If such interval doesn't exist, $MinInterval(s,w)=\infty$
%\item $W(G,\leq k)$ denote the set of all walk on graph $G$ with length $\leq k$
%\item Let $s\in V^{m}$ be any length-$m$ string on $V$. $MinInterval(s,w)$ is the smallest integer $a$ such that there exists $s_{i_{1}},\hdots,s_{i_{\ell(w)+1}}$ with $s_{i_{j}}=w_{j}$ and $|i_{j}-i_{j+1}|\leq a$. If such $s_{i_{1}},\hdots,s_{i_{\ell(w)+1}}$ doesn't exist, $MinInterval(s,w)=\infty$
    %\end{enumerate}
%\underline{File model}: Our file model is a Markov chain-type model. Given a graph G=(V,E) where V is the set of chunks. $W(G,\leq k)$ denote all length $\leq k$ walk on $G$ is the set of all possible files.    
%\end{definition}
%We are interested in finding an uncoded chunk store $s\in V^m$ or coded chunk store plan $s\in S^m_V$ that minimizes the stretch number in worst cases under the file model described by $G=(V,E)$ and $P(G,\le t)$. 
The \emph{stretch metric} of a file system $P(G,\leq t)$ is defined as
\begin{align*}
    S_t(G,m)&:=\min_{\cs}\max_{p\in P(G,\leq t)}\MinS(\cs,p),
    %CS_t(G,m)&:=\min_{s\in S^m_V}\max_{p\in P(G,\leq t)}\MinS(s,p).%\\
    %g_S(G,m,t)&:=\frac{S_t(G,m)}{CS_t(G,m)}
\end{align*}
where $\cs$ is constrained to be either a nonredundant store, or a store with repetitions or a coded chunk store, of length $m$. Whenever $m=n$, we are focusing on the fragmentation metric for nonredundant stores which are either permutations of the $n$ unique chunks or $n$ coded chunks. Whenever $m>n$, we are focusing on the fragmentation metric for stores that include redundant chunks, which can be repeated or coded chunks. To avoid confusion, we will add the superscript $u$ or $c$ to $S_t(G,m)$ to indicate that the store is uncoded or coded, respectively. Also, with a slight abuse of terminology, for a \emph{fixed} chunk store $s$ we will use the words \emph{stretch metric} to describe the maximum stretch number across files in the file system.

When the file retrieval process based on reading substrings of the chunk store is too costly due to unnecessary chunk readouts and parsings, one may opt instead to retrieve multiple substrings that jointly cover the file chunks via readout head repositioning.  In such a setting, the fragmentation metric of interest is the largest number of ``jumps'' the readout head has to make. Specifically, the \emph{jump metric} $\MinJ(\cs,p)$ for a file $p\in P(G,\le t)$ under a chunk store $\cs$ is the minimum number of substrings of $\cs$ whose union of chunks allows for reconstructing all the $\ell(p)$ unique chunks of file $p$ \emph{without including chunks that do not belong to the file and without exceeding a total number of chunks equal to $\ell(p)$}. For example, if the chunk store equals $\cs= (\dc_{5},\dc_{1},\dc_{2},\dc_{6},\dc_{3},\dc_{4})$, and the file to be retrieved is $p=(\dc_{1},\dc_{3},\dc_{4})$, then $\text{MinJ}(\cs,p)=2$. Furthermore, for the same file and $\cs=(\dc_{5},\dc_{1},\dc_{2},\dc_{6},\dc_{3},\dc_{3}\oplus \dc_{4},\dc_{4},\dc_{1})$, we have $\text{MinJ}(\cs,p)=1$ since we only need to access one substring $(\dc_{3}\oplus \dc_{4},\dc_{4},\dc_{1})$. However for the file $p=(\dc_{1},\dc_{3},\dc_{4},\dc_{6})$ and the same chunk store as immediately above, $\text{MinJ}(\cs,q)=2$ since the fragment $(\dc_{6},\dc_{3},\dc_{3}\oplus \dc_{4},\dc_{4},\dc_{1})$ contains $5>\ell(q)+1=4$ chunks and we hence need to retrieve two substrings, $(\dc_6,\dc_3)$ and $(\dc_4,\dc_1)$. The \emph{jump metric} of a file system $P(G,\leq t)$ is defined as
\begin{align*}
    J_t(G,m)&:=\min_{\cs}\max_{p\in P(G,\leq t)}\MinJ(\cs,p),
\end{align*}
where $\cs$ is constrained to be either a nonredundant store, or a store with repetitions or a coded chunk store, of length $m$. Whenever $m=n$, our focus is on the jump metric for nonredundant stores which are either permutations of the $n$ unique chunks or $n$ coded chunks. Whenever $m>n$, our focus is on the jump metric for stores that include redundant chunks which are repeated or coded chunks. To avoid confusion, we will add the superscript $u$ or $c$ to $J_t(G,m)$ to indicate that the store is uncoded or coded, respectively. 

The following variants of the above two fragmentation metrics are also of practical relevance, but left for future studies. The first variant involves replacing the worst-case fragmentation level with the average fragmentation level across files. For example, the \emph{average-case stretch metric} and \emph{average-case jump metric} take the form:
\begin{align*}
  \bar{S}_t(G,m)&:=\min_{\cs}\sum_{p\in P(G,\leq t)}\MinS(\cs,p)\\
    \bar{J}_t(G,m)&:=\min_{\cs}\sum_{p\in P(G,\leq t)}\MinJ(\cs,p)
\end{align*}
The second variant represents a weighted linear combination of the stretch and jump metric. For inline deduplication, a more relevant measure of fragmentation involves placing unique chunks from the same file in consecutive storage locations, following the order dictated by the file itself. In this case, the set of allowed storage plans $\cs$ is constrained by the data file system itself. Clearly, other constraints on $\cs$ may be considered as well.

It is worth pointing out that our definitions of fragmentation metrics are related to previously known concepts from graph theory. For example, the uncoded stretch metric $S_{t}(G,m)$ is closely related to the `bandwidth problem' for graphs~\cite{diaz2002survey,petit2013addenda}. Formally, we are given a graph $G=(V,E)$, and would like to find a permutation $\sigma^{*}$ in $\mathbb{S}_n,$ the symmetric group of order $n!$, that minimizes the worst-case edge stretch (also known as the bandwidth), defined as $B(G):=\min_{\sigma}\max_{uv\in E}|\sigma(u)-\sigma(v)|.$
%\end
Similarly, the average stretch metric is related to minimum linear arrangement for graphs~\cite{diaz2002survey}. 
See Section~\ref{sec_no_redundancy} for a detailed discussion.
On the other hand, to the best of our knowledge, there are no known closely related concepts in graph theory to the jump metric and average jump metric.

%\section{Related Works}\label{sec:relatedwork}
%\input{\srcpath sec_relatedwork.tex}

\section{Summary of Results}\label{sec:summary}
\begin{figure}[H]
    \centering  
    \vspace{-0.1in}
    \includegraphics[scale=0.28]{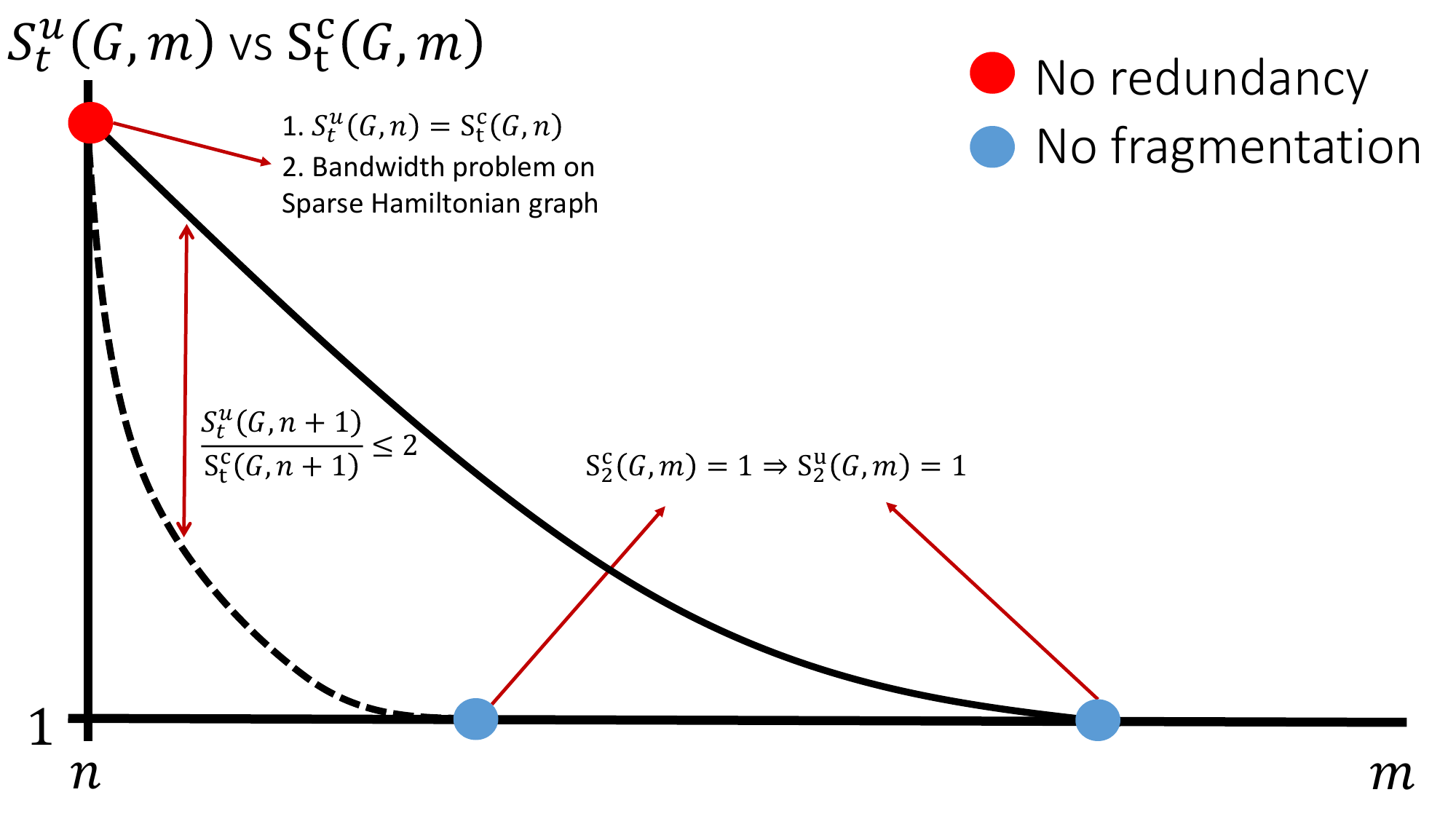}
    \caption{A possible pair of trade-off curves for the uncoded stretch metric $S^{u}_{t}(G,m)$ (solid line) and coded stretch metric $S^{c}_{t}(G,m)$ (dashed line) versus the chunk store size $m$. The two extreme ``corner points'' correspond to the case of no redundancy and no fragmentation.} \label{fig:summary_S}
\end{figure}

\begin{figure}[H]
    \centering    \includegraphics[scale=0.28]{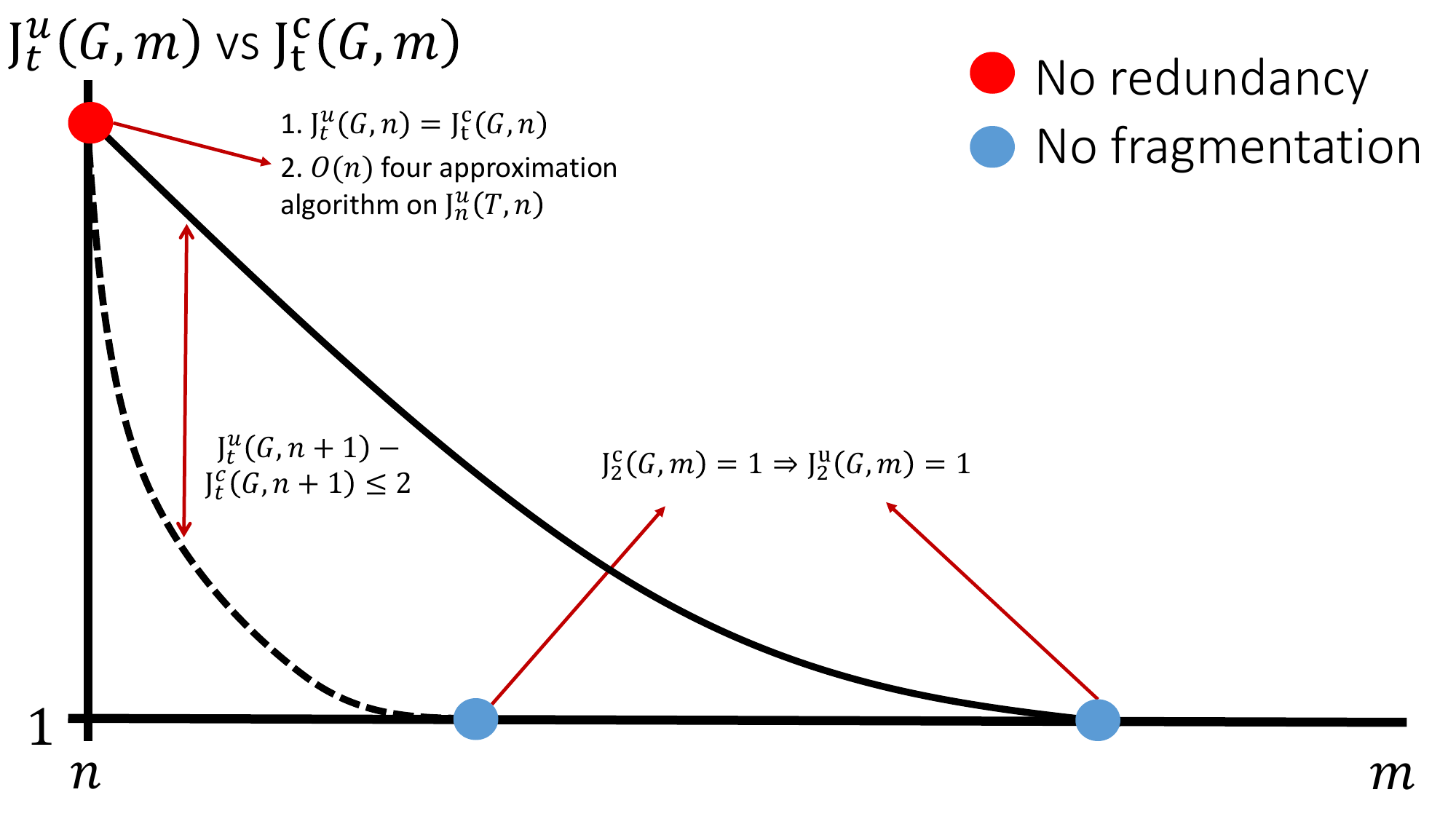}
    \caption{Trade-off curves for the uncoded stretch metric $J^{u}_{t}(G,m)$ (solid line) and coded stretch metric $J^{c}_{t}(G,m)$ (dashed line) versus the chunk store size $m$. The two extreme ``corner points'' correspond to the case of no redundancy and no fragmentation.} \label{fig:summary_J}
\end{figure}

Our results include both new bounds on the stretch metric $S^{u}_{t}(G,m)$, jump metric $J^{u}_{t}(G,m)$, and their coded counterparts $S^{c}_{t}(G,m),J^{c}_{t}(G,m),$ for the class of sparse Hamiltonian and tree-structured file-graphs, as well as algorithmic approaches for  constructing the resulting chunk stores. Most of the results hold for the special parameter values $t=2$ and $m=n+1$. These limitations are a consequence of the difficulty of the underlying graph-theoretic problems and the requirement to keep the number of redundant data chunks as small as possible, respectively. Notable general results that pertain to arbitrary values of $t$ and graphs $G$ establish that when no redundancy is allowed (i.e., when $m=n$), coding cannot reduce neither fragmentation metric. When one redundant chunk is allowed (i.e., $m=n+1$), coding can reduce the stretch metric by a factor of at most two, and the jump metric by an additive term of at most two.
%By definitions, all of them are non-increasing function of $m$ that depends on the specific file model \fm. Our results are two parts. Part one is \emph{how much can coding help compared to repetition?} 
Illustration of the results are provided in Figures~\ref{fig:summary_S} and~\ref{fig:summary_J}.

In summary, for the stretch metric and all possible file models \fm, we have the following results.
\begin{enumerate}
    \item We prove in Theorem~\ref{no_redundancy} of Section~\ref{subsect:no_re} that $S^{u}_{t}(G,n)=S^{c}_{t}(G,n)$, which establishes that coding cannot help when no  redundant chunks are allowed (and therefore, the two red points in the previous figures coincide). Moreover, since $S^{u}_{t}(G,n)$ is closely related to the classical graph bandwidth problem difficult to study for general graphs, we focus on bounding the stretch metric for sparse Hamiltonian graphs in Section~\ref{sec_no_redundancy}.
    \item We prove in Theorem~\ref{1bit_reduction} of Section~\ref{subsect:one_re} that $S^{u}_{t}(G,n+1)/S^{c}_{t}(G,n+1)\leq 2,$  which establishes that coding can at most halve the stretch metric with one redundant data chunk. We also present some explicit examples where coding reduces the stretch by a multiplicative factor of $2/3$.
    \item We prove in Theorem~\ref{no_fragmentation_coding} of Section~\ref{subsect:no_fra} that for $t=2$ coding cannot reduce fragmentation (i.e., the two blue points in the figures coincide). Moreover, Section~\ref{no_fragmentation}  establishes the connection between the problem in this setting and that of finding Eulerian paths in graphs. 
\end{enumerate}

For the jump metric and arbitrary file-graph models \fm, the results are similar to those presented for the stretch metric, as outlined below. 
\begin{enumerate}
    \item We prove in Theorem~\ref{no_redundancy_j} of Section~\ref{SvCS_j} that $J_{t}^u(G,n)=J_{t}^c(G,n)$, which establishes that coding cannot help when no  redundant chunks are allowed. Moreover, in Section~\ref{sec_no_redundancy_j} we establish bounds on $J_{t}(G,n)$ for trees $G$ and $t=n$ (i.e., for paths of all possible lengths in $G$), and the case when no redundant chunks are allowed. Our result is a $4$-approximation algorithm of complexity $O(n)$.
    \item In Section~\ref{SvCS_j}, Theorem~\ref{coding_gain}, we prove that for $m=n+1,$ $J^{u}_{t}(G,n+1)-J^{c}_{t}(G,n+1)\leq 2$, which implies that in this setting coding can reduce the jump metric by at most an  additive term equal to two. We also provide explicit examples of graphs for which coding can reduce the jump metric by one.
\end{enumerate}

\section{The Stretch Metric: Results}\label{sec:stretch}
We start by analyzing the uncoded stretch metric. It is important to note that by definition,
$S^{u}_{t}(G,m)$ is a nonincreasing function of $m$. 
%Therefore we can expect the trade-off curve for each $G$ as shown in Figure~\ref{fig:curve_detail}. 
Two cases to consider are when no redundancy is allowed ($m=n$) or when no fragmentation can be tolerated (in which case we seek for the smallest $m$ such that $S^{u}_{t}(G,m)=1$), which correspond to the two end points on the coding/redundancy-fragmentation trade-off curve as depicted in Fig.~\ref{fig:summary_J}. For the case of no redundancy, we start by analyzing sparse Hamiltonian graphs in Section~\ref{sec_no_redundancy}, and then proceed to analyze the same setting for general graphs in Section~\ref{no_fragmentation}. Determining intermediary points on the coding/redundancy-fragmentation trade-off curves in Figure \ref{fig:summary_J} is left for future work. We address the problem of whether coding helps reduce the stretch metric in Section~\ref{SvCS}.

%\begin{figure}[H]
%    \centering    \includegraphics[scale=0.28]{p3x2.pdf}
%    \caption{The trade-off curve $S^{u}_{t}(G,m)$ versus $m$ with two corner cases no redundancy and no fragmentation.} \label{fig:curve_detail}
%\end{figure}   

\subsection{Deduplication without Redundancy}\label{sec_no_redundancy}
The problem of computing the uncoded stretch metric when no redundancy is allowed, $S^{u}_{t}(G,n),$ is closely related to the extensively studied graph bandwidth problem~\cite{diaz2002survey,petit2013addenda} that may be succinctly described as follows. Given a graph $G=(V,E)$, find a permutation $\sigma^{*}$ in $\mathbb{S}_n,$ the symmetric group of order $n!$ that minimizes the worst-case edge stretch, defined as $B(G):=\min_{\sigma}\max_{uv\in E}|\sigma(u)-\sigma(v)|.$ The graph bandwidth problem is NP-hard for general graphs~\cite{diaz2002survey,petit2013addenda}. The best known results for general graphs with efficient placement algorithms ensure an approximation ratio of $O(\log^{3}(n)\sqrt{\log\log(n)})$ \cite{dunagan2001euclidean}. Tighter approximations results are known for certain subclasses of graphs including trees, grid graphs, interval graphs, and others~\cite{lai1999survey,diaz2002survey,petit2013addenda}. The fragmentation problem is also related to several related optimization problems on graphs, including the graph edgesum, profile width, pathwidth, cutwidth, cop and robber game, etc. The interested readers are referred to~\cite{lai1999survey,diaz2002survey,petit2013addenda} for comprehensive surveys of the problem domain. 

It is easy to prove that the bandwidth $B(G)$ represents a $2$-approximation for $S^{u}_{t}(G,n)$. Since the server chunk store $\cs$ represents a permutation of $n$ chunks, $\forall p\in P(G,\leq t)$ we have
\begin{align*}
    \MinS(\cs,p)&\leq \frac{1+\sum_{i=1}^{\ell(p)-1}|\cs^{-1}(p_{i})-\cs^{-1}(p_{i+1})|}{\ell(p)}\\
    &\leq \frac{1+(\ell(p)-1)B(G)}{\ell(p)}.    
\end{align*}
 
Furthermore, by the definition of the bandwidth, there exists an edge $\e{u}{v}\in E$ such that $|\cs^{-1}(\dc_{u})-\cs^{-1}(\dc_{v})|\geq B(G)$. 
%\textcolor{red}{Is our notation for edges consistent? Sometimes we have brackets, commas etc, sometimes not.}
Together, these results imply that $\frac{B(G)}{2} \leq S^{u}_{t}(G,n) < B(G)$. 
In particular when $t=2$, $\frac{B(G)}{2} = S^{u}_{2}(G,n)$.
Therefore, we need to optimize the bandwidth of sparse Hamiltonian graphs and trees. %Note that computing the bandwidth $B(G)$ for arbitrary graphs is computationally challenging, and existing general lower/upper bounds on $B(G)$ are mostly loose. This is why in this section we focus on trees and the sparse Hamiltonian graphs. 
For the latter case, many results are readily available from the literature~\cite{feige2009approximating,dubey2011hardness}. For example, it is NP-hard to approximate the bandwidth of trees within any constant $c \in \mathbb{N},$ even for a subclass of trees called `caterpillars', where every branching vertex (i.e., vertex of degree $>2$) is on the same line~\cite{feige2009approximating}. Moreover, the existence of a $c\sqrt{\frac{\log(n)}{\log\log(n)}}$-approximation, where $c$ is a constant, implies that the NP class admits quasipolynomial time algorithms. On the other hand, efficient $O\lp\frac{\log(n)}{\log\log(n)}\rp$-approximation algorithms for computing the bandwidth of caterpillars are readily available. 
Currently, no specialized results regarding the bandwidth of sparse Hamiltonian graphs exist, and we will show that the bandwidth can be bounded via a technique termed \emph{line folding}. Formal definitions used in our proof are provided next. 
%\textcolor{red}{Below needs fixing}
 \begin{definition}
    Graph $G=(V,\Eline\bigcup \Earc)$ is called a sparse Hamiltonian graph (Figure~\ref{fig:sham}) if 
    \begin{enumerate}
        \item $\Eline=\{\e{j}{j+1}:j\in[n-1],\dc_j\in V\}$ is a Hamiltonian path in $G$.
        \item $\Earc=\{\e{j^\ell_1}{j^\ell_2}:\ell\in [k],|j^\ell_{1}-j^\ell_2|>1\},$ where $j^\ell_1 \neq j^\ell_2$ for $\ell\in[k]$, is a set of edges (arcs) not in $\Eline$. We refer to  $\dc_{j^\ell_1},\dc_{j^\ell_2}$ as the ``feet of the arc'' and make the additional assumption that no two arcs share the same foot.
    \end{enumerate}
\end{definition}
%Let $j_{1}<j_{2}<\ldots<j_{2k}$ be all the $2k$ feet. Without loss of generality, we assume each $j_{i+1}-j_{i}$ is an even number. 
%Without specific clarification, in this subsection, $G$ is a sparse Hamiltonian graph. 
%To illustrate one of our key ideas for optimizing bandwidth in sparse Hamiltonian graphs, 
We find the following examples useful for our analysis of the folding method.
\begin{example}\label{cycle}[A cycle of odd length]
    Let $G=(V,\Eline\bigcup\{\e{1}{2n+1}\}),$ where $|V|=2n+1$.
    It is obvious that $B(G)>1$ since the cycle will have to be ``broken up'' somewhere in the linear chunk store. Folding the cycle at $n+1$ and interleaving the vertices using a zig-zag order, as shown in Figure~\ref{fig:cyclefolding}, results in the optimal bandwidth $B(G)=2$ since two vertices of the red edge are aligned and the fold has two layers, i.e. ``thickness'' equal to $2$ (this result matches the lower bound for the bandwidth in terms of the chromatic number of the graph, $B(G) \geq \chi -1,$ where $\chi$ is the chromatic number, i.e., the smallest number of colors that one can use to index the vertices while avoiding monochromatic neighbors). Note that some widely known upper bounds for the bandwidth of a graph, such as $B(G) \leq n-diam(G),$ where $diam(G)$ is the diameter of the graph $G$, defined as the length of the longest shortest path between pairs of vertices in $V$, are loose for cycles; the diameter bounds establishes that $B(G) \leq \frac{n}{2},$ while we know that the bandwidth is $=2$. 

    \begin{figure}[H]
        \centering
        \includegraphics[scale=0.25]{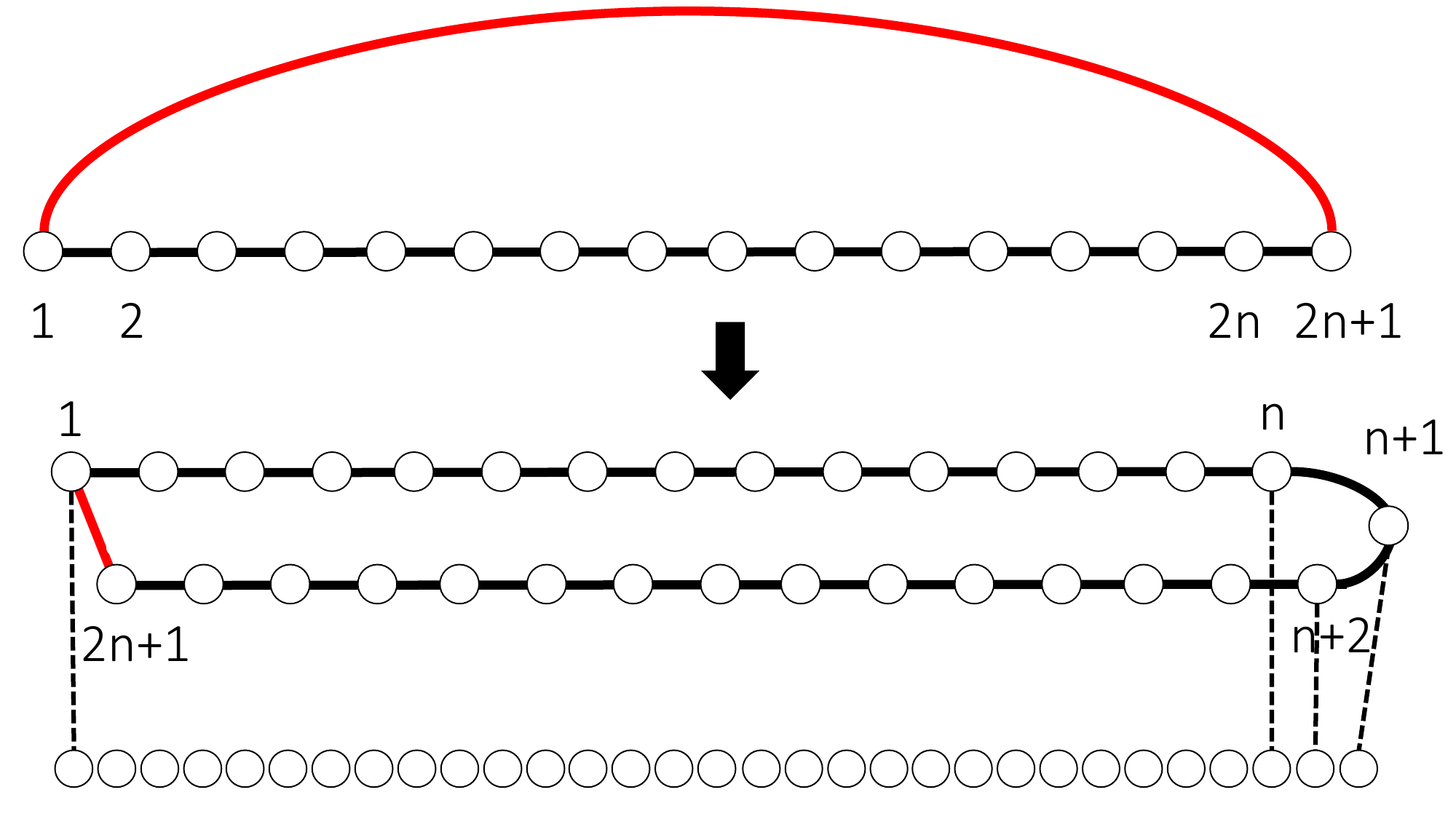}
        \caption{Folding a cycle of odd length and then writing the vertices in linear ordered dictated by a top-bottom (or vice versa) zig-zag curve results in a bandwidth $=2$. For simplicity, we only plot the indices of the data chunks.}
    \end{figure} \label{fig:cyclefolding}
\end{example}
%\end{tcolorbox}

\begin{example}\label{rainbow}[Nested sparse Hamiltonian graph]
    Let $G=(V,\Eline\bigcup \{\e{i}{2k+2-i}\}_{i=1}^{k})$ be a sparse Hamiltonian graph with $k$ arcs of the form $x_i x_{2k-i+2}, i=1,\ldots,k$, as shown in Figure~\ref{fig:nestedham}. Folding the first $2k+1$ vertices around the vertex $x_{k+1}$ offers a linearization with $B(G)=2,$ since every pair of vertices incident to a red arc is aligned and the ``thickness'' of the folding pattern $=2$.
    \begin{figure}[H]
        \centering
        \includegraphics[scale=0.25]{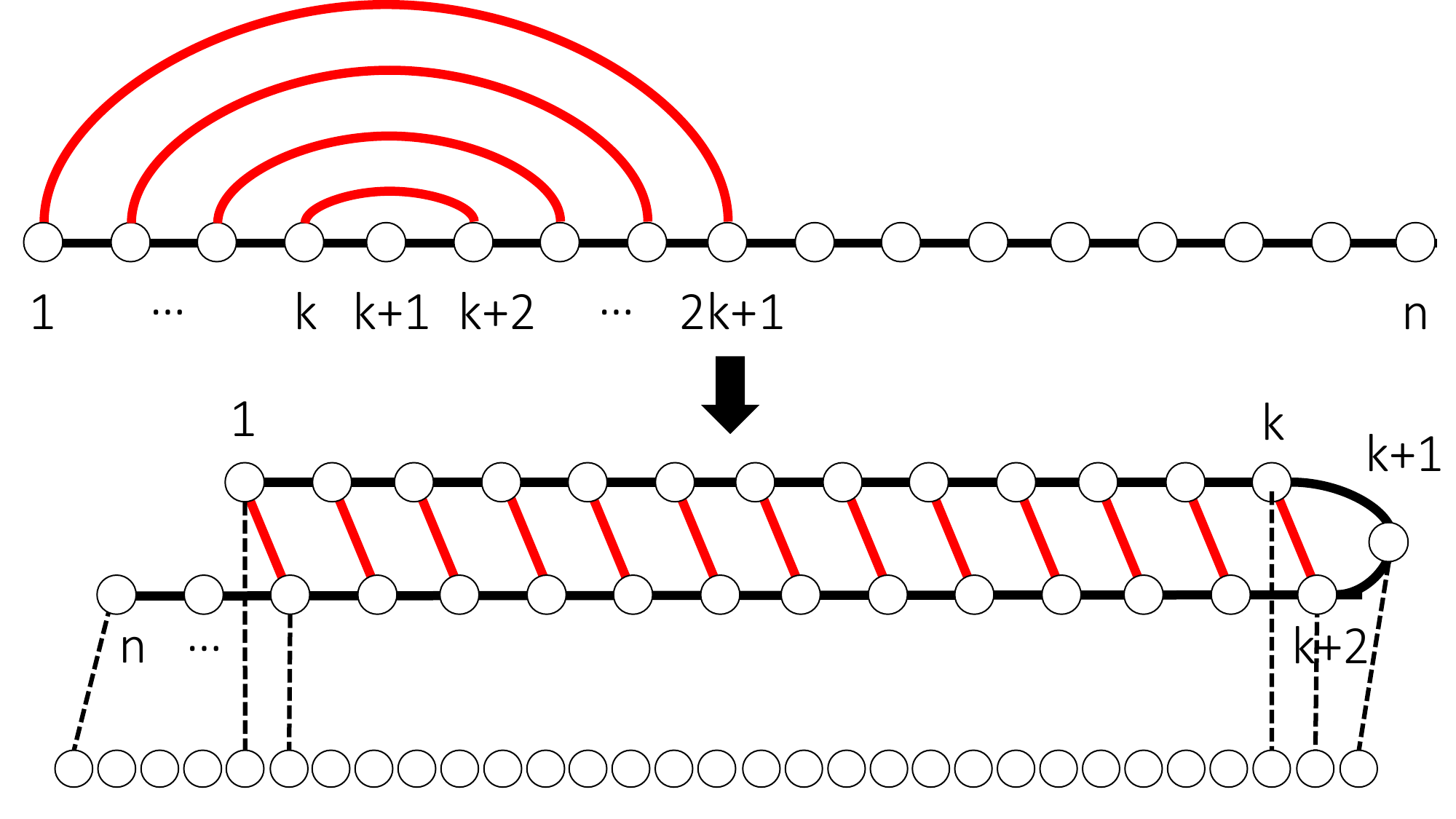}
        \caption{Folding a sparse Hamiltonian graph with nested arcs. Only the vertices incident to arcs are folded around the vertex labeled by $k$, while the arc-free vertices remain as a line. For simplicity, we only plot the indices of the data chunks.}
    \end{figure} \label{fig:nestedham}
\end{example} 

\begin{example}\label{multiply}
        [A long-arc sparse Hamiltonian graph] Let $G=(V,\Eline\bigcup \{\e{i}{\frac{in}{k}}\}_{i=1}^{k}),$ where $k=o(\sqrt{n})$ and $\frac{n}{k}$ is an integer. The graph is depicted in Figure~\ref{fig:thickfolding}. For this sparse Hamiltonian graph, unlike the previous two examples, we have $B(G)\geq |\Earc|=k$.
        \begin{figure}[H] 
         \vspace{-0.15in}
            \centering
            \includegraphics[trim={0 0cm 0 5cm},clip,scale=0.24]{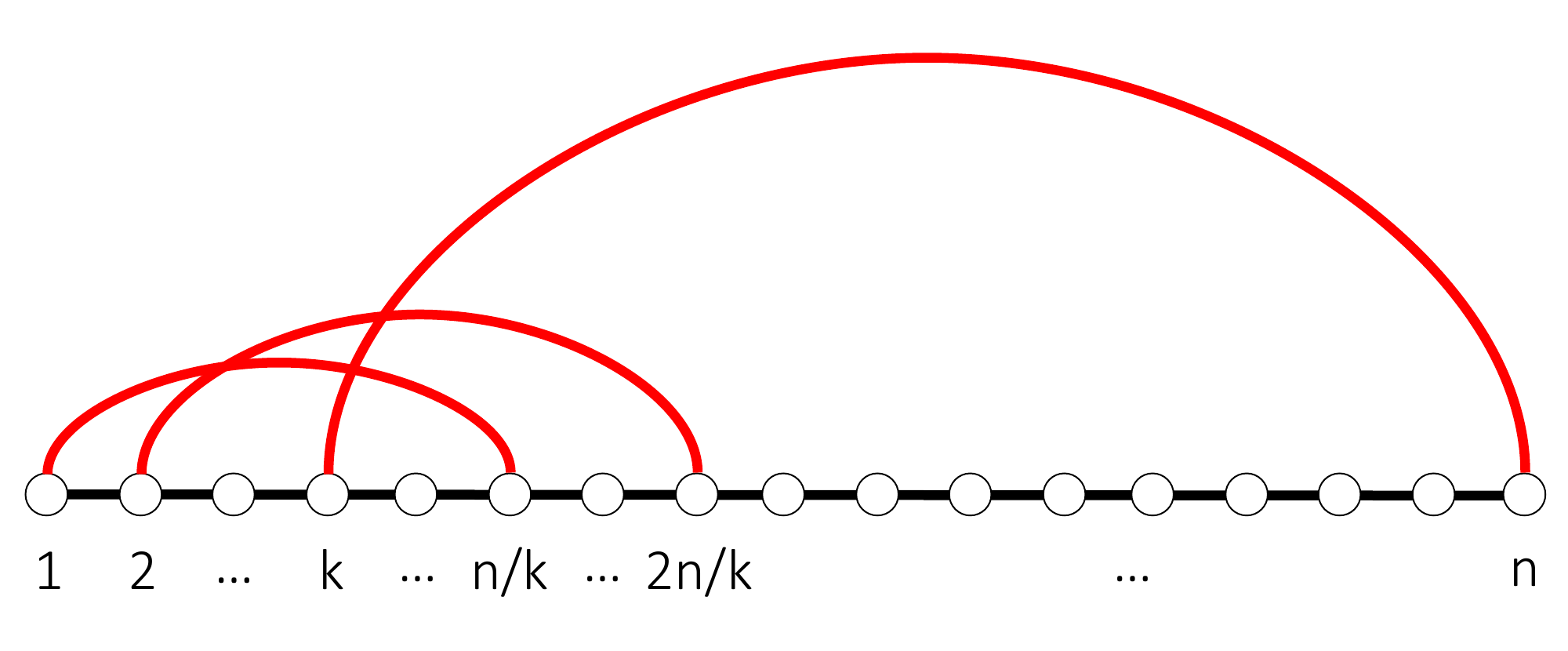} 
            \caption{A long-arc sparse Hamiltonian graph. For simplicity, we only plot the indices of the data chunks.} \label{fig:thickfolding}
        \end{figure}
        
        \begin{figure}[H]
         \vspace{-0.15in}
            \centering
            \includegraphics[trim={0 0cm 0 4cm},clip,scale=0.26]{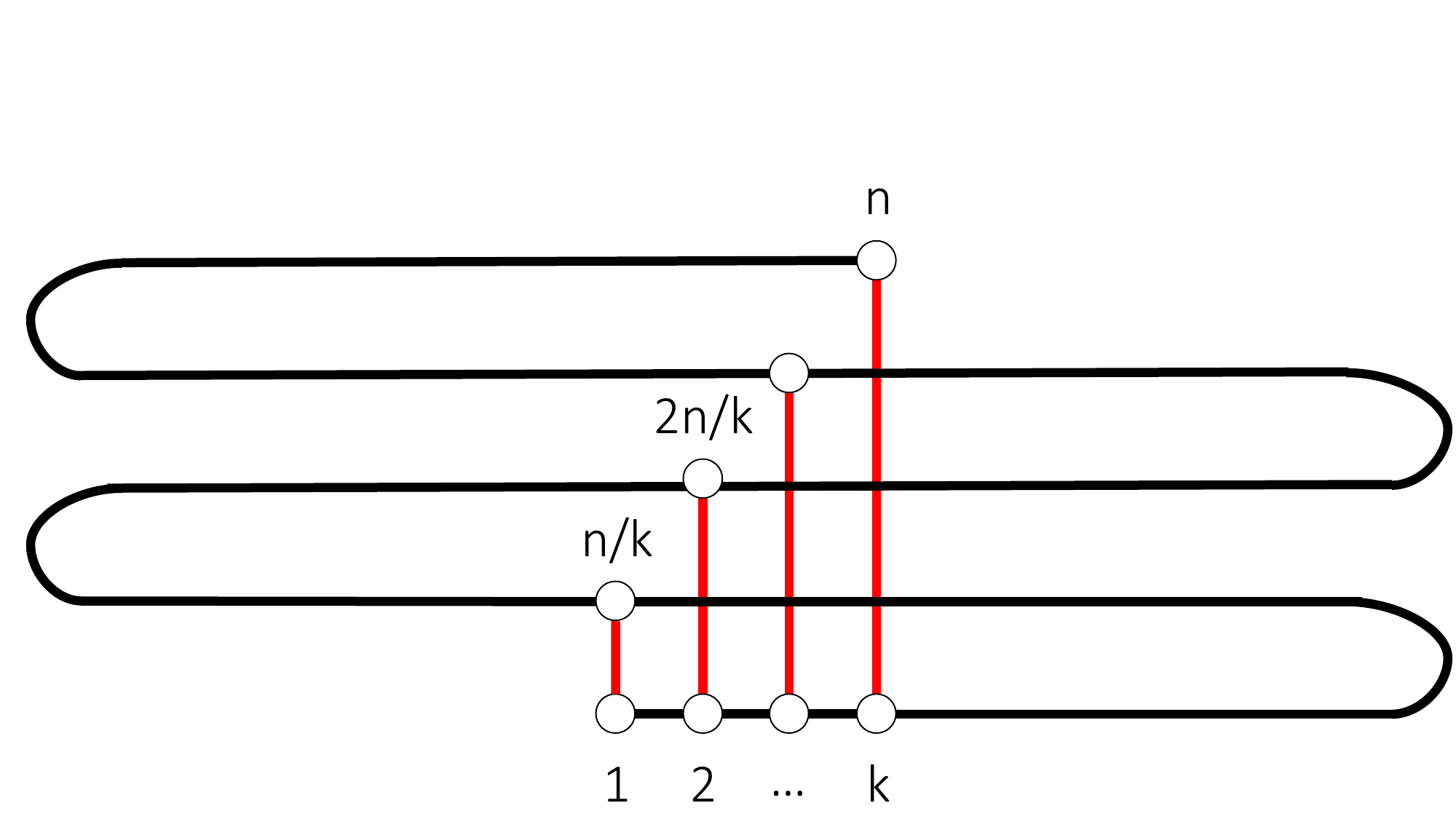}
            %\vspace{-0.39in}
            \caption{The folding scheme used for the long-arc sparse Hamiltonian graph, with a resulting thickness equal to $k+1$. For simplicity, we only plot the indices of the data chunks.} \label{fig:foldingk}
        \end{figure}
        The folding pattern used is depicted in Figure~\ref{fig:foldingk}. To show that $B(G)\geq |\Earc|$, we first note that all pairs of vertices $\dc_{i},\dc_{j},$ with $i,j\in [n],$ are within distance $n/k+k+1$ of each other in $G$. This follows since one can reach $\dc_{i}$ from $\dc_{j}$ (or vice versa) by first reaching the nearest vertex to $\dc_{j}$ with an index that is a multiple of $n/k$ using $\leq n/(2k)$ steps, then making one additional step to vertices with indices in $[k]$ via one of the arcs, then taking another $\leq k-1$ steps to another vertex in $[k]$, and repeating the trip back to $\dc_{i}$. The sum of the number of steps totals $\leq 2(n/(2k)+1)+k-1=n/k+k+1$.
        Let $\cs$ be one of the possible chunk stores that achieves the bandwidth $B(G)$. Then $\forall i,j\in [n], |\cs^{-1}(\dc_{i})-\cs^{-1}(\dc_{j})|\leq (n/k+k+1)B(G)$. But since there have to be two chunks at distance $n-1$ on $\cs$, we must have $B(G)\geq (n-1)/(n/k+k+1)$. Since $k=o(\sqrt{n})$, $B(G)/k\geq (n-1)/(n+k^{2}+k)\rightarrow 1$ as $n\rightarrow\infty$. %\textcolor{red}{We need to check which values of k are allowed. I wrote at several places that k=o(n), but is this true for every case/example we have? the asymptotic you mentioned holds for $k$ a constant, $\rightarrow k$ as $n\rightarrow\infty$.}
    \end{example}

The examples above showed that to construct an optimal or good chunk store via the bandwidth value, one needs to ``fold'' the Hamiltonian path so as to minimize the thickness of the folding, which is the number of linear line segments stacked on top of each other in the fold, but under the constraint that any pair of vertices in the arc set $\Earc$ is aligned (clearly, for some cases, this constraint may not be possible to impose). We make this high-level intuition more concrete as follows.

\textbf{Line folding.} A \emph{folding} of a sparse Hamiltonian graph $G=(V,\Eline\bigcup \Earc)$ with $|V|=n$ is a function $h: [n]\rightarrow [n]$ that has two properties, listed below:
\begin{enumerate}
    \item[\textbf{P1}] $h(i)-h(i+1)\in \{\pm 1\}$ for all $\e{i}{i+1}\in \Eline$;  
    \item[\textbf{P2}] $h(u)=h(v)$ for all $\e{u}{v}\in \Earc$.
\end{enumerate}    
The set of breakpoints of the folding are defined as %\textcolor{red}{[starts with $0$ or $1$? It starts with $1$ in Lemma 5.2]} \textcolor{blue}{changed to 1.} 
$1=d_0<d_1< d_2<\ldots<d_{L-1}<n=d_{L}$, where $h(d_{i})-h(d_{i}-1)=h(d_{i})-h(d_{i}+1),$ $\forall i \geq 1.$ Intuitively, $(d_i,h(d_i))$, $i\in\{0,\ldots,L\}$ are the ``extremal points'' of the piecewise linear function obtained by connecting $(i-1,h(i-1))$ and $(i,h(i))$ for $i\in [n]$. The thickness of a folding is defined as $T(h):=\max_{i\in[n]}|h^{-1}(i)|$, where for $i\in [n]$, let $h^{-1}(i)=\{u\in [n]:h(u)=i\}$. . An optimal folding has the minimum possible thickness across all foldings.
The idea related to graph folding also appeared in~\cite{feige2005approximating} (Definition 5) for general graphs under the name ``bucketwidth.'' Since we focus on sparse Hamiltonian graphs, our specialized definition allows us to enforce additional appropriate vertex alignment constraints and derive tighter bounds.
\begin{remark}
    The folding used in Example~\ref{cycle} is of the form $h(u)=u,$ for $u\leq n+1,$ and $h(u)=2(n+1)-u$ for $u>n+1$. The only breakpoint is $d_{1}=n+1$ and the thickness is $T(h)=2$.  
    %\textcolor{red}{you used h for folding in the definition and now switched to g. Why? Also, it illustrates the notion of breakpoint thickness through this function. By looking ahead, there are way too many g's to be corrected, so maybe switch h to g and keep g?}
\end{remark}
Note that if there exists $\e{u}{v}\in \Earc$ with $u-v$ being an odd number, then it is impossible to have a folding that satisfies \textbf{P1} and \textbf{P2}. To overcome this problem, for any sparse Hamiltonian graph $G=(V,\Eline\bigcup \Earc)$, let $G'=(V' ,\Eline' \bigcup \Earc' )$ be the modified sparse Hamiltonian graph obtained from $G$ by splitting each edge in $\Eline$ into two. More precisely, $V' =x_{[2n-1]}$, $\Eline' =\{\e{i}{i+1}\}_{i=1}^{2n-2}$, and $\Earc' =\{\e{2u-1}{2v-1}|\e{u}{v}\in\Earc\}$. Then 
\begin{equation*}
    h(i):=\begin{cases}1,&\text{ if }i\mod{2}=0,\\
                                2,&\text{ if }i\mod{2}=1,\end{cases}
\end{equation*} satisfies properties \textbf{P1} and \textbf{P2} and consequently, an optimal folding for $G'$ exists.
Then we have Lemma~\ref{folding_lemma}, which shows that any optimal folding plan for the modified graph $G'$ leads to a constant approximation of bandwidth of $G$.
%\textcolor{red}{I added some disclaimers above that we cannot always have a perfect folding with aligned points. Same here; we may not have the uniform difference result for all i. Please make sure to point out everywhere where needed that for some graphs, there may not be a perfect folding etc.}
%\textcolor{red}{the optimal folding plan for the modified graph of $G$ - clarify, this is hard to understand; and, is folding always going to produce the optimum?}
\begin{lemma}\label{folding_lemma}
     Let $h$ be any optimal 
     folding of the graph $G'=(V' ,\Eline' \bigcup \Earc' )$. Label the vertices of $G'$ according to $\cs:=h^{-1}(1),h^{-1}(2),\ldots,h^{-1}(n),$  and arrange vertices in the same group $h^{-1}(i)$ arbitrarily. Then, deleting all vertices $\{\dc_{i}:i\mod{2}=0\}$ and replacing each $\dc_{2i+1}$ by $\dc_{i}$ for all $i\in[n]$ in $\cs$ results in a linearization of $G=(V,\Eline\bigcup \Earc)$ with $\max_{\e{u}{v}\in E}|\cs^{-1}(\dc_{u})-\cs^{-1}(\dc_{v})|\leq 6B(G)$. 
\end{lemma}
\begin{IEEEproof}
    We first show there exists a folding $h$ for $G'$ with thickness $T(h)\leq 3B(G)$.
    Let $\cs$ be one chunk store order that achieves $B(G)$. Then, $|\cs^{-1}(\dc_{u})-\cs^{-1}(\dc_{v})|\leq B(G)$ for all $\e{u}{v}\in E$. Define $h$ for $G'$ as follows.
    Partition the vertex set $V$ (i.e., indices of the data chunks) into $\lceil\frac{n}{B(G)}\rceil$ sets $S_{1},S_{2},\ldots,S_{\lceil\frac{n}{B(G)}\rceil},$ with $S_{i}:=\{\cs(j)\}_{j=(i-1)B(G)+1}^{iB(G)}$ being the $i$th index set. Then the two vertices of each edge in $E$ must belong to the same $S_{i}$ or two sets $S_{i},S_{i+1}$ with adjacent labels. Furthermore, partition each $S_{i}$ into three sets, $L_{i},R_{i},M_{i},$ where $L_{i}\subset S_{i}$ is the set of vertices that are feet of arcs whose other foot is in $S_{i-1}$. %\textcolor{red}{this definition is not ok or precise; what about vertices that are not feet? can you even make it consistent by always placing the left foot in SL and the right foot in SR?}. 
    The set $R_{i}\subset S_{i}$ is the set of vertices that are feet of arcs whose other foot is in $S_{i+1}$. Note that $L_{i}\cap R_{i}=\emptyset$ since each foot belongs to one arc only. Finally $M_{i}=(S_{i}\backslash L_{i})\backslash R_{i}$. Next, define a new partition $S_{i}' $ according to
    \begin{equation*}
        S_{i}' :=\begin{cases}M_{1}\cup R_{1}\cup L_{2}\cup M_{2},&\text{ if }i=1\\
                                R_{2i-2}\cup L_{2i-1}\cup M_{2i-1} &\\
                                \cup R_{2i-1}\cup L_{2i}\cup M_{2i},&\text{ if }1<i\leq\lceil\frac{n}{2B(G)}\rceil\end{cases}
    \end{equation*}
    Then the two vertices incident to each edge in $E$ still belong to the same $S_{i}' $ or two adjacent sets $S_{i}' ,S_{i+1}' $. Moreover, the feet of each arc in $\Earc$ always belong to the same set  $S_{i}' $. The size of each of the above sets satisfies $|S_{i}' |\leq 3\max_{j}|S_{j}|=3B(G)$.
    
    Subsequently, define a function $f:[n]\rightarrow[n]$ such that $f(j)=i$ whenever $\dc_{j}\in S_{i}' $. Clearly, $f(u)=f(v)$ for all $\e{u}{v}\in \Earc$, and $f(u)-f(v)\in \{0,\pm 1\}$ for all $\e{u}{v}\in \Eline$. Define $h:[2n-1]\rightarrow [2n-1]$ based on $f$ as follows. Let $h(2i-1)=2f(i)-1$ for all $i \in [n]$, and 
    \begin{equation*}
        h(2i):=\begin{cases}\frac{h(2i-1)+h(2i+1)}{2},&\text{ if }h(2i-1)\neq h(2i+1),\\
                                h(2i-1)+1,&\text{ if }h(2i-1)= h(2i+1),\end{cases}
    \end{equation*}
    for all $i\in [n-1]$. Then $h$ is a folding for $G'$ that satisfies properties \textbf{P1}, \textbf{P2} and has thickness $T(h)\leq 3B(G)$.
    Let $\cs$ be labeling of groups of vertices of $G'$ according to $\cs:=h^{-1}(1),h^{-1}(2),\ldots,h^{-1}(2n-1),$ with vertices arranged as dictated by the labeling and vertices in the same group $h^{-1}(i)$ placed in arbitrary order. Then, $\max_{\e{u}{v}\in E' }|\cs^{-1}(\dc_{u})-\cs^{-1}(\dc_{v})|\leq 2T(h)\leq 6B(G)$.
    Since $G'$ is obtained from $G$ by splitting edges in $\Eline$ into two edges through the addition of a mid-vertex, after deleting all vertices $\{\dc_{i}:i\mod{2}=0\}$ and replacing each $\dc_{2i+1}$ by $\dc_{i}$ for all $i\in[n]$ in $\cs$, we have $\max_{\e{u}{v}\in E}|\cs^{-1}(\dc_{u})-\cs^{-1}(\dc_{v})|\leq 6B(G)$.
\end{IEEEproof}

Lemma~\ref{folding_lemma} shows that one can reduce the problem of finding or bounding the bandwidth of a sparse Hamiltonian graph $G$ to that of finding a folding for $G'$ with minimum thickness. The next result, presented in Lemma~\ref{lem:folding}, is a simple upper bound on the bandwidth based on folding arguments.
\begin{lemma}\label{lem:folding}
If there exists a piece-wise linear function $h$ with $L$ breakpoints $1=d_0<d_1< d_2<\ldots<d_{L-1}<n=d_{L}$ satisfying \textbf{P1}, \textbf{P2}, then $B(G)\le L$.
\end{lemma}
\begin{IEEEproof}
The key idea is to concatenate all the vertices whose integer indices $u$ have the same value $h(u)$ in increasing order of the value $h(u),$ $u \in [n]$. 
Let $\cs$ be the sequence of vertices obtained by concatenating $(I^{-1}(1),\ldots,I^{-1}(n))$, where $I^{-1}(y)$ stands for a linear arrangement of vertices in $h^{-1}(y)$ in increasing order.

We now show that $\max_{\e{u}{v}\in E}|\cs^{-1}(\dc_{u})-\cs^{-1}(\dc_{v})|\leq L$.
By \textbf{P1}, $h(u)=h(v)$ for all $\e{u}{v}\in \Earc$. Therefore $|\cs^{-1}(\dc_{u})-\cs^{-1}(\dc_{v})|<|h^{-1}(h(u))|\leq L$ since the function $h$ consists of $L$ linear segments.
By \textbf{P2}, $h(u)-h(u+1)\in\{\pm1\}$ for all $\e{u}{u+1}\in \Eline$. Without loss of generality, let $h(u)-h(u+1)=-1$. Then $|\cs^{-1}(\dc_{u})-\cs^{-1}(\dc_{u+1})|=|\{v\geq u|h(v)=h(u)\}|+|\{v\leq u+1|h(v)=h(u+1)\}|-1\leq L+1-1= L$. This follows from $h(u)\neq h(v)$ for all $d_{i-1}\leq u<v<d_{i}, i\in[L]$ (i.e., each linear segment contributes at most once).
\end{IEEEproof}
We have the following result based on Lemma~\ref{lem:folding}.
\begin{theorem}\label{lem:generalk}
For any sparse Hamiltonian graph $G=(V,\Eline\cup \Earc)$, we have $B(G)\le \lfloor \frac{9|\Earc|+1}{5} \rfloor$.
\end{theorem}
%\iscomment{Is the bound $\lfloor \frac{9|\Earc|+1}{5} \rfloor$ for $S_2^u$ of for $B(G) = 2 S_2^u$? The line after eq. (2) claims that we find a permutation with bandwidth 5, not stretch 5.}
\begin{IEEEproof}
    The detailed proof 
    %of Theorem~\ref{lem:generalk} 
    %for general sparse Hamiltonian graphs and the corresponding algorithms 
    is available in  Appendix~\ref{pf:generalk}.
\end{IEEEproof}
% \iscomment{This structure here is a bit confusing. Should this paragraph be inside a proof? 
% Isn't Proposition 5.1 just Theorem 5.1 for $|\Earc|=3$?
% In that case, shouldn't the next paragraph (starting with ``Let $x_ax_{b_1}$...'' be moved into the proof of Proposition 5.1?}
The idea for the proof of Theorem \ref{lem:generalk} is to construct a function $h$ that satisfies properties \textbf{P1} and  \textbf{P2}.
For simplicity, we illustrate our key ideas through the following proposition on a sparse Hamiltonian graph with $3$ arcs,
    for which $\lfloor \frac{9|\Earc|+1}{5} \rfloor=5$. 
\begin{proposition}\label{prop:k3}
For a sparse Hamiltonian graph $G=(V,\Eline\cup \Earc)$ where $|\Earc|=3$, we have $B(G)\le 5$.
\end{proposition}
\begin{IEEEproof}
%Let $\Earc=\{\e{a_1}{b_1},\e{a_2}{b_2},\e{a_3}{b_3}\}$ and
%Our proof of Proposition \ref{prop:k3} can be generalized to give an upper bound on $S^{u}_{2}(G,n)$ for arbitrary values of $|\Earc|$. Let $\Earc=\{(x_{a_i},x_{b_i}):i\in\{1,\ldots,|\Earc|\}\}$ where $b_i>a_i$ for $i\in [|\Earc|]$.  %To illustrate the key ideas of our algorithm and results, we consider a sparse Hamiltonian graph with $3$ arcs. 
Let $\e{a_1}{b_1}$, $\e{a_2}{b_2}$, and $\e{a_3}{b_3}$ (assuming $b_i>a_i$ for $i\in[3]$) be the arcs in $\Earc$. It is tacitly assumed that $a_i$, $b_i$, $i\in[3]$, are even. Otherwise, if some of the $a_i$, $b_i$, $i\in[3]$, are odd, it is always possible to create an augmented graph $G',$ as in the proof of Lemma~\ref{folding_lemma}, which does not have a smaller bandwidth nor an increased number of arcs, and hence also satisfies the upper bound $\lfloor \frac{9|\Earc|+1}{5} \rfloor$ for the bandwidth.
To construct the function $h$ satisfying the constraints in Lemma \ref{lem:folding}, we define functions $h$ for which there exist $L$ distinct even integers $1\le r_1<\ldots <r_L\le n,$ satisfying $h(r_i) = h(r_j)$, for $i,j \in [L]$. In addition, $h$ satisfies folding property $\textbf{P1}$, with the break points set to $d_1,\ldots,d_{L-1},$  
\begin{align}\label{eq:diri}
d_i = \frac{r_i+r_{i+1}}{2}
\end{align}
for $i\in\{1,\ldots,L-1\}$. Note that the $d_i$s, $i\in[L-1]$, are integers since the values $r_i$, $i\in[L]$, are even. Moreover, the choice of $d_i$ stated in~\eqref{eq:diri} is required for foldings $h$ that have thickness $L$.
% \iscomment{Are you defining $d_i$s in this way? Or are you saying that, if $h$ is a folding, the breakpoints must satisfy this?
% From the text above (``we start with functions $h$...'') it's not clear that this $h$ is a folding}
%In addition, we require that 
%\begin{align}\label{eq:aibieq}
%h(a_i)=h(b_i)    
%\end{align}
%for any $i\in\{1,\ldots,3\}$. 
Let integer $a_i$ belong to the $j^1_i$th linear piece of $h$
% , i.e., $d_{j^1_i-1}\le a_i<d_{j^1_i}$ 
and let integer $b_i$ belong to the $j^2_i$th linear piece of $h$, i.e., $d_{j^1_i-1}\le  a_i<d_{j^1_i}$ and $d_{j^2_i-1}\le  b_i<d_{j^2_i}$. Then, from $h(a_i)=h(b_i)$, we have that
\begin{align}\label{eq:decidingr}
    r_{j^1_i}-a_i=\begin{cases}r_{j^2_i}-b_i, & \text{if} j^1_i\equiv j^2_i\bmod 2,\\
    b_i-r_{j^2_i}, & \text{if} j^1_i\not\equiv j^2_i\bmod 2.
    \end{cases}
\end{align}
Next, we construct a function $h$ with $L=5$ for $|\Earc|=3$, which results in a storage (i.e., a permutation) of bandwidth $5$ according to Lemma~\ref{lem:folding}.

Rearrange $\{a_1,a_2,a_3,b_1,b_2,b_3\}$ into $\{c_1,\ldots,c_6\}$ so  that $c_1<\ldots<c_6$, i.e., the sequence is increasing. Let $$\{c^{\min}_{1},c^{\min}_{2}\}=\arg\min_{i,j\in\{1,\ldots,6\},i\ne j,\{c_i,c_j\}\notin \Earc}|c_i-c_j|,$$ 
i.e., let $(\dc_{c^{\min}_1},\dc_{c^{\min}_2})$ be  the ``closest'' pair of vertices that do not constitute an arc in $\Earc$. Then, we must have $c^{\min}_{1}=c_{i^*}$ and $c^{\min}_2=c_{i^*+1}$ for some $i^*\in [5].$ %\textcolor{red}{completely confused with these indices, - does this mean that the star indices just lie in {1,2,3,4,5|?}?}. %\textcolor{blue}{[$i^*$ is just a notation for the indices. To make it not confused with the star notation of $c^*_1$, the notations $c^*_1$ and $c^*_2$ are changed to $c^{\min}_1$ and $c^{\min}_2$ respectively]} 
We construct $h$ using the following two-step approach. 

\textbf{Step 1:} Partition the integers $c_1,\ldots,c_6$ into $\{c_1\},\ldots,\{c_{i^*-1}\},\{c_{i^*},c_{i^*+1}\},\{c_{i^*+2}\},\ldots,\{c_6\}$, i.e., place $c_{i^*}$ and $c_{i^*+1}$ in the same set and create singleton sets for all remaining elements in $\{c_1,\ldots,c_6\}.$ %\textcolor{red}{this does not reflect the true set membership since star-indexed c's are missing? or based on the comment above, you really have only five sets, one with 2 and four with singletons? Please answer my questions but *do not delete comments* so that I can track changes}. 
%\textcolor{blue}{[The c-* notations are replaced. $c^*_1$ was supposed to be one of $c_1,\ldots,c_6$, denoted as $c_{i^*}$. Here we have five sets. One with 2 and four singletons.]} 
The goal is to construct $h$ such that the entries in each subset belong to the same linear piece, i.e.,
\begin{align}\label{eq:group6}
&d_{i-1}\le c_i<d_i, \text{ for }i\in\{1,\ldots, i^*-1\},\nonumber\\
&d_{i^*-1}\le c_{i^*},c_{i^*+1}<d_{i^*},\nonumber\\
&d_{j-1}\le c_{j+1}<d_j,\text{ for }j\in\{ i^*+1,\ldots,5\}.
\end{align}

\textbf{Step 2:} Since there are $5$ subsets that correspond to $5$ linear pieces in Step $1,$ we need to find $5$ values $r_1,\ldots,r_5$ that satisfy~\eqref{eq:diri}, ~\eqref{eq:decidingr}, and ~\eqref{eq:group6}. Then a function $h$ that satisfies  \textbf{P1} and \textbf{P2}, and is such that $h(a_i)=h(b_i)$ for $i\in\{1,\ldots,3\}$ can be constructed based on the values of $r_1,\ldots,r_5$ and the values of $d_1,\ldots,d_4$ determined by~\eqref{eq:diri}. According to Lemma~\ref{lem:folding}, the function $h$ results in a store $s$ that has bandwidth equal to $5$.
%\iscomment{Lemma 5.2 is for $B(G)$ not stretch metric}

Therefore, all that is left to do is to find $r_1,\ldots,r_5$ satisfying~\eqref{eq:diri},~\eqref{eq:decidingr}, and~\eqref{eq:group6}. 
Without loss of generality, assume that $c_{i^*}\in\{a_1,b_1\}$ and $c_{i^*+1}\in\{a_2,b_2\}$. Let $$j_i=\begin{cases}i,&\text{ if }i\le i^*,\\
i+1,&\text{ if }i^*<i,\end{cases}$$
$i\in[5]$, 
be such that $c_{j_i}$ is the smallest integer in the $i$th group (with groups arranged in increasing order) among $\{c_1\},\ldots,\{c_{i^*-1}\},\{c_{i^*},c_{i^*+1}\},\{c_{i^*+2}\},\ldots,\{c_6\}$. 
For $i=1,\dots,5$, let
\begin{align}\label{eq:defri1}
r_i=c_{j_i},\text{ if }c_{j_i}\in\{a_3,b_3\} \text{ or }c_{j_i}\in\{a_1,b_1\}, 
\end{align}
and 
\begin{align}\label{eq:defri2}
&r_i=\begin{cases}c_{j_i}-(c_{i^*+1}-c_{i^*}), & \text{if } i\equiv i^*\bmod 2,\\
    c_{j_i}+(c_{i^*+1}-c_{i^*}), & \text{if } i\not\equiv i^*\bmod 2,
    \end{cases}\nonumber\\
    &\text{ if }c_{j_i}\in\{a_2,b_2\}\backslash\{c_{i^*+1}\}.
\end{align}

\begin{figure}[H]
\centering
\includegraphics[width=1\linewidth]{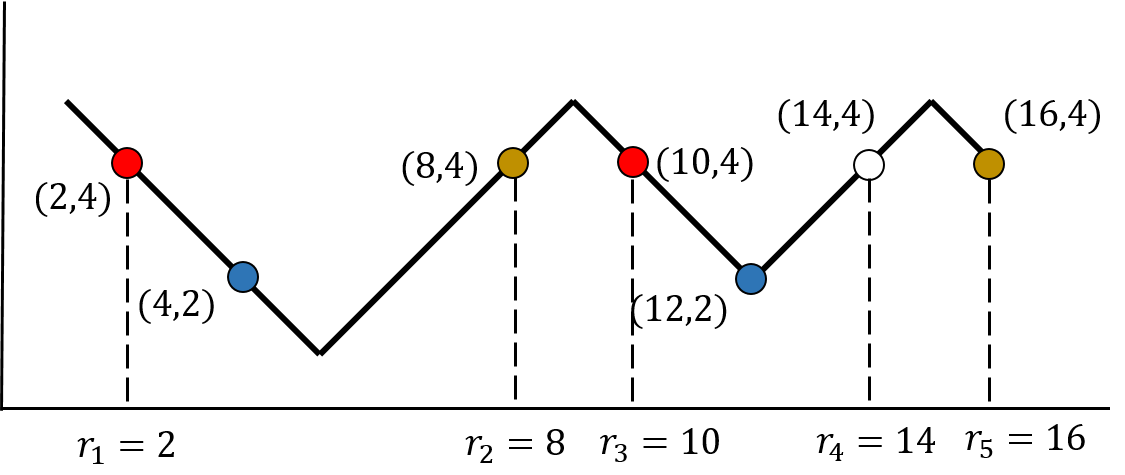}
\caption{A function $h$ for sparse Hamiltonian graph with $3$  arcs, where $a_1=2,b_1=10,a_2=4,b_2=12,a_3=8$, and $b_3=16$. In this case, $c_1=2,c_2=4,c_3=8,c_4=10,c_5=12$, and $c_6=16$. We also have $i^*=1$. According to~\eqref{eq:defri1} and ~\eqref{eq:defri2},  $r_1=2,r_2=8,r_3=10,r_4=14$, and $r_5=16$, which implies that $d_1=5,d_2=9,d_3=12$, and $d_4=15$. Note that there can be multiple choices of $h$ given $h(r_i)=h(r_j)$ for $i,j\in[5]$. The function $h$ becomes completely determined when we fix the values of $h(2)=4$ and $h(4)=2$.}
\label{fig:choiceofh}
\end{figure}

Note that~\eqref{eq:defri2} is based on the constraints from~\eqref{eq:decidingr}. For an illustrative example, refer to Figure~\ref{fig:choiceofh}.
%\iscomment{This is quite difficult to see without a figure. A figure could be very helpful for this proof.}

Using~\eqref{eq:defri1} and~\eqref{eq:defri2} it can be shown that $r_1,\ldots,r_5$ satisfy~\eqref{eq:decidingr}. We show next that $r_1,\ldots,r_5$ also satisfy~\eqref{eq:group6}. Note that $r_i\ne c_{j_i}$ only when $c_{j_i}\in\{a_2,b_2\}$. 
%\iscomment{It says ``For any $i\in\{1,\ldots,5\}$,'' followed by a statement that does not involve $i$}
 If $c_{i^*+2}\in\{a_2,b_2\}$, we have that 
\begin{align}\label{eq:distar61}
d_{i^*}=&\frac{r_{i^*}+r_{i^*+1}}{2}\\
\overset{(a)}{=}&\frac{a_2+b_2}{2}\in (c_{i^*+1},c_{i^*+2}),
\end{align}
where $(a)$ follows from \eqref{eq:decidingr}. Furthermore, for any $i\in[5]$, we have that
\begin{align}\label{eq:distar62}
d_{i}=&\frac{r_{i}+r_{i+1}}{2}\\
\overset{(b)}{=}&\frac{c_{j_i}+c_{j_i+1}\pm (c_{i^*+1}-c_{i^*})}{2}\overset{(c)}{\in} (c_{j_i},c_{j_i+1}),
\end{align}
for $i\ne i^*$ satisfying $c_{j_i}\in\{a_2,b_2\}$ or $c_{j_i+1}\in\{a_2,b_2\}$ (Note that at least one of $r_i=c_{j_i}$ and $r_{i+1}=c_{j_i+1}$ hold for $i\ne i^*$). Equality $(b)$ follows from~\eqref{eq:defri2}, while $(c)$ is a consequence of the definitions of $c_{i^*+1}$ and $c_{i^*}$. In addition, we have 
\begin{align}\label{eq:distar63}
d_{i}=&\frac{r_{i}+r_{i+1}}{2}\\
\overset{(d)}{=}&\frac{c_{j_i}+c_{j_i+1}}{2}\in (c_{j_i},c_{j_i+1}),
\end{align}
for $i\ne i^*$ satisfying $c_{j_i},c_{j_i+1}\notin\{a_2,b_2\}$. The results in~\eqref{eq:distar61}, ~\eqref{eq:distar62}, and~\eqref{eq:distar63} imply~\eqref{eq:group6} when  $c_{i^*+2}\in\{a_2,b_2\}$. Similarly, when  $c_{i^*+2}\notin\{a_2,b_2\}$, we have that either~\eqref{eq:distar62} or~\eqref{eq:distar63} hold, which implies~\eqref{eq:group6}. This completes the proof.
%\iscomment{Even though the writing is fine, overall, this proof is difficult to follow. A figure illustrating an example with $a_i$s, $b_i$s, $c_i$s, $r_i$s, $d_i$s would help the reader visualize some of the statements in the proof}
\end{IEEEproof} 
%\iscomment{DISCUSS: Is there a general folding algorithm that can be stated as an algorithm, and works for the proof of Lemma 4.3 and Proposition 4.1?}

\subsection{Zero-Fragmentation Stores with Repetitions}\label{no_fragmentation}
The next result investigates another extreme point in the design of chunk stores, where we require that no file is fragmented. One obvious solution, as already mentioned, is to avoid deduplication altogether and concatenate all files in the store. The next straightforward result establishes the smallest redundancy needed to ensure that there is no fragmentation. 
\begin{theorem}
    For arbitrary graph $G=(V,E)$, let $m^{*}(t)$ be the minimum $m$ such that $S^{u}_{t}(G,m)=1$. Then $m^{*}(2)=|E|+1+\lceil \frac{|V_{odd}|}{2}\rceil,$ where $V_{odd}$ is the set of odd-degree vertices in $V$. Moreover, for all $t$, 
    $$|P(G, \leq t)|/t \leq m^{*}(t)\leq t|P(G, \leq t)|.$$
\end{theorem}
\begin{IEEEproof}
    For the case $t=2$, the length of each file is either one or two. Since the case of length-one only requires a chunk store to include each original chunk at least once, we focus on the case of length-two files, where each file is an edge. Let $\cs^{*}\in V^{m^{*}(2)}$ be an optimal chunk store of $m^{*}(2)$ chunks. Then $E\subset \{\cs^{*}_{i}\cs^{*}_{i+1}\}_{i=1}^{m^{*}(2)-1}:=E_{new}$, and $\cs^{*}$ is an Eulerian path of $G_{new}:=(V,E_{new})$. Therefore $m^{*}(2)-|E|-1$ is the smallest number of additional edges we need to put in $E$ such that $G$ has an Eulerian path, which is $\lceil \frac{|V_{odd}|}{2}\rceil$.

    For general $t$, simply enumerating paths in $P(G, \leq t)$ in arbitrary order gives a chunk store with no fragmentation and length $\leq t|P(G, \leq t)|$. On the other hand, there exists $1\leq t'\leq t$ 
    %\iscomment{write $t'$ not $t' $}
    such that $|P(G, =t')|\geq |P(G, \leq t)|/t$. For any chunk store $\cs$ with no fragmentation and length $\ell(\cs)$, the maximum number of length-$t'$ files with no fragmentation on $\cs$ is $\ell(\cs)-t'+1\leq \ell(\cs)$. Therefore $\ell(\cs)\geq |P(G, \leq t)|/t$. 
\end{IEEEproof}
In what follows, we turn our attention to investigating if coded chunk stores can outperform chunk stores that use repetitions only in terms of fragmentation reduction. 
%\iscomment{DISCUSS: Can we have a (suboptimal) algorithm based on the idea above that generalizes to settings with fragmentation and redundancy?}

\subsection{Coding Versus Repetition}\label{SvCS}
%\iscomment{Story-wise, this transition is a bit abrupt. We dealt with ``No redundancy'' and then with ``No fragmentation''. At this point, the reader would want to see something about the middle of the feasibility curves}
Next we address the question of \emph{whether coding can reduce the stretch metric}. Specifically, we derive bounds for the coding gain, defined as $S^{u}_{t}(G,m)/S^{c}_{t}(G,m),$ and present some extreme-case examples. The problem of finding an optimal coded chunk store is highly challenging and left for future work.

We start with the setting in which no redundant chunks are allowed ($m=n$), but the chunks themselves may be encoded. The results are given in Subsection~\ref{subsect:no_re}. We then turn our attention to the case when only one redundant chunk is allowed (i.e., $m=n+1$) in Subsection~\ref{subsect:one_re}. Finally, we revisit the problem from Subsection B, in which no fragmentation is allowed, except that the stored chunks may be encodings of user chunks in Subsection~\ref{subsect:no_fra}.

\subsubsection{Case 1 -- No Redundant Chunks}
\label{subsect:no_re} Our main result is described in the following theorem.
\begin{theorem}\label{no_redundancy}
   For any file model \fm, we have $S^{u}_{t}(G,n)=S^{c}_{t}(G,n)$, which establishes that coding (linear or nonlinear) cannot reduce the stretch metric when no redundancy is allowed (i.e., when $m=n$). 
\end{theorem}
Before we prove Theorem~\ref{no_redundancy}, we need the following definitions and Lemma~\ref{coreset}.

Let $\{\dc_{i}\}_{i=1}^{n}$ be the $n$ original data chunks. Let $\cs=\cs_{1},\cs_{2},\hdots,\cs_{n}$ be a chunk store with $n$ chunks, where each $\cs_{i}=f_{i}(\dc_{1},\dc_{2},\hdots,\dc_{n})$ is an arbitrary (and possibly different) function of $\{\dc_{i}\}_{i=1}^{n}$. A $\dc_{i}$-recovery set of a chunk store $\cs$ is a set of positions $\plnR=\{i_{1}, i_{2}, \hdots\}\subset[n],$ listed in ascending order, such that $\dc_{i}$ can be uniquely recovered from $\cs_{\plnR}:=\cs_{i_{1}},\cs_{i_{2}},\hdots$. For any $X\subset\{\dc_{i}\}_{i=1}^{n}$, a $X$-recovery set allows for reconstructing \emph{all} $\dc_{i}\in X$. Moreover, a recovery set is minimal if any strict subset of its elements is not a recovery set.

\begin{lemma}\label{coreset}
    For any nonredundant chunk store $\cs=\cs_{1},\cs_{2},\hdots,\cs_{n}$, each $\{\dc_{i}\}_{i=1}^{n}$ has only one minimal $\dc_{i}$-recovery set $\plnR_{\dc_{i}}$. %\textcolor{red}{This implies that $I_{\dc_{i}}^{\cs}$ is the `core' of all $\dc_{i}$-recovery set on $\cs$. - this sentence is completely uninformative and should not be in a formal result statement - remove.} 
    Consequently, any subset of chunks $X\subset\{\dc_{i}\}_{i=1}^{n}$ also has only one minimal $X$-recovery set $\plnR_{X}:=\bigcup_{\dc_{i}\in X}\plnR_{\dc_{i}}$.
\end{lemma}
\begin{IEEEproof}
    We prove the lemma by contradiction using a probabilistic approach. Let $\{\dc_{i}\}_{i=1}^{n}$ be $n$ discrete, i.i.d random variables with distribution $\rm Uniform(\{0,1\}^d)$, and hence, Shannon entropy of $d$ bits. Since a chunk store is not allowed to compromise the information of any user data chunk, the potentially coded chunks in $\cs$ should allow for reconstructing all $\{\dc_{i}\}_{i=1}^{n}$, which is equivalent to requiring that $H(\cs)=nd$. Suppose next that there exists an $\dc_{i}$ with two minimal recovery sets, say $\plnR_{\dc_{i}}^{1},\plnR_{\dc_{i}}^{2}$. 
    Then
    \begin{align*}
        &H\lp\oneutwo\rp-H\lp\one\rp\\
        &=H\lp\twosone\middle|\one\rp\\
        &\overset{(a)}{=}H\lp\twosone\middle|\one,\dc_{i}\rp\\
        &\leq H\lp\twosone\middle|\oneatwo,\dc_{i}\rp\\
        &=H\lp \dc_{i},\twosone\middle|\oneatwo\rp-H\lp \dc_{i}\middle|\oneatwo\rp\\
        &\overset{(b)}{<}H\lp \dc_{i},\twosone\middle|\oneatwo\rp\\
        &=H\lp\twosone\middle|\oneatwo\rp+H\lp \dc_{i}\middle|\twosone,\oneatwo\rp\\
        &\overset{(c)}{=}H\lp\twosone\middle|\oneatwo\rp\\
        &\leq H\lp\twosone\rp,
    \end{align*}
    where ($a$) and ($c$) follow from $\plnR_{\dc_{i}}^{1},\plnR_{\dc_{i}}^{2}$ being $\dc_{i}$-recovery sets, and ($b$) follows from the sets being minimal. %\textcolor{red}{I would also add an explanation for the equality between (b) and (c) since this one is in my opinion the least obvious one. Others are more than clear.} \textcolor{blue}{I change the inequality between (b),(c).}
    Consequently,
    \begin{align*}
        H\lp \cs\rp&\leq H\lp\oneutwo\rp + H\lp \cs_{[1:n]\setminus (\plnR_{\dc_{i}}^{1}\bigcup \plnR_{\dc_{i}}^{2})}\rp\\
        &<H\lp\one\rp+H\lp\twosone\rp+H\lp \cs_{[1:n]\setminus (\plnR_{\dc_{i}}^{1}\bigcup \plnR_{\dc_{i}}^{2})}\rp\\
        &\leq nd,
    \end{align*}
    which implies that some chunk from $\{\dc_{i}\}_{i=1}^{n}$ cannot be reconstructed based on $\cs$.
\end{IEEEproof}
%\textcolor{red}{I am wondering if this is needed at all as a proof. You know that s are encodings/functions of n iid RVs,  so the data processing inequality should immediately give you this result. Why the messy chain of bounds? Am I missing something? Basically, the above inequality is true no matter what kind of reconstruction sets we have. Also, upon reading further, all we had to show is that no function of the n chunk variables can have more than a total of nd bits of information, which means that coding cannot help.}\textcolor{blue}{we need strictly less than nd bits.}

We are now ready to prove Theorem~\ref{no_redundancy}. 
\begin{IEEEproof}[Proof of Theorem~\ref{no_redundancy}]
    By Lemma~\ref{coreset}, every $\dc_{i}$ has a unique minimal recovery set $\plnR_{\dc_{i}}$ for any chunk store $\cs$. Define the \emph{reconstruction  graph} for $\cs$ to be a bipartite graph $G_{\cs}:=(\{\dc_{i}\}_{i=1}^{n},[1:n],E),$ with $E=\{(\dc_{i},j)|j\in \plnR_{\dc_{i}}\}$. Then, for any $X\subset \{\dc_{i}\}_{i=1}^{n}$, the set of neighbors of $\dc_{i}\in X$ on $G_{\cs}$ is $G_{\cs}(X)=\bigcup_{\dc_{i}\in X}\plnR_{\dc_{i}}=\plnR_{X}$, which is the unique minimal $X$-recovery set of $\cs$. Given that $\dc_{i}$ has a uniform distribution with entropy equal to $d$ bits, $|G_{\cs}(X)|\geq |X|$. But this condition can be recognized as Hall's condition for the existence of a perfect matching in a balanced bipartite graph~\cite{hall}; therefore, $G_{\cs}$ has a perfect matching $M\subset E$, which is a bijection between $\{\dc_{i}\}_{i=1}^{n}$ and $[1:n]$. We can then define an uncoded chunk store as
    \begin{equation*}
        \cs^{u}:= \cs_{M^{-1}(1)},\cs_{M^{-1}(2)},\hdots,\cs_{M^{-1}(n)},
    \end{equation*}
    with reconstruction graph $G_{\cs^{u}}=(\{\dc_{i}\}_{i=1}^{n},[1:n],M)$. Then, since $M\subset E$, 
    \begin{equation*}
        G_{\cs^{u}}(X)\subset G_{\cs}(X), \forall X\subseteq\{\dc_{i}\}_{i=1}^{n}.
    \end{equation*}
    %\iscomment{One thing that needs to be clarified here: suppose you have a file $p$ with set of chunks $X_p \subset \{x_i\}_{i=1}^n$. Could it be that the shortest substring of $s$ that reconstructs $X_p$ does not contain $G_s(X_p)$? In other words, there exists a non-minimal reconstruction set for $X_p$ that corresponds to the minimal substring of $s$?
    %Yun-Han, I believe you had this notion of ``core'' when you first presented this proof, which may be related to this question. Is it no longer needed?
    
    %\textcolor{red}{But $G_{\cs}(X)$ is the unique minimal $X$-recovery set for $\cs$. Therefore any $X$-recovery set for $\cs$ must contain $G_{\cs^{u}}(X)$ as a subset.} 
    Consider a file $p$ with chunks $X_p \subset \{x_i\}_{i=1}^n$. 
    Since $G_s(X_p)$ is the unique minimal $X_p$-reconstruction set, any $X_p$-reconstruction set must contain $G_s(X_p)$.
    Since $G_{s^u}(X_p) \subset G_s(X_p)$, the stretch of $p$ in $s^u$ cannot be larger than the stretch of $p$ over $s$. Therefore the uncoded chunk store $\cs^{u}$ is as good as $\cs$ for all $X\subset\{\dc_{i}\}_{i=1}^{n}$, and  consequently, all file models \fm.
\end{IEEEproof}

%\textcolor{red}{\sout{We should mention that we seek examples with the smallest amount of redundancy - one or two chunks - since after all we want to deduplicate. Also, it seems pretty clear that "central" vertices in the graph - the ones contained in the largest number of paths/files should be duplicated for robustness but are also the ones that will help you the most for fragmentation if duplicated.}}

\subsubsection{Case 2 -- One Redundant Chunk}\label{subsect:one_re}
Theorem~\ref{no_redundancy} shows that coding cannot reduce fragmentation when no redundancy is allowed.  On the other hand, when redundant data chunks can be included in the store, Example~\ref{example1} and Example~\ref{example2} show that coding can help even when only one redundant chunk is included (i.e., when $m=n+1$). These examples also suggest that ``central'' vertices in the graph -- the ones contained in the largest number of paths/files that should be duplicated for robustness -- are also the ones that help the most in reducing stretch fragmentation. The analysis of high-redundancy regimes in which the number of redundant chunks is allowed to grow with $n$ is left for future work.

\begin{example}\label{example1} The file model \fm ~arises from the graph $G$ depicted in  Figure~\ref{fig:example1}. %\textcolor{red}{We assume that $t=2$, so that all files comprise two chunks connected by an edge. --Li:But we have files with length $1$ and length $2$?} 
First, we show the uncoded stretch metric $S^{u}_{2}(G,n+1)$ equals $(3N+2)/2$. 
%\iscomment{is this technically $(3N+2)/t = (3N+2)/2$? If so, we may want to clarify the writing below}
Since we can only repeat one chunk, at least one of the chunks $\dc_{a},\dc_{b}$ is not duplicated. By construction, both of these chunks have $6N+1$ neighbors. Therefore, in any uncoded chunk store, at least one length-$2$ files including the nonredundant chunk has to have a stretch $>(3N+1)/2$.
Combined with the fact that the chunk store described below with $n$ chunks has the worst stretch metric for files of  length two, which equals $(3N+2)/2$, we have $(3N+1)/2<S^{u}_{2}(G,n+1)\leq S^{u}_{2}(G,n)\leq (3N+2)/2$, which implies $S^{u}_{2}(G,n+1)=(3N+2)/2$.
 
Consider the following nonredundant chunk store  
\begin{equation*}    \cs=\dc_{[1:3N]}\dc_{a}\dc_{[3N+1:5N]}\dc_{b}\dc_{[5N+1:8N]}.
\end{equation*}
It is easy to verify that the stretch metric for $t=2$ is $(3N+2)/2$. On the other hand, the following coded chunk store of length $n+1$ has a stretch metric for $t=2$ equal to $(2N+2)/2,$
\begin{align*}
    \cs=\dc_{[1:2N]}\dc_{a}\dc_{[2N+1:4N]}(\dc_{a}\oplus \dc_{b})\dc_{[4N+1:6N]}\dc_{b}\dc_{[6N+1:8N]}.&
\end{align*}
Therefore $S^{c}_{2}(G,n+1)\leq (2N+2)/2$. 
It suffices to check all edges incident to $\dc_{a}$ or $\dc_{b}$ can be decoded from some length $\leq 2N+2$ segment. The recovery procedure is described below.

\begin{align*}
    \text{Recover }    &   \{\e{a}{i}\}_{i=1}^{2N}     ~&   \text{ from }~& \dc_{[1:2N]}\dc_{a};\\
    \text{Recover }    &   \{\e{a}{i}\}_{i=2N+1}^{4N}  ~&   \text{ from }~& \dc_{a}\dc_{[2N+1:4N]};\\
    \text{Recover }    &   \{\e{b}{i}\}_{i=4N+1}^{6N}  ~&   \text{ from }~& \dc_{[4N+1:6N]}\dc_{b};\\
    \text{Recover }    &   \{\e{b}{i}\}_{i=6N+1}^{8N}  ~&   \text{ from }~& \dc_{b}\dc_{[6N+1:8N]};\\
    \text{Recover }    &   \{\e{a}{i}\}_{i=4N+1}^{6N}  ~&   \text{ from }~& (\dc_{a}\oplus \dc_{b})\dc_{[4N+1:6N]}\dc_{b};\\
    \text{Recover }    &   \{\e{b}{i}\}_{i=2N+1}^{4N}  ~&   \text{ from }~& \dc_{a}\dc_{[2N+1:4N]}(\dc_{a}\oplus \dc_{b});\\
    \text{Recover }    &   \{{x_{a}x_{b}\}}                    ~&   \text{ from }~& \dc_{a}\dc_{[2N+1:4N]}(\dc_{a}\oplus \dc_{b}).
\end{align*}

\begin{figure}[H]
 \vspace{-5mm}
    \centering
    \includegraphics[scale=0.24]{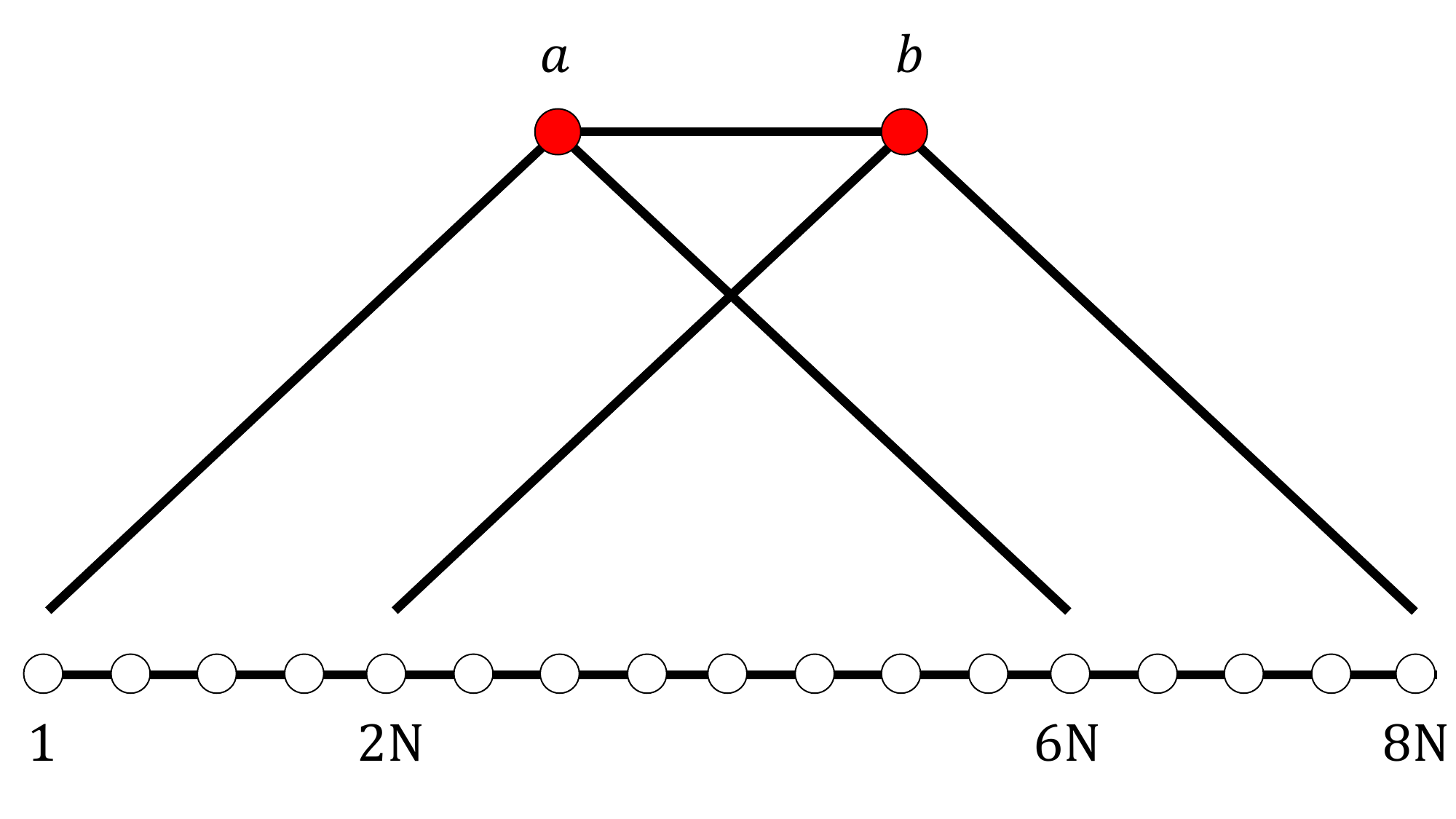}
    \caption{A graph $G=(V,E)$ with $V=\dc_{[1:8N]}\bigcup\{\dc_{a},\dc_{b}\}$ comprising a total of  $|V|=n=8N+2$ vertices. The vertices $\dc_{[8N]}$ form a line-graph, while $\dc_{a}$ is connected to all vertices in $\dc_{[6N]}$, and $\dc_{b}$ is connected to all vertices in $\dc_{[2N+1:8N]}$. The vertices $\dc_{a}$ and $\dc_{b}$ are also connected by an edge.} \label{fig:example1}
\end{figure}

It is worth reiterating that all store fragments used for recovery are of length $\leq 2N+2$. Furthermore, the coding gain is
\begin{equation*}
    \frac{S^{u}_{2}(G,n+1)}{S^{c}_{2}(G,n+1)}\geq\frac{(3N+2)/2}{(2N+2)/2}\approx\frac{3}{2}.
\end{equation*}

%The stretch reduction from coding is $S^{u}_{1}(G,n+1)-S^{c}_{1}(G,n+1)=3N+2-(2N+2)=N\approx \frac{n}{8}$. 
\end{example}

In summary, Example~\ref{example1} shows that coding can help reduce the stretch metric by a nontrivial factor even when only one redundant chunk is included (i.e., when $m=n+1$). However, the following uncoded chunk store with \emph{two} redundant chunks also has a stretch metric $(2N+2)/2$:
\begin{equation*}
\cs=\dc_{[1:2N]}\dc_{a}\dc_{b}\dc_{[2N+1:6N]}\dc_{a}\dc_{b}\dc_{[6N+1:8N]}
\end{equation*}
Therefore, a natural question to ask is: \emph{Can coded stores with one redundant chunk always be matched in performance by repeating a small number of chunks?} As we will see, this is not always the case. Example~\ref{example2} shows that there exists a graph $G$ such that any uncoded chunk store needs $\Theta(n)$ redundant chunks to outperform a coded chunk store with a single redundant chunk that also allows coding.
\begin{example}\label{example2}
Consider the graph shown in Figure~\ref{fig:example2}. There, the vertices are partitioned into two groups, one forming a line graph, and another connecting vertices within a complete subgraph. Each of the vertices in the complete graph, $\dc_{b_{i}}, i\in [N]$ is connected to a total of $4N+1$ neighbors on the line graph and within the clique. Therefore, in order to keep the stretch metric bounded from above by $(2N+1)/2$, one must duplicate every $\dc_{b_{i}}$. Even if we duplicate $N-1$ of the vertices, the stretch metric will still be greater than $(2N+1)/2$, which implies $S^{u}_{2}(G,n+N-1)>(2N+1)/2$. %\textcolor{red}{again, my basic math says different - duplicating N nodes should lead to n+N, not n+N-1?}. %Therefore $S^{c}_{2}(G,n+1)\leq 2N+1.$ 

\begin{figure}[H]
 \vspace{-1mm}
    \centering
    \includegraphics[scale=0.24]{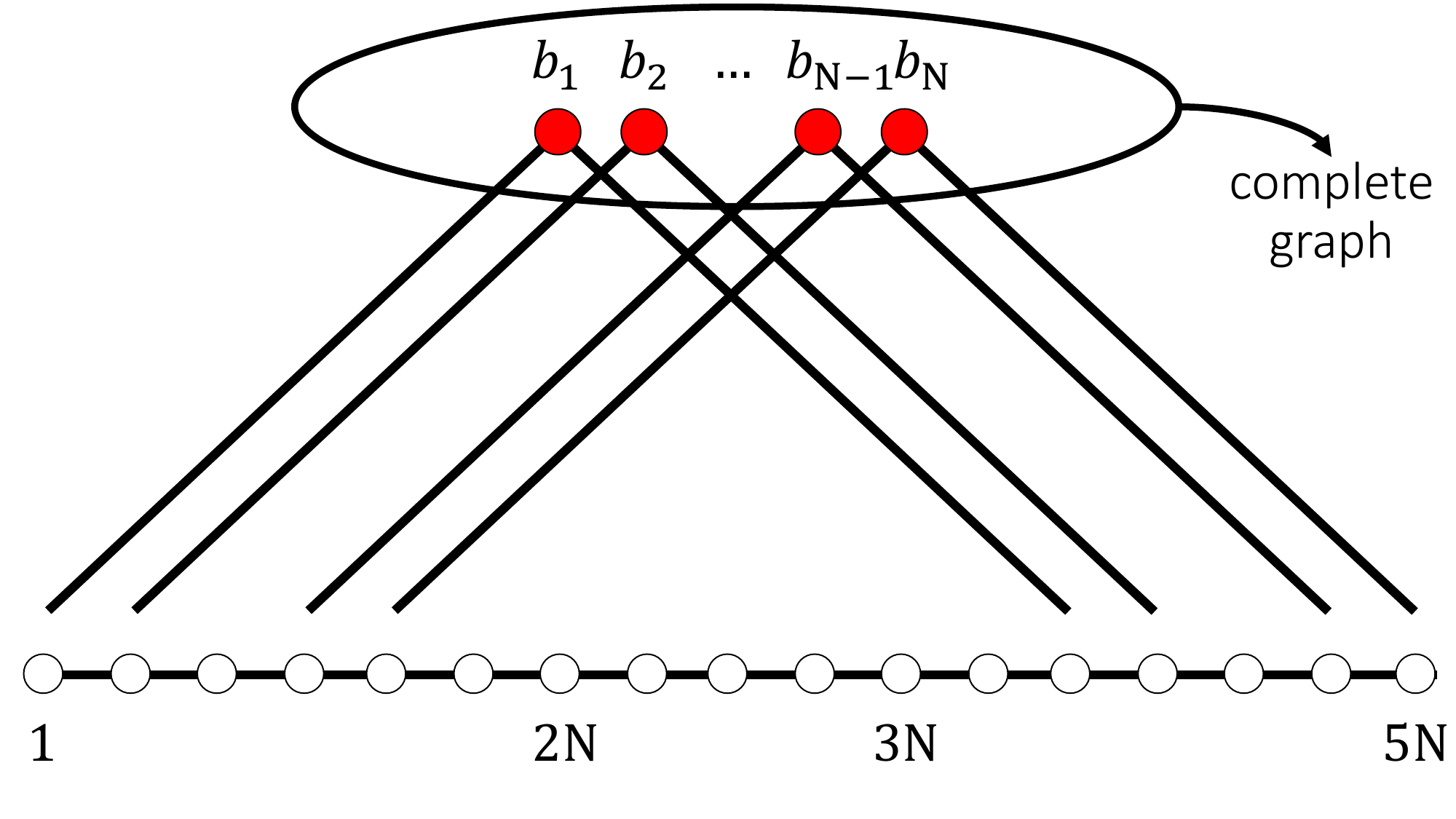}
    \caption{A graph $G=(V,E)$ with vertex set $V=\dc_{[1:5N]}\bigcup\{\dc_{b_{1}},\ldots,\dc_{b_{N}}\}$ and a total number of $n=6N$ vertices. Here, the vertices  $\dc_{[1:5N]}$ form a line graph, $\dc_{b_{i}}$ is connected to $\dc_{[2i-1:3N+2i]}$, and all $\dc_{b_{i}}'$s are connect to each other (i.e., they form a complete graph with $N$ vertices).} \label{fig:example2}
\end{figure}

At the same time, the following coded chunk store has stretch metric $(2N+1)/2$ and uses only one redundant chunk while allowing coding: 
\begin{align*}   \cs=\dc_{[1:2N]}\;y_{1}\dc_{2N+1}\;y_{2}\dc_{2N+2}\ldots y_{N}\dc_{3N}\;y_{N+1}\dc_{[3N+1:5N]},
\end{align*}
where $$y_{i}= 
    \begin{cases}
        \dc_{b_{1}}, ~&\text{ for } i=1;\\
        \dc_{b_{N}}, ~&\text{ for } i=N+1;\\
        \dc_{b_{i-1}}\oplus \dc_{b_{i}}, ~&\text{ for } 1<i<N+1.
    \end{cases}$$    
It suffices to check that the files $\{\e{b_{i}}{j}\}_{j=2i-1}^{3N+2i}$ and $\{\e{b_{i}}{b_{j}},i\neq j\}$ can be reconstructed from some length $\leq 2N+1$ fragment. Note that there are two disjoint sets of coded chunks that can be used to solve for each of the $\dc_{b_{i}}$, which are $\{y_{j}\}_{j=1}^{i}$, and $\{y_{j}\}_{j=i+1}^{N+1}$. This follows from $\sum_{j=1}^{i}y_{j}=\sum_{j=i+1}^{N+1}y_{j}=\dc_{b_{i}}$. In summary, the recovery map reads as follows:
\begin{align*}
    \text{Recover }    &\{\e{b_{i}}{j}\}_{j=2i-1}^{2N}\\
    &\text{ from } \dc_{[2i-1:2N]}y_{1}\dc_{2N+1}\ldots y_{i}\\
    \text{Recover }    &\{\e{b_{i}}{j}\}_{j=3N+1}^{3N+2i}\\
    &\text{ from } y_{i+1}\dc_{2N+i+1}\ldots y_{N+1}\dc_{[3N+1:3N+2i]}\\
    \text{Recover }    &\{\e{b_{i}}{j}\}_{j=2N+1}^{3N}\\
    &\text{ from } y_{1}\dc_{2N+1}\ldots y_{N+1}\\
    \text{Recover }    &\{\e{b_{i}}{b_{j}},i\neq j\}\\
    &\text{ from } y_{1}\dc_{2N+1}\ldots y_{N+1}.
\end{align*}
As noted, all fragments listed above have length $2N+1$.
%\iscomment{they are all exactly $2N+1$, no?}
\end{example}

The next result shows that for any file model \fm, all linearly coded stores with one single redundant chunk reduce to stores similar to the ones described in Example~\ref{example1} and Example~\ref{example2}. 

\begin{theorem}\label{1bit_reduction}
    For any file model \fm, there always exists an \emph{optimal} linearly coded store with one redundant chunk%\textcolor{red}{We need to be careful with the use of the word optimal - optimal for one redundant chunk, optimal in general, optimal when all coded chunks include sums of two chunks etc}
    ,  $\cs=\cs_{1},\ldots,\cs_{n+1},$ that can be obtained by interleaving two sequences $\dc_{a_{1}}\ldots\dc_{a_{n-N}}$ and $y_{1},\ldots,y_{N+1}$. Here, $\{\dc_{a_{i}}\}_{i=1}^{n-N}$ and $\{\dc_{b_{i}}\}_{i=1}^{N}$ are a partition of $V$, while $\{y_{i}\}_{i=1}^{N+1}$ is a set of coded chunks of the form 
    $$y_{i}= 
        \begin{cases}
            \dc_{b_{1}}, ~&\text{ for } i=1,\\
            \dc_{b_{N}}, ~&\text{ for } i=N+1,\\
            \dc_{b_{i-1}}\oplus \dc_{b_{i}}, ~&\text{ for } 1<i<N+1,
        \end{cases}$$ 
    where as before, $\oplus$ stands for the bitwise xor function. 
    %\textcolor{red}{Q: Are we sure that this holds for any t? Somehow sounds hard to digest? Any conjectures about what happens when we have more than one redundant chunk - it almost seems that for two chunks an optimal store may use linear combinations of three etc?}
    %\textcolor{orange}{A: This holds for any t,G with one redundant chunk. The way it works is by repeatdly finding a new code that is universally better than the current one. The definition of universality doesn't utilize t,G at all. Therefore, the result holds for any t,G. The difficult part (utilizing structure of t,G) would be the main focus in the future. Since when we have $>>1$ redundancy, ignore structure of t,G would probability leads to a very suboptimal linearization. It is very likely that once we have more than one redundant chunk,an optimal linear code need to mix more than two chunks.}
    %\textcolor{red}{\sout{Q: I see you call it 1 redundancy, 2 redundancy, properly, refer to it one redundant chunk, two redundant chunks. In a text, we usually spell out numbers $<10$ and use numerical symbols only for values $\geq 10$.}}
\end{theorem}
\begin{IEEEproof}
    %The proof is delegated to Subsection~\ref{proof_1bit_reduction}. 
We start with some definitions. We assume that chunks are represented as binary strings, and that the ``sums of chunks'' in the linear codes correspond to bitwise xors. Due to the assumption, in the proof we can assume that each chunk $x_i$ is just one bit.
Next, let $\dc\in\{0,1\}^{n}$ be $n$ user bits, and let $\cs \in \{0,1\}^{m}$ be a length-$m$ binary coded bit string. Let $G\in\{0,1\}^{n\times m}$ be the generator matrix of a linear code, so that $\cs=\dc G$. For convenience, throughout the proof, we represent a linear code by a full rank matrix $\begin{bmatrix}\HH\\\KK\end{bmatrix}$ with $\KK\in\{0,1\}^{n\times m}$ such that $G\KK^{T}=I_{n}$, and where $\HH\in\{0,1\}^{(m-n)\times m}$ stands for a basis of the null space of $G$ (i.e., a parity-check matrix of the code).   Note that a coded store of the form given in Theorem~\ref{1bit_reduction} has an $(\HH,\KK)$ such that each $\KK(i)$ is either an indicator vector or there exists a $t\in[n+1]$ such that the set of values equal to $1$ in $\KK(i)$ are either the first or the last $t$ values equal to $1$ within $\HH$. As an example, the coded chunk store $(\dc_{1},\dc_{2},\dc_{4},\dc_{5},\dc_{2}\oplus\dc_{3},\dc_{3})$ corresponds to 
\begin{equation*}
    \begin{bmatrix}\HH\\\KK\end{bmatrix}=\begin{bmatrix}0 & 1 & 0 & 0 & 1 & 1\\
    1 & 0 & 0 & 0 & 0 & 0\\
    0 & 1 & 0 & 0 & 0 & 0\\
    0 & 1 & 0 & 0 & 1 & 0\\
    0 & 0 & 1 & 0 & 0 & 0\\
    0 & 0 & 0 & 1 & 0 & 0\end{bmatrix}.
\end{equation*}
% \iscomment{Here, it may be a good idea to explain what a coded store with  form given in Theorem 5.4 would look like in this $(K,H)$ notation.}

We say that a subset $\{\cs_{j_1},\cs_{j_2},\ldots|0<j_{1}<j_{2}<\ldots\leq m\}$ of coded bits
\emph{reconstructs} a user bit $\dc_{i}$ if $\dc_{i}$ is a linear combination of these coded bits dictated by the code at hand.
Let $\pln\in \{0,1\}^{m}$ be the indicator vector of this subset, and let $\cs_{\pln}$ stand for the subset itself. Then, it suffices to consider vectors $\pln$ that satisfy $\dc_{i}=\pln \cs^{T}$, which implies that $\pln\in\KK(i)+span\{\HH\}:=\{\KK(i)+v|v\in span\{\HH\}\},$ where $\KK(i)$ stands for the $i$-th row of $\KK$. For any file $p\in$\fm, let $X_{p}$ denote the set of indices of user bits in $p$.
Then, we say that a subset of coded bits $\cs_{\pln}$ reconstructs $p$ if it reconstructs all $\dc_{i}\in X_{p}$. By the previous discussion, it suffices to consider those $\cs_{\pln}=\bigcup_{\dc_{i}\in X_{p}}\cs_{\pln_{i}},$ where each $\pln_{i}\in \KK(i)+span\{\HH\}$ is an indicator vector of a subset $\cs_{\pln_{i}}$ that can be used to reconstruct $\dc_{i}$. For simplicity, we also write $\pln=\bigcup_{\dc_{i}\in X_{p}}\pln_{i}$ and equate the vector $\pln$ with its support set. The stretch metric of $\cs_{\pln}$ is the length $\ell\lp\itv{\pln}\rp$ of the interval $\itv{\pln}:=[\min(\pln),\max(\pln)]$, where $\min(\pln)$ and $\max(\pln)$ stand for the position of the leftmost and rightmost $1$ in $\pln$, respectively. 
%For any $X\subset \{\dc_{i}\}_{i=1}^{n}$, a $X$-solving set solve all $\dc_{i}\in X$.
Now, our goal is as follows: given a file model \fm, find a length-$n+1$ chunk store $\cs$ that minimizes
\begin{equation*}
    \max_{p\in P(G,\leq t)}\min_{\pln\in\{0,1\}^{n+1}: \cs_{\pln}\text{ reconstructs } 
   p}\ell\lp \itv{\pln}\rp.
    %&=\max_{X\in P(G,\leq t)}\min_{\cs_{\pln_{i}} \text{ solve }\dc_{i}}\ell\lp\bigcup \pln_{i}\rp\\
\end{equation*}
By the previous discussion, this is equivalent to minimizing
\begin{equation*}
    \max_{p\in P(G,\leq t)}\min_{\dc_{i}\in X_{p}
    \atop \pln_{i}\in \KK(i)+span\{\HH\}}\ell\lp \itv{\bigcup \pln_{i}}\rp.
\end{equation*}
Our proof is as follows. Given a linear code $\begin{bmatrix}\HH\\\KK\end{bmatrix}$, we are going to ``improve'' it according to the optimization metric iteratively, via a process similar to Gaussian elimination, until it reduces to a code of the simple form described in Theorem~\ref{1bit_reduction}. At each step, we construct a new code $\begin{bmatrix}\HH_{new}\\\KK_{new}\end{bmatrix}$ that offers as good a recovery performance as $\begin{bmatrix}\HH\\\KK\end{bmatrix}$ for each $\{\dc_{i}\}_{i=1}^{n}$, and consequently, for all files $p\in P(G,\leq t)$. More precisely, with a slight abuse of notation, we use $(\KK_{new}(i),\HH_{new})\leq (\KK(i),\HH)$ to indicate that %$\begin{bmatrix}\KK_{new}\\\HH_{new}\end{bmatrix}$ 
%$\begin{bmatrix}\HH\\\KK\end{bmatrix}$ for recovering $\dc_{i}$, which means
\begin{align*}
    \forall \pln_{b}\in \KK(i)+ span\{\HH\}&\,\exists \pln_{a}\in \KK_{new}(i)+ span\{\HH_{new}\}\\
    &\text{s. t. } \itv{\pln_{a}}\subset \itv{\pln_{b}}
\end{align*}
%A plan $\pln_{a}$ is as good as a plan $\pln_{b}$ if $[\min(\pln_{a}),\max(\pln_{a})]\subset[\min(\pln_{b}),\max(\pln_{b})]$. We denote it by $\pln_{a}\leq \pln_{b}$. 
%A code $(\KK_{1},\HH_{1})$ is as good as $(\KK_{2},\HH_{2})$ for solving $\dc_{i}$ if $\forall \pln_{b}\in \KK_{2}(i)+ span\{\HH_{2}\}\exists \pln_{a}\in \KK_{1}(i)+ span\{\HH_{1}\}$ such that $\pln_{a}\leq \pln_{b}$. We denote it by $(\KK_{1}(i),\HH_{1})\leq(\KK_{2}(i),\HH_{2})$. For the special case that $\HH_{1}=\HH_{2}$, we denote it by $\KK_{1}(i)\leq^{\HH}\KK_{2}(i)$. A code $(\KK_{1},\HH_{1})$ is universally as good as $(\KK_{2},\HH_{2})$ if $(\KK_{1}(i),\HH_{1})\leq(\KK_{2}(i),\HH_{2})$ for all $i\in[n]$. We denote it by $(\KK_{1},\HH_{1})\leq(\KK_{2},\HH_{2})$.
Before we describe the reduction process, we need Lemma~\ref{subsapce}. 
Let $e_{i}\in\{0,1\}^{m}$ be the indicator vector of position $i$. Also, let $Min(\KK(i)):=\max_{\pln\in \KK(i)+span\{\HH\}}\min(\pln)$ and $Max(\KK(i)):=\min_{\pln\in \KK(i)+span\{\HH\}}\max(\pln)$. 
%\textcolor{red}{right-most left boundary and left most right boundary across $\KK(i)+span\{\HH\}$ - do not understand at all, use math or better explanation}. 
Also, for a given $\HH\in \{0,1\}^{(m-n)\times m}$ and $h\in\{0,1\}^{m}$, define $T_{h}:=\{g\in\{0,1\}^{m}|(g,\HH)\leq (h,\HH)\}$, where we use the previously defined $(\cdot,\cdot) \leq (\cdot,\cdot)$ notation.
%\begin{align*}
%    \forall \pln_{b}\in g+ &span\{\HH\}\,\exists \pln_{a}\in h+ span\{\HH\}\\
%    &\text{s. t. } \itv{\pln_{a}}\subset \itv{\pln_{b}}
%\end{align*}}
\begin{lemma}\label{subsapce} 
    If $Min(h)\leq Max(h)$, then $e_{Min(h)}\in T_{h}$ and $h+e_{Min(h)}\in T_{h}$. Moreover, $Min(h+e_{Min(h)})>Min(h)$. 
\end{lemma}
\begin{IEEEproof}
    We first show that $T_{h}$ is a subspace of $\{0,1\}^m$.
    Let $g_{1},g_{2}\in T_{h}$. Then, by definition, $\forall p_{h}\in h+ span\{\HH\}$,  $\exists p_{g_{1}}\in g_{1}+ span\{\HH\}$ and $p_{g_{2}}\in g_{2}+ span\{\HH\}$ such that $\itv{p_{g_{1}}}\subset \itv{p_{h}}$ and $\itv{p_{g_{2}}}\subset \itv{p_{h}}$.  
    But then $\itv{p_{g_{1}}+p_{g_{2}}}\subset \itv{p_{h}}$ with $p_{g_{1}}+p_{g_{2}}\in (g_{1}+g_{2})+span\{\HH\}$. Therefore $g_{1}+g_{2}\in T_{h}$, which implies that $T_{h}$ is a subspace. Since $h\in T_{h}$, it follows that $h+e_{Min(h)} \in T_{h}$ if $e_{Min(h)}\in T_{h}$. By the definition of $Min(h)$, $Max(h)$, and the assumption that $Min(h)\leq Max(h)$, all $p_{h}\in h+span\{\HH\}$ satisfy $\min(p_{h})\leq Min(h)\leq Max(h)\leq \max(p_{h})$. Therefore $\itv{e_{Min(h)}}=Min(h)\subset \itv{p_{h}}$. Consequently $e_{Min(h)}\in T_{h}$.

    Let $p_{h}^{*}\in h+span\{\HH\}$ be such that  $\min(p_{h}^{*})=Min(h)$. Then $p_{h}^{*}+e_{Min(h)}\in (h+e_{Min(h)})+span\{\HH\}$, which implies $Min(h+e_{Min(h)})\geq \min(p_{h}^{*}+e_{Min(h)})>Min(h)$, as claimed.
\end{IEEEproof}
Since we assumed that $m=n+1$, $\HH$ is a vector and $\KK(i)+span\{\HH\}=\{\KK(i),\KK(i)+\HH\}$. Given $\KK$ and $\HH$, the replacement approach consists of the following two steps.

In Step $1$, we repeatedly find a $\KK(i)$ 1) which is not an indicator vector; and 2) that satisfies $Min(\KK(i))\leq Max(\KK(i))$. Then, replace $\KK(i)$ by $e_{Min(\KK(i))}$ if $e_{Min(\KK(i))}\in span\{\HH,\KK(j)\},$ where $j\neq i$, and otherwise, replace it by $\KK(i)+e_{Min(\KK(i))}$. In both cases, $\begin{bmatrix}\HH\\\KK\end{bmatrix}$ retains its full rank. Moreover, by Lemma~\ref{subsapce}, $e_{Min(\KK(i))}\in T_{\KK(i)}$ and $\KK(i)+e_{Min(\KK(i))}\in T_{\KK(i)}$. Replacing $\KK(i)$ by $e_{Min(\KK(i))}$ basically introduces a new indicator vector. On the other hand, if $\KK(i)$ is replaced by $\KK(i)+e_{Min(\KK(i))}$, then $Min(\KK(i)+e_{Min(\KK(i))})>Min(\KK(i))$. Therefore, the successive replacement approach eventually stops with all $\KK(i)$ either equal to indicator vectors or when $Min(\KK(i))>Max(\KK(i))$. For those $\KK(i)$ with $Min(\KK(i))>Max(\KK(i))$, there does not exist a $j$ such that $\HH(j)=0$ and $\KK(i,j)=1$. Otherwise we would have $\KK(i,j)=\KK(i,j)+\HH(j)=1$, which would imply that $Min(\KK(i))\leq j\leq Max(\KK(i))$. Moreover there cannot exist $j_{1}<j_{2}<j_{3}\in [n+1]$ such that $\HH(j_{1})=\HH(j_{2})=\HH(j_{3})=1$ and $\KK(i,j_{1})=\KK(i,j_{3})\neq \KK(i,j_{2})$. Otherwise $Min(\KK(i))\leq j_{2}\leq Max(\KK(i))$. 
These two claims together imply that there exists a $t\in[n+1]$ such that the set of values equal to $1$ in $\KK(i)$ are either the first or the last $t$ values equal to 1 within $\HH$. By definition, replacing $\KK(i)$ by $\KK(i)+\HH$ still results in the same code. Therefore, we can assume that the set of positions occupied by $1$s in $\KK(i)$ -- i.e., the support of $\KK(i)$-- matches the first $t$ positions with values equal to $1$ within $\HH$.
 
After Step $1$, all $\KK(i)$ are either indicator vectors or there exists a $t\in[n+1]$ such that the support of $\KK(i)$ is contained in the set of the positions of the first $t$ $1$s in $\HH$. In Step $2$, whenever there exist indices $i$ and $j$ such that $\KK(i)=e_{j}$ with $j\neq\min(\HH),\max(\HH)$, set $\KK(i' ,j)=0$ for all $i' \neq i$, and $\HH(j)=0$. Let $\KK_{new},\HH_{new}$ denote $\KK,\HH$ after the above modification. Then, we have that 1) $(\KK_{new}(i),\HH_{new})\leq (\KK(i),\HH)$ since $\itv{\KK_{new}(i)}=\itv{\KK(i)}$, and $\itv{\KK_{new}(i)+\HH_{new}}=\itv{\KK(i)+\HH}$; 2) $(\KK_{new}(i'),\HH_{new})\leq (\KK(i'),\HH)$ for all $i'\neq i$ since $\itv{\KK_{new}(i')}\subset\itv{\KK(i')}$, and $\itv{\KK_{new}(i')+\HH_{new}}\subset\itv{\KK(i')+\HH}$.
%Moreover $(\KK_{new}(i),\HH_{new})\leq (\KK(i),\HH)$ since $\KK(i)=\KK_{new}(i)=e_{j}$ and $j\neq\min(\HH),\max(\HH)$. Consequently,  $\itv{\KK_{new}(i)}=j=\itv{\KK(i)}$ and $\itv{\KK_{new}(i)+\HH}= \itv{\HH} =\itv{\KK(i)+\HH}$. Finally, for those $\KK(i)$ with $t\in[n+1]$ such that \textcolor{red}{positive bits of $\KK(i)$ is the first $t$ bits of positive bits of $\HH$, this structure still remains in $\KK_{new},\HH_{new}$.???} 

Upon completion of Step $2$, all $\KK(i)$ satisfy 1) there exists a $t\in[n+1]$ such that the support of $\KK(i)$ is confined to the set of positions of the first $t$ values equal to $1$ in $\HH$; or 2) $\KK(i)=e_{j}$ with $\HH(j)=0$. Let $\lnorm\HH\rnorm_{0}$ denote the number of values equal to $1$ in $\HH$ (i.e., the support size or $\ell_0$ norm of $\HH$).
Since $\begin{bmatrix}\HH\\\KK\end{bmatrix}$ is full rank, we conclude that the number of  $\KK(i)'$s that satisfy the first condition is $\lnorm\HH\rnorm_{0}-1$,  with the set of values equal to $1$ for the $\KK(i)$ 
being exactly the first $1$, first $2$, $\hdots$, first $\lnorm\HH\rnorm_{0}-1$ values equal to $1$ in $\HH$. This gives us a generator matrix of the form $(\KK,\HH)$ described in Theorem~\ref{1bit_reduction}.
%\iscomment{Overall, this proof is very interesting, but long and very dense. A figure showing a small example, showing the initial generator matrix $(K,H)$ that doesn't have the desired form, the final generator matrix that does have the desired form, and the corresponding coded store, would help the reader quite a bit.}
\end{IEEEproof}

From Theorem~\ref{1bit_reduction}, we can deduce the following corollary. 
\begin{corollary}
    For linearly coded stores with one redundant chunk and for any file model \fm, the optimal coding scheme described in Theorem~\ref{1bit_reduction} involves at most $2$ user chunks in each coded chunk; i.e.,  xoring more than $2$ chunks does not reduce fragmentation compared to a code that xors at most two chunks.
\end{corollary}
\begin{remark} An intuitive explanation of why coding helps reducing fragmentation is as follows. The chunks $\{\dc_{b_{i}}\}_{i=1}^{N}$ may be viewed as the ``popular'' chunks used in many files. We therefore encode them so that each $\dc_{b_{i}}$ can be reconstructed using two sets of coded chunks $\sum_{j=1}^{i}y_{j}=\sum_{j=i+1}^{N+1}y_{j}=\dc_{b_{i}}$ that are arranged within disjoint store fragments. The optimal coded chunk store described in Theorem~\ref{1bit_reduction} effectively ``repeats'' all chunks $\{\dc_{b_{i}}\}_{i=1}^{N}$ twice via one redundant chunk. In contrast, any uncoded chunk store with one redundant chunk can clearly only repeat one user chunk.  
\end{remark}

The next Proposition~\ref{coding_gain} shows that despite the previous results, one cannot expect the coding gain to exceed a factor of $2$. 

\begin{proposition}\label{coding_gain}
    One has $S^{u}_{t}(G,n+1)/S^{c}_{t}(G,n+1)\leq 2$ for any \fm. %This means linear code reduces stretch from repetition at most half with one redundant chunk.   
\end{proposition}
\begin{IEEEproof}
    Let $\cs=\cs_{1},\ldots,\cs_{n+1}$ be the optimal coded chunk store of Theorem~\ref{1bit_reduction}, with stretch $S^{c}_{t}(G,n+1)$. Let $\cs^{-1}(y_{i})$ denote the position of $y_{i}$ in $\cs$. There are two ways to reconstruct each $\{\dc_{b_{i}}\}_{i=1}^{N}$, which include using $\{y_{j}\}_{j=1}^{i}$ or $\{y_{j}\}_{j=i+1}^{N+1}$. 
    If $\cs^{-1}(y_{N+1})-\cs^{-1}(y_{2})\geq S^{c}_{t}(G,n+1)$, then $\{y_{j}\}_{j=2}^{N+1}$ cannot be used to reconstruct $\dc_{b_{1}}$. 
    By replacing $y_{2}=\dc_{b_{1}}\oplus \dc_{b_{2}}$ with $\dc_{b_{2}}$, the resulting code is still of the form described in Theorem~\ref{1bit_reduction}; and, the only remaining way to reconstruct $\dc_{b_{1}}$ is to use $\{y_{j}\}_{j=1}^{i}$.
    Moreover, the two ways to perform reconstruction from $\{y_{j}\}_{j=2}^{i}$ or $\{y_{j}\}_{j=i+1}^{N+1}$ for each $\{\dc_{b_{i}}\}_{i=2}^{N}$ based on the new code actually involves going through a subset of the two recovery sets, $\{y_{j}\}_{j=1}^{i}$ and $\{y_{j}\}_{j=i+1}^{N+1},$ of the ``old'' code. The above argument also holds when reconstructing $\dc_{b_{N}}$. Therefore, it suffices to consider the case when both $\cs^{-1}(y_{N+1})-\cs^{-1}(y_{2})< S^{c}_{t}(G,n+1)$ and $\cs^{-1}(y_{N})-\cs^{-1}(y_{1})< S^{c}_{t}(G,n+1)$ are satisfied. These two inequalities  together imply $\cs^{-1}(y_{i+1})-\cs^{-1}(y_{i})< S^{c}_{t}(G,n+1),$ $\forall \, 1\leq i\leq N$. Then, the set of chunks on $\cs$ covered by some length-$\leq S^{c}_{t}(G,n+1)$ fragment capable of recovering $\dc_{b_{i}}$ is $[\cs^{-1}(y_{i})-S^{c}_{t}(G,n+1)+1,\cs^{-1}(y_{i+1})+S^{c}_{t}(G,n+1)-1]$, with all chunks within distance $S^{c}_{t}(G,n+1)+\cs^{-1}(y_{i+1})-\cs^{-1}(y_{i})< 2S^{c}_{t}(G,n+1)$ from the position of $y_{i}$. Therefore, by replacing each $y_{i}$ by $\dc_{b_{i}}$ and deleting 
    $y_{s+1}$ we get a length-$n$ uncoded chunk store with stretch metric $< 2S^{c}_{t}(G,n+1)$. This implies $S^{u}_{t}(G,n+1)\leq S^{u}_{t}(G,n)< 2S^{c}_{t}(G,n+1)$, as claimed.
\end{IEEEproof}

\begin{figure}[H]
 \vspace{-1mm}
    \centering
    \includegraphics[scale=0.24]{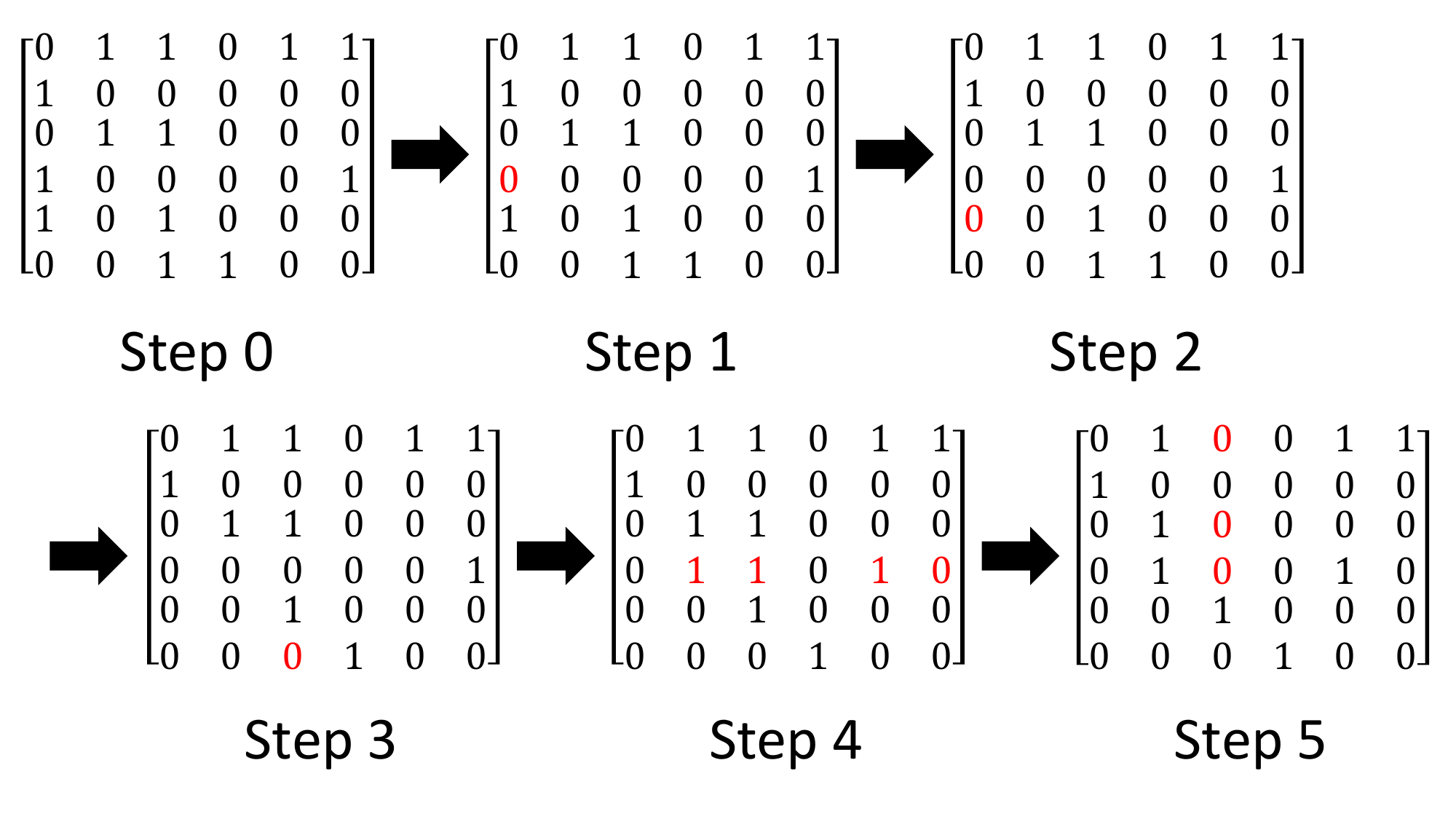}
    \caption{The matrix $\begin{bmatrix}\HH\\\KK\end{bmatrix}$, top row, left, has full rank but is not of the desired form outlined in Theorem 5.4. The first row is $\HH$. The remaining rows correspond to $\KK(1),\ldots,\KK(5)$. The elements updated at each step are marked in red. Note  that $\KK(1)$ is an indicator vector, and that $\KK(2)$ is such that $Min(\KK(2))> Max(\KK(2))$. Therefore, in Step 1, we start with $\KK(3)$. Also, $Min(\KK(3))=1$, but we cannot change $\KK(3)$ to $e_{1}$ since $\KK(1)=e_{1}$. Therefore, we change $\KK(3)$ to $\KK(3)+e_{1}=e_{6}$. This strictly improves  $Min(\KK(3)+e_{1})=2$. A similar procedure is applied to $\KK(4)$ in Step 2 and $\KK(5)$ in Step 3. At the end of Step 3, each $\KK(i)$ is either an indicator vector or there exists an $t\in[n+1]$ such that the set of values equal to $1$ in $\KK(i)$ are either the first or the last $t$ values equal to 1 within $\HH$. In Step 4, we replace those $\KK(i)$ whose set of values equal to $1$ are the last $t$ values equal to $1$ within $\HH$ by $\KK(i)+\HH$. Finally, in Step 5, we have that $\KK(4)=e_{3}$ and $min(\HH)=2\neq 3$ and $max(\HH)=6\neq 3$. Therefore, we set all elements except $\KK(4,3)$ in the third column to $0$. At the end of Step 5, we have reduced  $\begin{bmatrix}\HH\\\KK\end{bmatrix}$ to the desired form. The corresponding coded chunk store equals $(\dc_{1},\dc_{2},\dc_{4},\dc_{5},\dc_{2}\oplus\dc_{3},\dc_{3})$.}\label{fig:example_reduction}
\end{figure} 

By Proposition~\ref{coding_gain}, we also have the result from the next corollary. 
\begin{corollary}
    The problem of finding an optimal coded chunk store with one redundant chunk for general file systems \fm \, is NP-hard. 
\end{corollary}
Suppose that we are given an optimal coded chunk store $s^{*}$ with one redundant chunk for a general file systems \fm. By the reduction algorithm from the proof of Theorem~\ref{1bit_reduction}, we can find another optimal coded chunk store with one redundant chunk $s^{**}$ such that it has the specific form outlined in Theorem~\ref{1bit_reduction}, using a polynomial time reduction. Then, by the reduction process from Proposition~\ref{coding_gain}, we can find another uncoded chunk store with no redundant chunks $s$ with stretch metric $\leq 2S^{c}_{t}(G,n+1)\leq 2S^{c}_{t}(G,n)\leq 2S^{u}_{t}(G,n),$ 
also in polynomial time. Together, these reduce the problem of finding a $2$-approximation (or any constant approximation) for $S^{u}_{t}(G,n)$ (the bandwidth, when $t=2$) to the problem of finding an optimal coded chunk store with one redundant chunk. 
This follows since the graph bandwidth problem is NP-hard to approximate within any constant independent on $n$.
%Although we have a negative result for finding the best coded chunk store with one redundant chunk, when redundancy $>>1$, this might not be the case, which is left for future work.

\subsubsection{The case of no fragmentation, with coded chunk stores}\label{subsect:no_fra}
In Subsection~\ref{subsect:no_re}, we showed that coding does not help when no redundancy is allowed. We next briefly comment on the case when no fragmentation is allowed. 
\begin{theorem}\label{no_fragmentation_coding}
    For any file model \fm \, with $t=2$, let $m^{*}$ be the smallest $m$ such that $S^{c}_{2}(G,m^{*})=1$. Then, $S^{u}_{2}(G,m^{*})=1$, which means coding does not help if we wish to completely eliminate fragmentation, when $t=2$.
\end{theorem}
\begin{IEEEproof}
    Let $\{\dc_{i}\}_{i=1}^{n}$ be $n$ user data chunks, and let  $\cs^*=\cs_{1},\cs_{2},\hdots,\cs_{m^{*}}$ be an optimal coded chunk store with each $\cs_{i}$ equal to a linear combination of elements in $\{\dc_{i}\}_{i=1}^{n}$. Let $\cs_{i}$ be a coded chunk. We can delete it if it involves more than two data chunks because each file comprises at most two chunks (i.e., no file will contain elements in the linear combination that is $\cs_{i}$). For simplicity, let $\cs_{i}=\dc_{a}+\dc_{b}$, which does not correspond to a length-one file. Also, it suffices to consider only files of length two. Since we require that there is no fragmentation, we are only allowed to use either $\cs_{i-1},\cs_{i}$ and $\cs_{i},\cs_{i+1}$. It suffices to focus on just one of them since if both of them are used, then they can lead to the reconstruction of the same set of chunks $\{\dc_{a},\dc_{b}\}$. Therefore, say that $\cs_{i-1},\cs_{i}$ is used to reconstruct $\{\dc_{a},\dc_{b}\}$; then, $\cs_{i-1}\in \{\dc_{a},\dc_{b}\}$. In this case, replacing $\cs_{i}$ by $\cs_{i-1}\oplus \cs_{i}$ still works, with $\cs_{i}$ now reducing to an uncoded chunk. By repeating this procedure, we conclude that uncoded chunk stores produce the same result as coded ones. As a final remark, note that the above arguments do not easily generalize to the case $t>2$.     
\end{IEEEproof}

\section{The Jump Metric: Results}\label{sec:jump}
We start this section with a discussion of the uncoded jump metric.  
%Therefore we can expect the trade-off curve for each $G$ as shown in Figure~\ref{fig:curve_detail_j}. 
We consider two extreme cases when there is no redundancy ($m=n$) and no fragmentation. In the later case, we seek the smallest $m$ such that $J^{u}_{t}(G,m)=1$. To position our ideas and approaches, we analyze trees in Section~\ref{sec_no_redundancy_j}, and then proceed to investigate fragmentation-free scenarios in Section~\ref{no_fragmentation_j}. Afterwards, we focus on the question of \emph{whether coding can improve the jump metric} in Section~\ref{SvCS_j}. The  cases where both redundancy and fragmentation are allowed are left for future analyses. 

\subsection{Deduplication Without Redundancy}\label{sec_no_redundancy_j}
As described in Section~\ref{sec:formulation}, to compute the uncoded jump metric $J^{u}_{t}(G,n)$ for the case when no redundancy is allowed, we proceed as follows. Given a graph $G=(V,E)$, we seek to find a length-$n$ chuck store $\cs$ that minimizes the number of runs of $1$s  in the binary vector $(\mathbbm{1}\{\cs_{1}\in p\},\mathbbm{1}\{\cs_{2}\in p\},\ldots,\mathbbm{1}\{\cs_{n}\in p\})$ for all paths $p\in P(G,\leq t)$. Clearly, $J^{u}_{t}(G,m)$ is a nonincreasing function of $m$. To the best of our knowledge, this problem was not previously considered in the  literature. 

We start with the case where $G=T$ is a tree and $t=n$, which means that the file system $P(T,\le n):=P(T)$ includes all self-avoiding paths in $T$. We provide an algorithm with time complexity $O(n)$ for finding an uncoded chunk store that offers a $4$-approximation for the jump metric $J^{u}_{n}(T,n)$. We conjecture that the problem of finding the exact jump metric $J^{u}_{n}(T,n)$ for trees is not NP-hard (unlike the problem of finding the stretch metric, which is NP-hard even for subclasses of trees such as caterpillars.)

\begin{figure}[H]
    \centering    \includegraphics[scale=0.24]{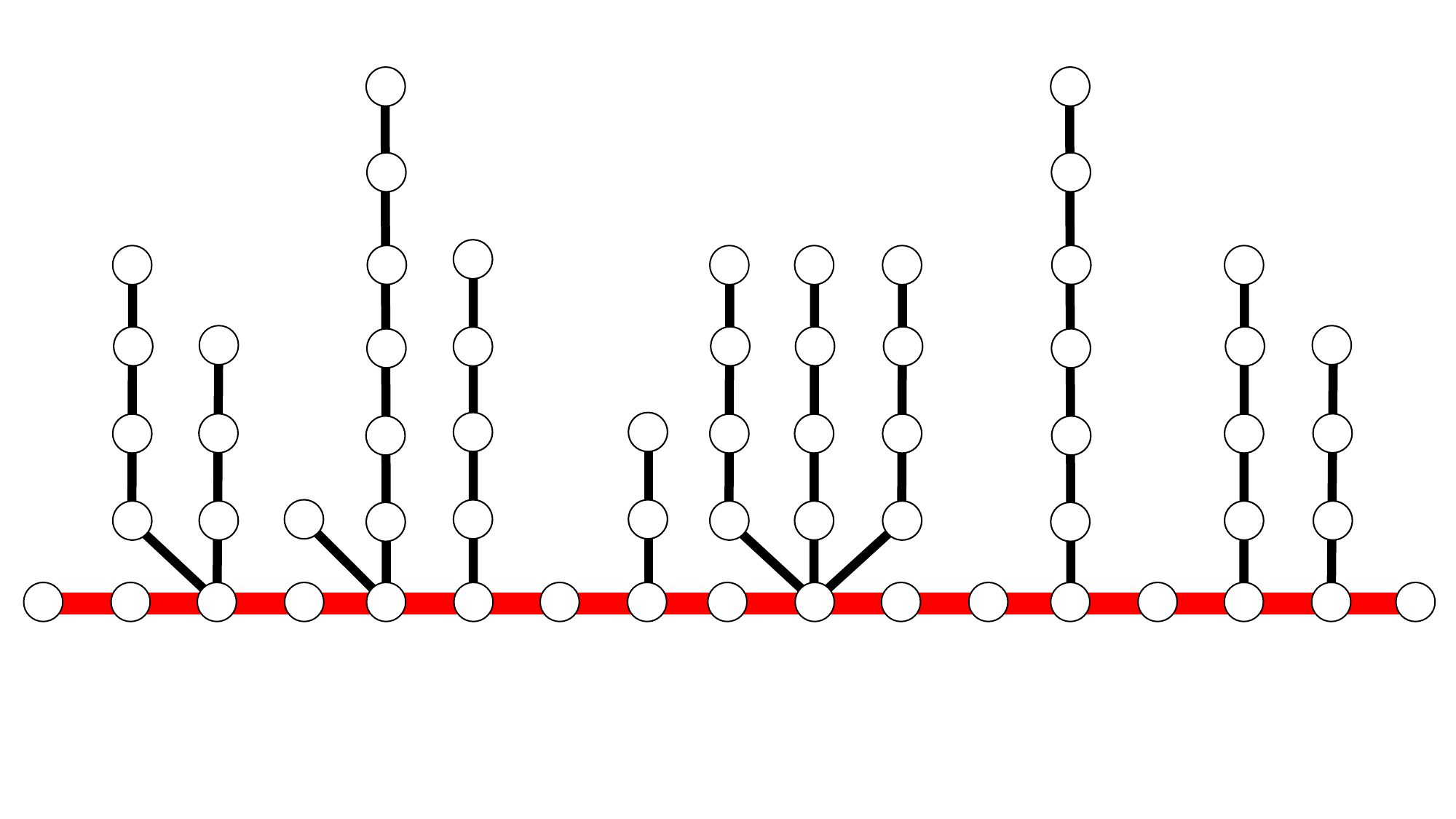}
    \vspace{-0.25in}
    \caption{A caterpillar tree with $12$ hairs.} \label{fig:tree1}
    \vspace{-0.15in}
\end{figure}

%\vspace{-0.1in}
Figure~\ref{fig:tree1} depicts a \emph{caterpillar} graph. A caterpillar is a tree in which every branching node (i.e., a node of degree $>2$) lies on the same path called the body (marked in red). The remaining paths, which intersect with the body at only a branching node, are called hairs. The example in Figure~\ref{fig:tree1} has $12$ hairs. We can construct a linearization $s$ of the caterpillar examining the hairs in sequential order (say, from left to right), then writing out the nodes in each hair from the top (the node furthest away from the body) to the root of the hair, but excluding the branching node. Then, $\MinJ(s,p)\leq 3$ for all $p\in P(T)$ since each $p$ intersects at most two hairs and the body. This  implies that $J^{u}_{n}(T,n)\leq 3$. However, there are examples for which $J^{u}_{n}(T,n)<3$, as shown next.

\begin{figure}[H]
    \centering    \includegraphics[scale=0.24]{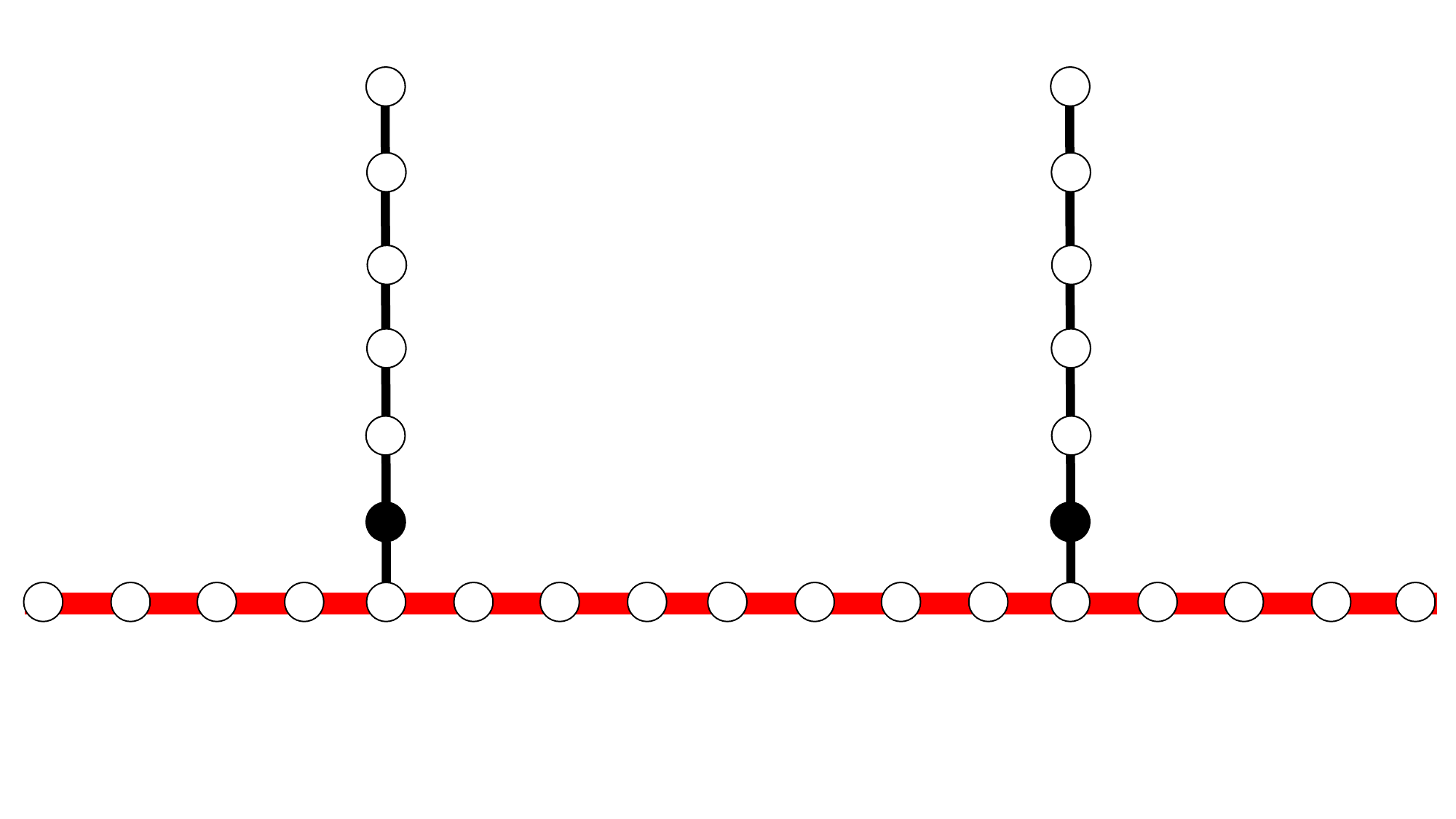}
    \vspace{-0.2in}
    \caption{A caterpillar tree with $2$ hairs.}
    \vspace{-0.1in}
    \label{fig:tree2}
\end{figure}

The caterpillar in Figure~\ref{fig:tree2} has two hairs. A jump metric value of $J^{u}_{n}(T,n)=2$ can be achieved through the following linearization procedure. We first enumerate the left hair from the top to the root (the black node), the right hair from the root to the top, and finally, the path (body) of the caterpillar.

We start with some illustrative examples.

 The only possible paths corresponding to $3$ jumps are those starting from the left hair, going through the body, and then continuing on with the right hair. Their intersection with the two hairs corresponds to two paths starting from the black node and ending somewhere on the hair. These two paths end up being merged into a single run of $1$s in the linearization.   

To generalize these ideas for more general tree structures, we study another quantity termed the \emph{unidirectional path number}. For a rooted tree $(T,r)$ with $r\in V$, let $UP(T,r)\subset P(T)$ be the set of unidirectional paths on $T$. A path $p=v_{1},v_{2},\dots$ with $v_{i}\in V$ is called a unidirectional path if $v_{i}$ is a child of $v_{i+1}$ for all $i$. As an example, for a tree with three nodes, say $a,b,c$, where $b$ is the root node and $a$ and $c$ are its children, the path $p=abc$ is not unidirectional. The terminal nodes of the unidirectional path at the child and parent side are for simplicity called the bottom and the top of the path, respectively. %$MUP(T,r)\subset UP(T,r)$ is the set of maximal unidirectional paths (the bottom is a leaf and the top is the root).

An \emph{unidirectional path decomposition}, denoted by $\upd,$ is a set of disjoint paths $p\in UP(T,r)$ such that every node in $T$ is included in exactly one path. Let $UD(T,r)$ be the set of all unidirectional path decompositions. For any $\upd\in UD(T,r)$ and a file path $p\in UP(T,r)$, the \emph{unidirectional path number} $\F(\upd,p)$ is the number of paths in $\upd$ that intersect with $p$. The min-max unidirectional path number is defined as $$UF(T,r)=\min_{\upd\in UD(T,r)}\max_{p\in UP(T,r)}\F(\upd,p).$$
The following theorem shows that an optimal unidirectional path decomposition leads to a $4$-approximation for $J^{u}_{n}(T,n)$.
\begin{theorem}\label{main_theorem}
Let $T=(V,E)$ be a tree with jump metric $J^{u}_{n}(T,n)$, rooted at an arbitrary node $r\in V$. Let $\upd^{*}=\argmin_{\upd\in UD(T,r)}\max_{p\in UP(T,r)}\F(\upd,p)$ be an optimal unidirectional path decomposition, and let $\cs^{*}\in V^{n}$ be a linearization of the tree obtained by enumerating paths in $\upd^{*}$ in arbitrary order and direction (note that there are two possible directions to enumerate nodes on a path). Then, 
$$\max_{p\in P(T)}\MinJ(\cs^{*},p)\leq 4 J^{u}_{n}(T,n).$$ 
\end{theorem}
\begin{IEEEproof}
We have 
    \begin{align*}
        \max_{p\in P(T)}\MinJ(\cs^{*},p)&\overset{(a)}{\leq} \max_{p\in P(T)}\F(\upd^{*},p)\\
        &\overset{(b)}{\leq} 2 \max_{p\in UP(T,r)}\F(\upd^{*},p)\\
        &\overset{(c)}{=} 2 \min_{\upd\in UD(T,r)}\max_{p\in UP(T,r)}\F(\upd,p)\\
        %&\overset{(d)}{=} 2 \min_{d\in UD(T,r)}\max_{p\in MUP(T,r)}\F(d,p)\\
        &\overset{(d)}{\leq} 4 \min_{\cs\in V^{n}}\max_{p\in UP(T,r)}\MinJ(\cs,p)\\
        &\leq 4 \min_{\cs\in V^{n}}\max_{p\in P(T)}\MinJ(\cs,p)\\
        &=4 J^{u}_{n}(T,n),
    \end{align*}
    where 
    $(a)$ follows from the fact that $\cs^{*}\in V^{n}$ comprises paths in $\upd^{*}$. Since the intersection of two paths on a tree is still a path, for any file path $p\in P(T)$, the intersection of $p$ and each path in $\upd^{*}$ contributes at most one single run of $1$s on $\cs^{*}$ (two runs could merge into one if they are next to each other on $\cs^{*}$ like in the example of Figure~\ref{fig:tree2}). Consequently, the number of runs of $1$s on $\cs^{*}$ for each $p$ is $\MinJ(\cs^{*},p)\leq \F(\upd^{*},p)$; 
    $(b)$ follows from the fact that any path on $T$ can be split into two unidirectional paths. Let $p_{xy}\in P(T)$ with $x,y\in V$ be the path on $T$ from $x$ to $y$. Let $z\in V$ be the lowest common ancestor of $x,y$.  Then $p_{xy}=p_{xz}p_{zy}$ with $p_{xz}\in UP(T,r)$ and $p_{yz}\in UP(T,r)$. Since $\F(\upd^{*},p)$ is the number of paths in $\upd^{*}$ that intersect with $p$, $\F(\upd^{*},p_{x,y})=\F(\upd^{*},p_{x,z})+\F(\upd^{*},p_{z,y})-1\leq 2\max_{p\in UP(T,r)}\F(\upd^{*},p)$;
    $(c)$ follows from the fact that $\upd^{*}$ is optimal; %$(d)$ follows from the worst $\F(d,p)$ must from some maximal path. 
    $(d)$ follows from the following auxiliary  Theorem~\ref{main_tech}.
\end{IEEEproof} 
%$MUP(T)\subset UP(T)$ be the set of maximal unidirectional paths (start from a leaf to the root $r$). The minimum worst-case path number is defined as $$F_{r}(T)=\min_{d\in D_{r}(T)}\max_{p\in P_{r}(T)}\F(d,p).$$ Results in this section heavily rely on the following technical lemma.
\begin{theorem}\label{main_tech}
    For any rooted tree $(T,r)$, and linearization $\cs$ over $\mathbb{S}_n$, there exists a unidirectional path decomposition $\upd(\cs)\in UD(T,r)$ such that $\max_{p\in UP(T,r)}\F(\upd(\cs),p)\leq 2 \max_{p\in UP(T,r)}\MinJ(\cs,p)$.
\end{theorem}
\begin{IEEEproof}
Let $\cs$ be any chunk store of length $n$, and let $\cs_{i}$ denote the $i$-th entry of $\cs$. Furthermore, let $\cs_{i}^{j}$ stand for the substring of $\cs$ starting at  position $i$ and ending at $j$. Recall that $\MinJ(\cs,p)$ is the number of runs of $1$s in the binary vector $(\mathbbm{1}\{\cs_{1}\in p\},\mathbbm{1}\{\cs_{2}\in p\},\ldots,\mathbbm{1}\{\cs_{n}\in p\})$. 
%\iscomment{write $\cs'$ instead of $\cs^{'}$}
Similarly, for any subsequence $\cs'$ of $\cs$, $\MinJ(\cs',p)$ equals the number of runs of $1$s in $(\mathbbm{1}\{\cs'_{1}\in p\},\mathbbm{1}\{\cs'_{2}\in p\},\ldots,\mathbbm{1}\{\cs'_{|\cs'|}\in p\})$.

Let $1\leq i_{1}\leq n$ be the first position such that there does not exist a $p\in UP(T,r)$ that satisfies $\{\cs_{i}\}_{i=1}^{i_{1}}\subset p$; refer to this position as a break-point. Next, split $\cs=\cs_{1}^{i_{1}-1}\cs_{i_{1}}^{n}$ into two parts and repeat the procedure for $\cs_{i_{1}}^{n}$ to get a second break-point $i_{2}$, and continue until $\nb$ break-points $1<i_{1}<i_{2}<\ldots<i_{\nb}\leq n$ are obtained, so that $\cs=\cs_{1}^{i_{1}-1}\cs_{i_{1}}^{i_{2}-1}\ldots \cs_{i_{\nb-1}}^{i_{\nb}-1}\cs_{i_{\nb}}^{n}$ is partitioned into $\nb+1$ parts. Note that the segments $s_{i_j}^{i_{j+1}}$ cannot be covered by a single $p\in UP(T,r)$, although segments of the form $s_{i_j}^{i_{j+1}-1}$ can. 
%\iscomment{At this point, you should clarify to the reader what is the special property of this partition. Is it that each segment $s_{i_j}^{i_{j+1}}$ cannot be covered by a single $p$ (but each segment $s_{i_j}^{i_{j+1}-1}$ can)?}
%\iscomment{At this point, you should clarify to the reader what is the special property of this partition. Is it that each segment $s_{i_j}^{i_{j+1}}$ cannot be covered by a single $p$ (but each segment $s_{i_j}^{i_{j+1}-1}$ can)?}
Let $p_{0}\in UP(T,r)$, $p_{\nb}\in UP(T,r)$, and let the $p_{j}\in UP(T,r)$ be the shortest unidirectional paths that contain all vertices in
$\cs_{1}^{i_{1}-1},\cs_{i_{\nb}}^{n}$, and each $\cs_{i_{j}}^{i_{j+1}-1},$ respectively. Let $\upd(\cs):=\{p_{0},p_{1},\ldots\}$ be the collection of these paths.

We first observe  that for any break point $b$, $\MinJ(\cs,p)=\MinJ(\cs_{1}^{b},p)+\MinJ(\cs_{b}^{n},p)-\MinJ(\cs_{b},p)$. 
By repeatedly applying this equality to all breakpoints $\{i_{j}\}$ we arrive at
\begin{align*}
\MinJ(\cs,p)&=\MinJ(\cs_{1}^{i_{1}},p)+\sum_{j=1}^{\nb-1}\MinJ(\cs_{i_{j}}^{i_{j+1}},p)\\
    &+\MinJ(\cs_{i_{\nb}}^{n},p)-\sum_{j=1}^{\nb}\MinJ(\cs_{i_{j}},p)
\end{align*}
However, 
\begin{align*}
    2\sum_{j=1}^{\nb}\MinJ(\cs_{i_{j}},p)&\overset{(a)}{\leq}\MinJ(\cs_{1}^{i_{1}},p)+\MinJ(\cs_{i_{\nb}}^{n},p)\\
    &+\sum_{j=1}^{\nb-1}\lp\MinJ(\cs_{i_{j}},p)+\MinJ(\cs_{i_{j+1}},p)\rp\\
    &\overset{(b)}{\leq}\MinJ(\cs_{1}^{i_{1}},p)+\MinJ(\cs_{i_{\nb}}^{n},p)\\
    &+\sum_{j=1}^{\nb-1}\MinJ(\cs_{i_{j}}^{i_{j+1}},p),
\end{align*}
where 
$(a)$ follows from the fact that $\cs_{i_{1}},\cs_{i_{\nb}}$ are substrings of $\cs_{1}^{i_{1}},\cs_{i_{\nb}}^{n}$, respectively; $(b)$ follows from the fact that if $\MinJ(\cs_{i_{j}},p)=\MinJ(\cs_{i_{j+1}},p)=1$, then $\MinJ(\cs_{i_{j}}^{i_{j+1}},p)>1$.
Otherwise, $p$ would cover  $\cs_{i_{j}}^{i_{j+1}}$ and, by construction, no  $p\in UP(T,r)$ can cover $\cs_{i_{j}}^{i_{j+1}}$.
Consequently we have
\begin{align*}
    2&\MinJ(\cs,p)\\
    &\geq \MinJ(\cs_{1}^{i_{1}},p)+\sum_{j=1}^{\nb-1}\MinJ(\cs_{i_{j}}^{i_{j+1}},p)+\MinJ(\cs_{i_{\nb}}^{n},p)\\
    &\geq \mathbbm{1}\{\cs_{1}^{i_{1}-1}\cap p\}+\sum_{j=1}^{\nb-1}\mathbbm{1}\{\cs_{i_{j}}^{i_{j+1}-1}\cap p\}+\mathbbm{1}\{\cs_{i_{\nb}}^{n}\cap p\}.
\end{align*}
By the definition of the jump metric, 
\begin{align*}
    &2\max_{p\in UP(T,r)}\MinJ(\cs,p)\\
    &\geq \max_{p\in UP(T,r)}\mathbbm{1}\{\cs_{1}^{i_{1}-1}\cap p\}+\sum_{j=1}^{\nb-1}\mathbbm{1}\{\cs_{i_{j}}^{i_{j+1}-1}\cap p\}+\mathbbm{1}\{\cs_{i_{\nb}}^{n}\cap p\}\\
    &\overset{(a)}= \max_{p\in MP(T,r)}\mathbbm{1}\{\cs_{1}^{i_{1}-1}\cap p\}+\sum_{j=1}^{\nb-1}\mathbbm{1}\{\cs_{i_{j}}^{i_{j+1}-1}\cap p\}+\mathbbm{1}\{\cs_{i_{\nb}}^{n}\cap p\}\\
    &\overset{(b)}=\max_{p\in MP(T,r)}\mathbbm{1}\{p_{0}\cap p\}+\sum_{j=1}^{\nb-1}\mathbbm{1}\{(p_{j}\cap p)\}+\mathbbm{1}\{p_{\nb}\cap p\} \\
    &=\max_{p\in MP(T,r)} F(\mathcal{D}(s),p)=\max_{p\in UP(T,r)} F(\mathcal{D}(s),p)
\end{align*}
where $MP(T,r)\subset UP(T,r)$ is the set of maximal unidirectional paths, with the bottom corresponding to a leaf and the top to the root. Then, $(a)$ follows by the fact that $p$ should be a maximal unidirectional path; $(b)$ follows by that fact that if a maximal unidirectional path $p$ intersects with $p_{j}$, 
then $p$ must contain the node in $p_{j}$ that is ``closest'' to the root, which also belongs to $\cs_{i_{j}}^{i_{j+1}-1},$ since $p_{j}$ is the shortest unidirectional path that covers  $\cs_{i_{j}}^{i_{j+1}-1}$. 

The set of paths $\upd(\cs):=\{p_{0},p_{1},\ldots\}$ may not be a path decomposition yet since some of the paths may overlap. To resolve this issue, we repeatedly identify two intersecting paths $p_{i}$ and $p_{j}$. 
Let $c$ be the node at which they intersect that is closest to the root. Then, $c$ must be at the top of at least one of $p_{i}$ or $p_{j}$, say $p_{j}$. Otherwise, the parent of $c$ belongs to both $p_{i}$ and $p_{j}$, consequently being an intersecting node that is closer to the root than $c$.
Then, after removing $c$ from $p_{j}$, every node is still covered by $\upd(\cs)$, without an increase of $\F(\upd(\cs),p)$ for all $p\in UP(T)$. By repeating this procedure, $\upd(\cs)$ becomes a path decomposition with $\max_{p\in UP(T,r)}\F(\upd(\cs),p)\leq 2 \max_{p\in UP(T,r)}\MinJ(\cs,p)$. 
%\iscomment{from the previous equation, shouldn't the first max be $\max_{p\in MP(T,r)}$? but then this is not exactly the theorem statement?}
%\textcolor{red}{I increase a new line at the end of last chain of inequalities to make it into $\max_{p\in UP(T,r)}\F(\upd(\cs),p)$. }
%Subsection~\ref{mt_proof}.
\end{IEEEproof}
Theorem~\ref{main_theorem} reduces the problem of finding a $4$-approximation for the jump metric of trees to that of finding an optimal unidirectional path decomposition $\upd^{*}$. %Before we propose the algorithm, we further simplify the problem by the following observations. 
%\begin{enumerate}
    %\item Since no two paths in a unidirectional path decomposition can be merged into a longer one, the bottom of each path $p\in d^{*}$ must be a leaf. Otherwise, there must exist another $q\in d^{*}$ with the top being a child of the bottom of $p$. Then $pq$ is a longer unidirectional path. On the other hand, an unidirectional path cannot intersect with two leaves. Consequently, the number of paths in $d^{*}$ is exactly the number of leaves.
    %\item Since $\F(d^{*},p)$ is defined as the number of paths in $d^{*}$ that $p$ intersect with, $\max_{p\in UP(T)}\F(d^{*},p)=\max_{p\in MUP(T)}\F(d^{*},p)$. Therefore it suffices to consider the maximal file path (start from a leaf and end at root $r$).
%\end{enumerate} 
Algorithm~\ref{alg:path_number_rooted_tree} has complexity $O(n)$ and finds $\upd^*$ recursively. 

Let $\{r_{i}\}_{i=1}^{c}$ be the set of children of $r$, and let each $r_{i}$ be the root of a subtree denoted by $T_{i}$. Given a path decomposition $\upd\in UD(T,r)$, let $p=p'r\in \upd$ be the path in $\upd$ containing $r$. 
Then,  $(\upd\backslash \{p\})\cup \{p'\}$ is the union of path decompositions of the subtrees $(T_{i},r_{i})$. 
Hence, it suffices to first compute an optimal path decomposition for each of the subtrees, $\upd_{(T_{i},r_{i})}^{*}\in UD(T_{i},r_{i})$, evaluate the min-max unidirectional path number $UF(T_{i},r_{i})$ for each subtree $(T_{i},r_{i})$, and then extend one path among $\{p_{i}=p_{i}'r_{i}\in \upd_{(T_{i},r_{i})}^{*}\}$ to include $r$. Which path to extend can be determined as follows.  Without loss of generality, let $UF(T_{1},r_{1})\geq UF(T_{2},r_{2})\geq\ldots\geq UF(T_{c},r_{c})$. 
Suppose we chose to extend $p_{i}$. Then,  $\upd=\lp\lp\bigcup_{i=1}^{c}\upd_{(T_{i},r_{i})}^{*}\rp\backslash \{p_{i}\}\rp\bigcup \{p_{i}r\}$, and $\max_{p\in UP(T,r)}\F(\upd,p)=\max(UF(T_{1},r_{1})+1,UF(T_{2},r_{2})+1,\ldots,UF(T_{i},r_{i}),\ldots,UF(T_{c},r_{c})+1)$. Therefore the best choice for a path to be extended is $p_{1},$ and consequently,  $UF(T,r)=\max(UF(T_{1},r_{1}),UF(T_{2},r_{2})+1)$.   
%By the definition of path decomposition, our final path decomposition for $(T,r)$ must contain a path from some leaf to $r$. If the leaf is in $T_{1}$, then the worst case path number $F_{r}(T)=$
\begin{algorithm}
  \caption{Finding a min-max unidirectional path decomposition} 
  \label{alg:path_number_rooted_tree}
  \begin{algorithmic}[1]
    \STATE \textbf{Input:} $(T,r)$.
    \STATE \textbf{Output:} $(\upd_{(T,r)}^{*},UF(T,r))$.
    \STATE For each $(T_{i},r_{i})$, find $\lp \upd_{(T_{i},r_{i})}^{*},UF(T_{i},r_{i})\rp$. 
    \STATE $\upd_{(T,r)}^{*}\leftarrow \lp\bigcup_{i=1}^{c} \upd_{(T_{i},r_{i})}^{*} \backslash \{p_{1}\} \rp \bigcup \{p_{1}r\}$.\\ 
    Note that by definition, $p_i$ is the path in the decomposition $\upd_{(T_{i},r_{i})}^{*}$ that passes through node $r_i$. Since we found the decomposition, we know the path $p_i$.
    \STATE $UF(T,r)\leftarrow \max(UF(T_{1},r_{1}),UF(T_{2},r_{2})+1)$
    \STATE \textbf{Return} $(\upd_{(T,r)}^{*},UF(T,r))$.
  \end{algorithmic}
\end{algorithm}

\begin{figure}[H]
     \vspace{-0.1in}
        \centering
        \includegraphics[trim={0 3cm 0 1cm},clip,scale=0.26]{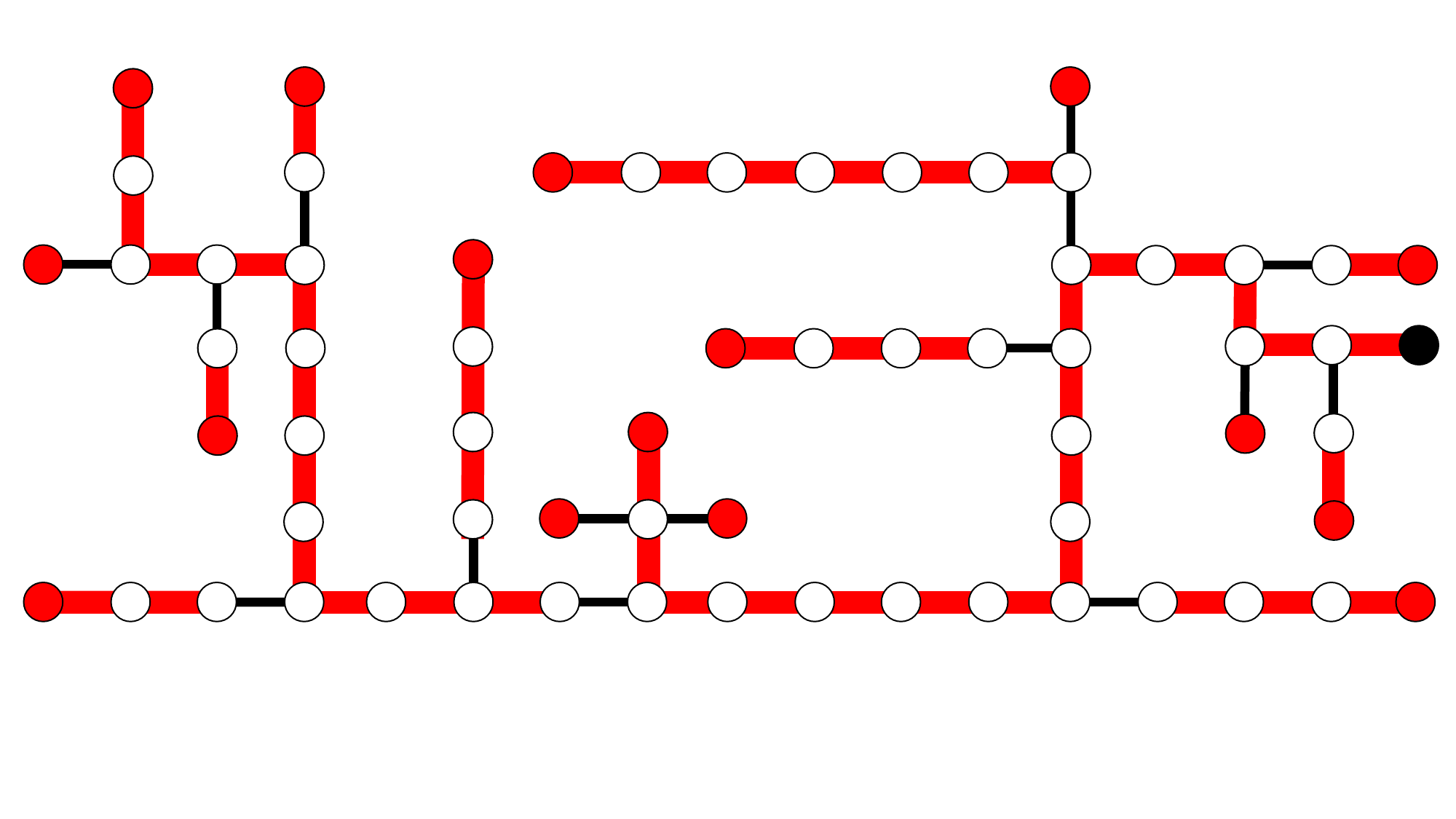}
        \vspace{-0.1in}
    \caption{An example illustrating the process described in   Algorithm~\ref{alg:path_number_rooted_tree}. The black node is the chosen root node. The $16$ red nodes are leaves. The corresponding $16$ paths in the path decomposition are wide red lines.}
    %\label{psht}
\end{figure}

%Finally plug in Algorithm~\ref{alg:path_number_rooted_tree} into Algorithm~\ref{alg:path_number_tree}, we have an algorithm that can find an optimal path decomposition for tree $T$, and consequently a $\log(n)$ approximation of the jump metric $J^{u}_{n}(T,n)$ due to Theorem~\ref{thm:lognapproximation}. 

\subsection{Zero-Fragmentation Deduplication}\label{no_fragmentation_j}
By the definitions of the stretch and jump metrics, the smallest value of $m^{*}$ that allows one to achieve $J^{u}_{t}(G,m^{*})=1$ also forces $S^{u}_{t}(G,m^{*})=1,$ and vice versa. Therefore, the zero-fragmentation results for both metrics are the same.

\subsection{Coding Versus Repetitions}\label{SvCS_j}
In this section we address the question of whether coding can help to reduce the jump metric. Specifically, we derive bounds on the difference $J^{u}_{t}(G,m)-J^{c}_{t}(G,m)$ and construct an example that shows that  $J^{u}_{t}(G,m)-J^{c}_{t}(G,m)>0$. The problem of finding the best coded chunk stores is difficult and left for future work. Also, we first consider chunk stores without redundancy ($m=n$) and then redirect our attention to stores with one redundant chunk ($m=n+1$).
\begin{theorem}\label{no_redundancy_j}
    We have $J^{u}_{t}(G,n)=J^{c}_{t}(G,n)$ for any file model \fm, which means that coding cannot improve the jump metric when no redundancy is allowed (i.e., when $m=n$). The result holds for both potential linear and nonlinear codes.
\end{theorem}
\begin{IEEEproof}
    The proof is similar to the proof of the results for the   stretch metric in  Theorem~\ref{no_redundancy} and is therefore omitted.
\end{IEEEproof}
When redundancy is allowed, Example~\ref{example1j} shows that coding can improve the jump metric even when only one redundant chunk ($m=n+1$) is inserted into the store.

\begin{example}\label{example1j}
Consider the file model $P(T,\leq n),$ with the generating graph $T$ depicted in Figure~\ref{fig:example1j}.
\begin{figure}[H]
 \vspace{-4.5mm}
    \centering
    \includegraphics[scale=0.24]{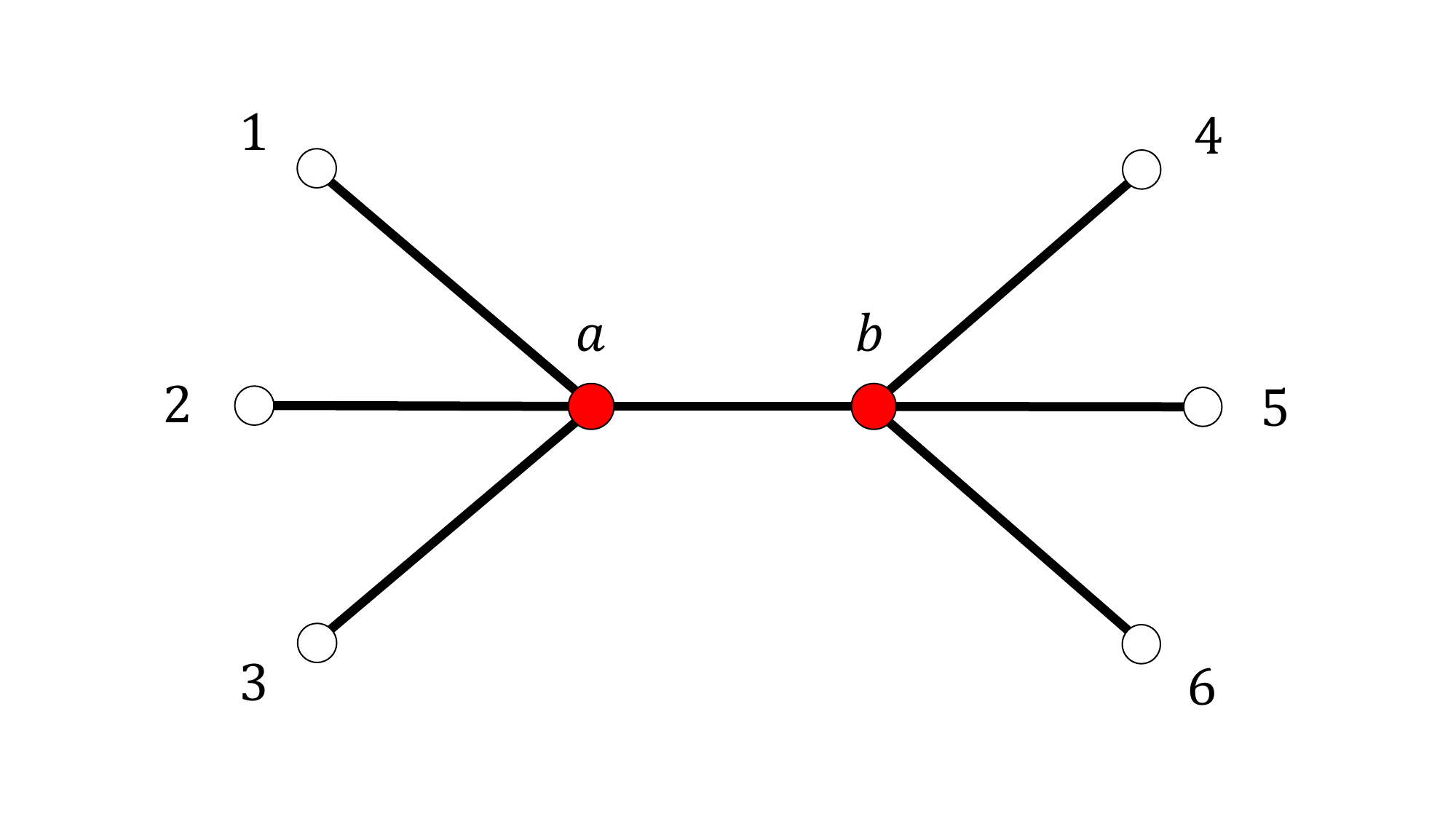}
    \caption{A tree $T=(V,E)$ with $8$ nodes.}\label{fig:example1j}
\end{figure}    

We start with the case of a chunk store with one redundant chunk ($m=n+1$). In this case, the uncoded jump metric satisfies $J^{u}_{n}(T,n+1)>2$. 
Suppose that there exists a length-$n+1$ chunk store $\cs$ with jump metric $\leq 2$. Since we can only duplicate one chunk, at least one of the sets of chunks from $\{\dc_{a},\dc_{1},\dc_{2},\dc_{3}\},\{\dc_{b},\dc_{4},\dc_{5},\dc_{6}\}$ (refer to the figures for chunk labels) is not duplicated in $\cs$, say $\{\dc_{a},\dc_{1},\dc_{2},\dc_{3}\}$. Let $\cs'$ denote the store $\cs$ after removing all replicas of $\dc_{b}$ (of which there may be at most two). Then, all $\{\dc_{i}\dc_{a}\dc_{j}\}_{i\neq j\in[3]}$ and $\{\dc_{i}\dc_{a}\dc_{b}\dc_{j}\}_{i\in[3],j\in[4:6]}$ must have a jump number $\leq 2$ with respect to $\cs$. This implies that all $\{\dc_{i}\dc_{a}\dc_{j}\}_{i\neq j\in[3]}$ and $\{\dc_{i}\dc_{a}\dc_{j}\}_{i\in[3],j\in[4:6]}$ must have a jump number $\leq 2$ with respect to $\cs'$ (since removing nodes does not lead to an increase in the jump metric). 
%\iscomment{but $\{\dc_{i}\dc_{a}\dc_{j}\}_{i\in[3],j\in[4:6]}$ are not paths on $T$?} 
There are two cases to consider for $\cs'$. The first case is that $\{\dc_{1},\dc_{2},\dc_{3}\}$ are all on the left of $\dc_{a}$ or all are on the right of $\dc_{a}$. The second case is that two of the three are on the left of $\dc_{a}$ and one at the right of $\dc_{a},$ or two of the three are on the right side of $\dc_{a}$ and one at the left side of $\dc_{a}$. We proceed with the first case, as the second case can be examined similarly.  Without loss of generality let $\cs'=\ldots\dc_{1}\ldots\dc_{2}\ldots\dc_{3}\ldots\dc_{a}\ldots$. If there exists any other chunk between $\dc_{3},\dc_{a}$, then $\dc_{1}\dc_{a}\dc_{3}$ has jump number $=3$ with respect to $\cs'$. 
Therefore $\cs'=\ldots\dc_{1}\ldots\dc_{2}\ldots\dc_{3}\dc_{a}\ldots$. Next, we show that all the elements in $\{\dc_4,\dc_5,\dc_6\}$ are adjacent to either $\dc_1$ or $\dc_a$. Suppose on the contrary, that one of $\{\dc_{4},\dc_{5},\dc_{6}\}$, say $\dc_{4}$, is not adjacent to $\dc_{1}$ or $\dc_{a}$ with respect to $\cs'$. Then $\dc_{1}\dc_{a}\dc_{4}$ has jump number equal to $3$ with respect to  $\cs'$. 
%\iscomment{but $\dc_{1}\dc_{a}\dc_{4}$ is not a path on $T$?}
Therefore,  without loss of generality, we can assume that $\cs'=\ldots\dc_{4}\dc_{1}\dc_{5}\ldots\dc_{2}\ldots\dc_{3}\dc_{a}\dc_{6}\ldots$. In this case, $\dc_{1}\dc_{a}\dc_{2}$ has jump number equal to $3$ with respect to  $\cs'$. Consequently,  $J^{u}_{n}(T,n+1)>2$.   
%By construction, $\dc_{a}$ has $3$ neighbors among $\{\dc_i\}_{i\in[6]}$. Therefore in any uncoded chunk store, at least one of them is not adjcient to $\dc_{a}$ on the chunk store, say $\dc_{1}$. 
%In addition, one of them must stand between another one and $\dc_{a}$. Consider $3$ different paths $\{\dc_{1}\dc_{a}\dc_{b}\dc_{i}\}_{i=4,5,6}$ on $T$. Notice  that all of them include $\dc_{1}$ and $\dc_{b}$. Since we have only one redundant chunk, at least one of them is not duplicated, say $\dc_{b}$. \textcolor{blue}{[previously we assume $x_a$ is not duplicated]} Then by the same reason, one of $\dc_{4},\dc_{5},\dc_{6}$ must stand between another one and $\dc_{b}$. Therefore there is always one path among three $\{\dc_{1}\dc_{a}\dc_{b}\dc_{i}\}_{i=4,5,6}$ that has jump metric $3$. 

On the other hand, the following coded chunk store has a jump number equal to $2$: 
\begin{equation*}
    \cs=\dc_{1}\dc_{a}\dc_{2}\dc_{3}(\dc_{a}\oplus \dc_{b})\dc_{4}\dc_{5}\dc_{b}\dc_{6}
\end{equation*}
Therefore, $J^{c}_{n}(T,n+1)\leq 2$.

To verify the claim, it suffices to check all paths of length $3$ and $4$ that can be recovered with $2$ jumps. The case of paths of length $3$ is easy to check. For paths of length $4$, we observe that we can
\begin{align*}
    \text{Recover }    &   \{\dc_{1},\dc_{2}\}\dc_{a}\dc_{b}\{\dc_{5},\dc_{6}\}\\
    ~&   \text{ from } \dc_{1}\dc_{a}\dc_{2}            ~&\text{ and } ~&\dc_{5}\dc_{b}\dc_{6}\\
    \text{Recover }    &   \dc_{3}\dc_{a}\dc_{b}\{\dc_{5},\dc_{6}\}\\
    ~&   \text{ from } \dc_{3}(\dc_{a}\oplus \dc_{b})   ~&\text{ and } ~&\dc_{5}\dc_{b}\dc_{6}\\
    \text{Recover }    &   \{\dc_{1},\dc_{2}\}\dc_{a}\dc_{b}\dc_{4}\\
    ~&   \text{ from } \dc_{1}\dc_{a}\dc_{2}            ~&\text{ and } ~&(\dc_{a}\oplus \dc_{b})\dc_{4}\\
    \text{Recover }    &   \dc_{3}\dc_{a}\dc_{b}\dc_{4}\\
    ~&   \text{ from } \dc_{a}              ~&\text{ and } ~&\dc_{3}(\dc_{a}\oplus \dc_{b})\dc_{4}
\end{align*}
The reduction of the jump metric achieved by coding is therefore 
\begin{equation*}
    J^{u}_{n}(T,n+1)-J^{c}_{n}(T,n+1)\geq 1.
\end{equation*}
\end{example}
The next result shows that for any file model and one redundant chunk, one can reduce the jump metric via coding by at most $2$. 
\begin{theorem}\label{coding_gain_j}
    We have $J^{u}_{t}(G,n+1)-J^{c}_{t}(G,n+1)\leq 2$ for any \fm, meaning that linear codes with one redundant chunk improve the uncoded jump metric by at most $2$.   
\end{theorem}
\begin{IEEEproof}
%We first show that all linear codes with one redundant chunk have at most one coded chunk as in  Example~\ref{example1j}.
Let $\cs=\cs_{1},\hdots,\cs_{n+1}$ be a length-$n+1$ coded chunk store in which each $\cs_{i}$ is a linear combination of a subset of user data chunks $\{\dc_{i}\}_{i=1}^{n}$. By the definition of the jump metric, it is required that for every file $p\in$ \fm\, comprising chunks in $X_{p}\subset \{\dc_{i}\}_{i=1}^{n}$, there exists a set of chunks $\cs_{X_{p}}\subset \cs_{[n+1]}$ of size $|\cs_{X_{p}}|=|X_{p}|$ such that $\cs_{X_{p}}$ can be used to recover $X_{p}$. Since all single chunks $\{\dc_{i}\}$ 
 belong to \fm, there are at least $n$ chunks in $\cs$ that are uncoded. Combining this claim with the fact that $m=n+1$, we conclude that there is at most one coded chunk in $\cs$.
 
Let $\cs^*$ be an optimal coded chunk store that achieves the jump metric $J^{c}_{t}(G,n+1)$. Let $\{\dc_{i_{1}},\dc_{i_{2}},\ldots\}\subset\{\dc_{i}\}_{i=1}^{n}$ and let, without loss of generality, $\cs^*_{j}=\sum_{\dc\in \{\dc_{i_{1}},\dc_{i_{2}},\ldots\}}\dc$ 
%\iscomment{This notation is strange. What is $\{a_i\}$? Is it a subset of $\{x_i\}_{i=1}^n$?}
be the only coded chunk in $\cs^*$. To recover each $X_{p}\subset\dc_{[n]}$, we can use the $n$ uncoded chunks in $\cs^*$ directly. If $\{\dc_{i_{1}},\dc_{i_{2}},\ldots\}\subset X_{p}$, then there exist $|\{\dc_{i_{1}},\dc_{i_{2}},\ldots\}|$ additional ways to recover $X_{p}$ by replacing one of the $\{\dc_{i_{1}},\dc_{i_{2}},\ldots\}$ in $\cs$ by $\cs_{j}$. Since we only remove one chunk and add another chunk, the increase of the jump metric is at most $2$. Therefore, we can create another uncoded chunk store $\cs'$ by removing $\cs_{j}$ from $\cs$. Then, all recovery sets for $\cs$ that do not use the coded chunk can also be used for $\cs'$, resulting in the same or smaller jump metric. Consequently $\cs'$ is an uncoded chunk store with jump metric $\leq J^{c}_{t}(G,n+1)+2$, which implies that $J^{u}_{t}(G,n+1)-J^{c}_{t}(G,n+1)\leq 2$.
\end{IEEEproof}
%\begin{corollary}
%    The proof above provides an intuition as of why coding helps. The set of original chunks $\{a_{i}\}$ can be understood as popular chunks shared by many files. We therefore apply coding on them so that $\{a_{i}\}$ can be reconstructed by $|\{a_{i}\}|+1$ different ways.
%\end{corollary}
%Finally, though jump metric for linear codes with one redundant chunk is only a base case, it might be more interesting when redundancy $\gg1$. 
Note that if we were to change the file model so that single-chunk files are prohibited, the arguments above for linear codes with one redundant chunk would no longer hold. %For example, an alternative file model can be an arbitrary collection of hyper edges (which are subsets of $\dc_{[n]}$).

\section{Related Works}\label{sec:related}
%\subsection{Information Retrieval}
The work closest in terms of its approach to data storage is~\cite{coffey2000fundamental},  which established ultimate performance limits of information retrieval with and without data redundancy. One simple example to illustrate the problem is as follows. Assume that one wants to store a large image (i.e., a 2D grid of pixels) on a tape (i.e., a 1D storage media). The question of interest is how to store the pixels on the tape in a linear order to minimize the average access time of collections of pixels of walks/paths on the grid. Intuitively, the question asks to determine the effective reduction of 2D neighborhoods of pixels stored on a 1D tape. 

More precisely, assume that the image grid comprises $n\times n$ pixels. The system is initially in a state in which one of the pixels is displayed for the user, and the read–write head of the tape is located at a copy of the block stored on the tape.
The user generate a random sequence of requests for accessing pixels on the tape dictated by moves to one of the four neighboring blocks in the grid with uniform probability, and for each of the requests. After each request, the read head of the tape moves from its current location on the tape to the new position of the requested pixel, with access time measured by the moving distance.
The goal is to arrange all $n^2$ pixels on the tape in a manner that minimizes the average moving distance per time steps across all users generating random requests.

The problem therefore involves constructing a mapping between vertices of two graphs such that a given random walk on the first graph maps to an induced random walk on the second. The aim is to design the mapping so that the induced random walk on the second graph has as low an expected cost as possible. Specifically, let $V$ be the set of nodes of the $n \times n$ grid-graph, and let the graph be endowed with transition probabilities between pairs of adjacent states in $V$. The cost function of interest reads as 
\begin{equation*}
    E_C=\sum_{i,j\in V}p(i)\Pr(i,j)c(i,j), 
\end{equation*}
where $Pr(i,j)$ is the transition probability from state $i$ to state $j$ of the underlying irreducible Markov chain, and where $p(i)$ is the unique invariant probability distribution of the chain. The cost $c(i,j)$ is the distance between the vertices $i,j \in V$ on the tape (i.e., in the linear order).

\begin{figure}[H]
     \vspace{-0.1in}
        \centering
        \includegraphics[trim={0 3cm 0 4cm},clip,scale=0.26]{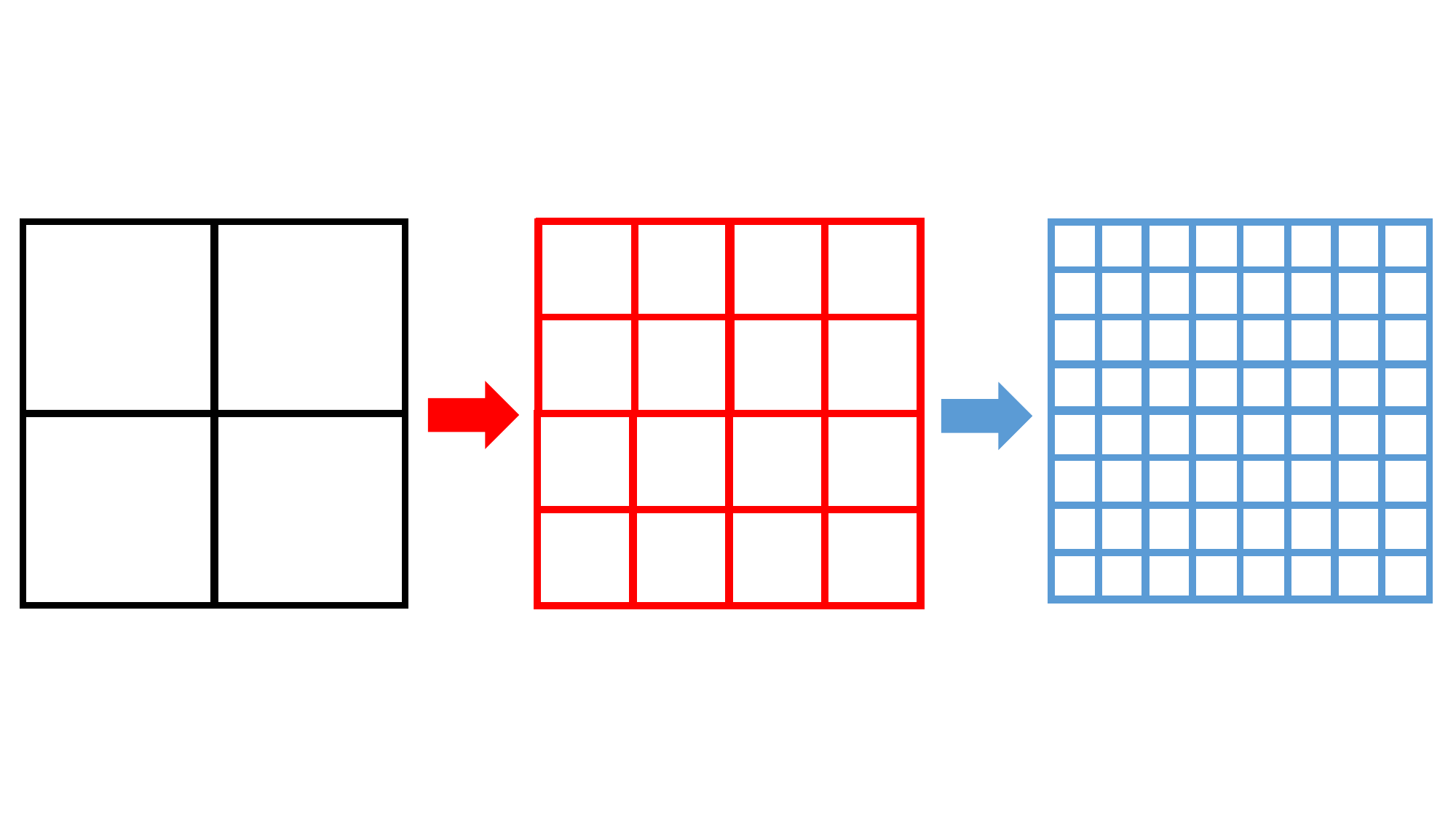}
        \vspace{-0.1in}
    \caption{The first step is to recursively partition the torus into $g\times g$ subgrids until reaching an  individual pixel (in this example, via a $g=2\times 2$ partition).}
    \label{coffee1}
\end{figure}

\begin{figure}[H]
     \vspace{-0.1in}
        \centering
        \includegraphics[trim={0 3cm 0 4cm},clip,scale=0.26]{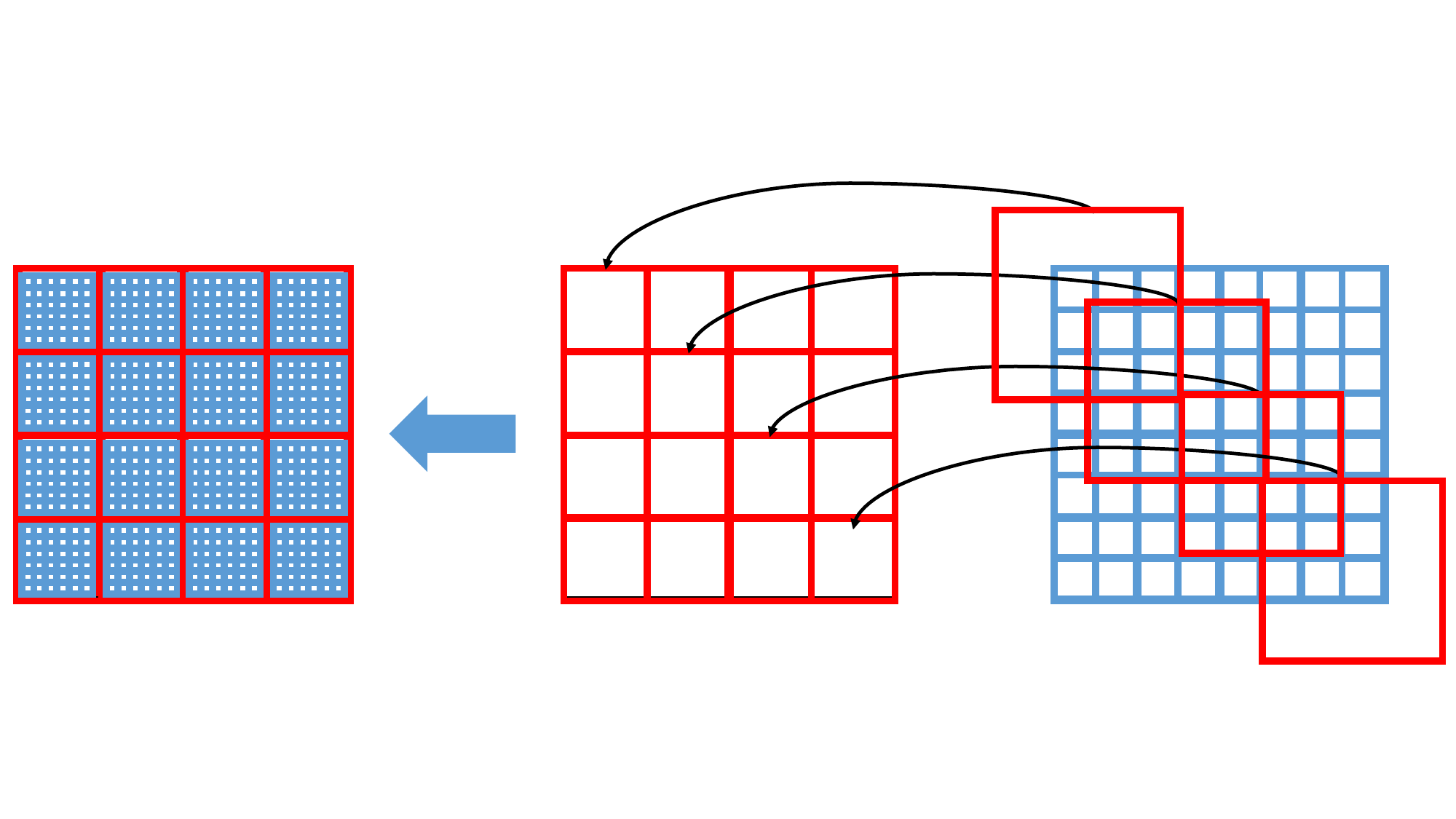}
        \vspace{-0.1in}
    \caption{The second step is to create a redundant square grid in the first layer, by expanding the $g \times g$ grids into $(g+2)\times(g+2)$ grids, wrapped around at the boundaries according to the torus neighborhoods. For the example, the redundant subgrids are of size $4 \times 4$.}
    \label{coffee2}
\end{figure}

\begin{figure}[H]
     \vspace{-0.1in}
        \centering
        \includegraphics[trim={0 3cm 0 4cm},clip,scale=0.26]{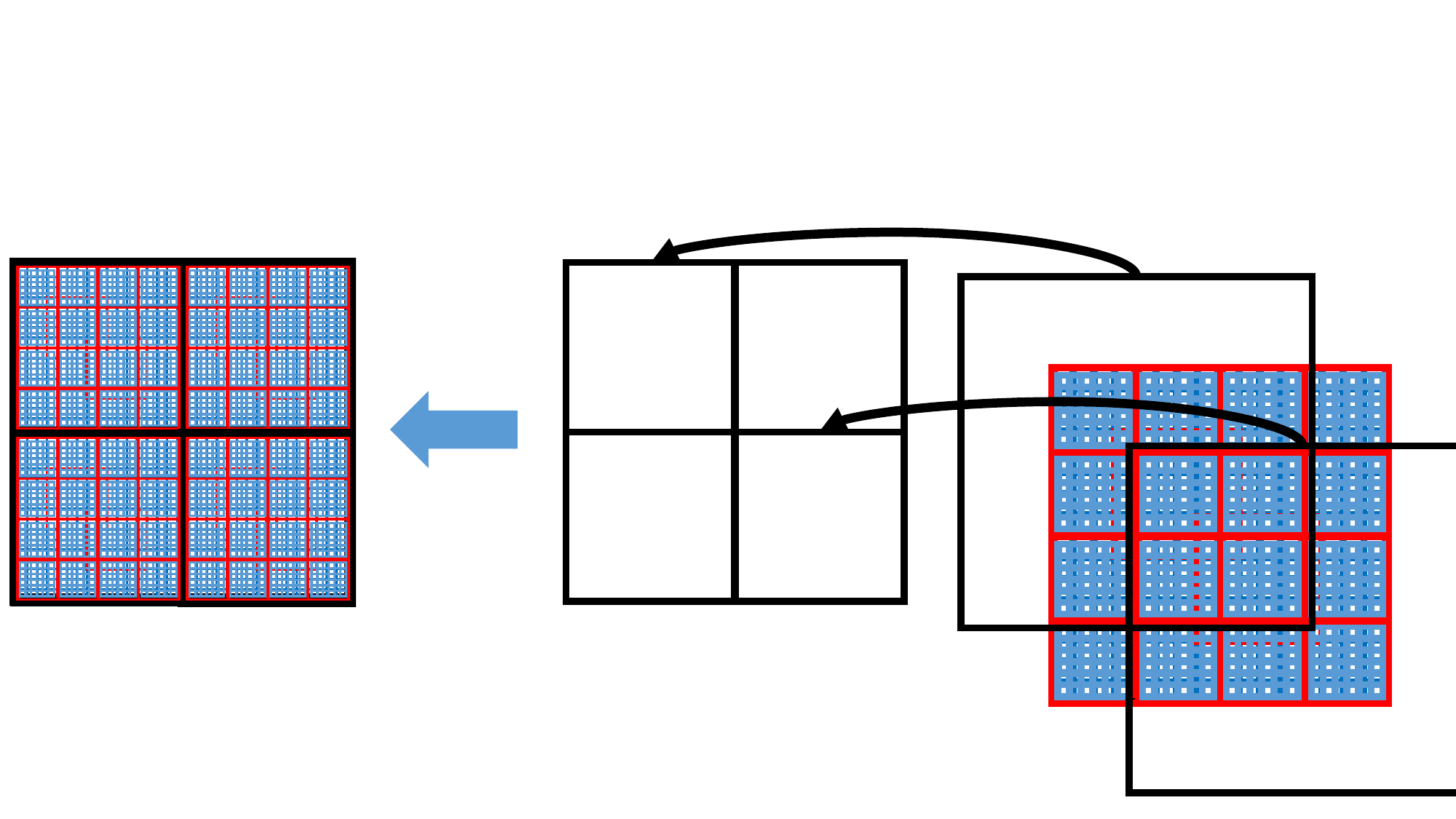}
        \vspace{-0.1in}
    \caption{In the next steps, the procedure described in the second step is repeated for the second, third etc layer, again using pixel subgrids of size $(g+2)\times(g+2)$.}
    \label{coffee3}
\end{figure}

\begin{figure}[H]
     \vspace{-0.1in}
        \centering
        \includegraphics[trim={0 3cm 0 4cm},clip,scale=0.26]{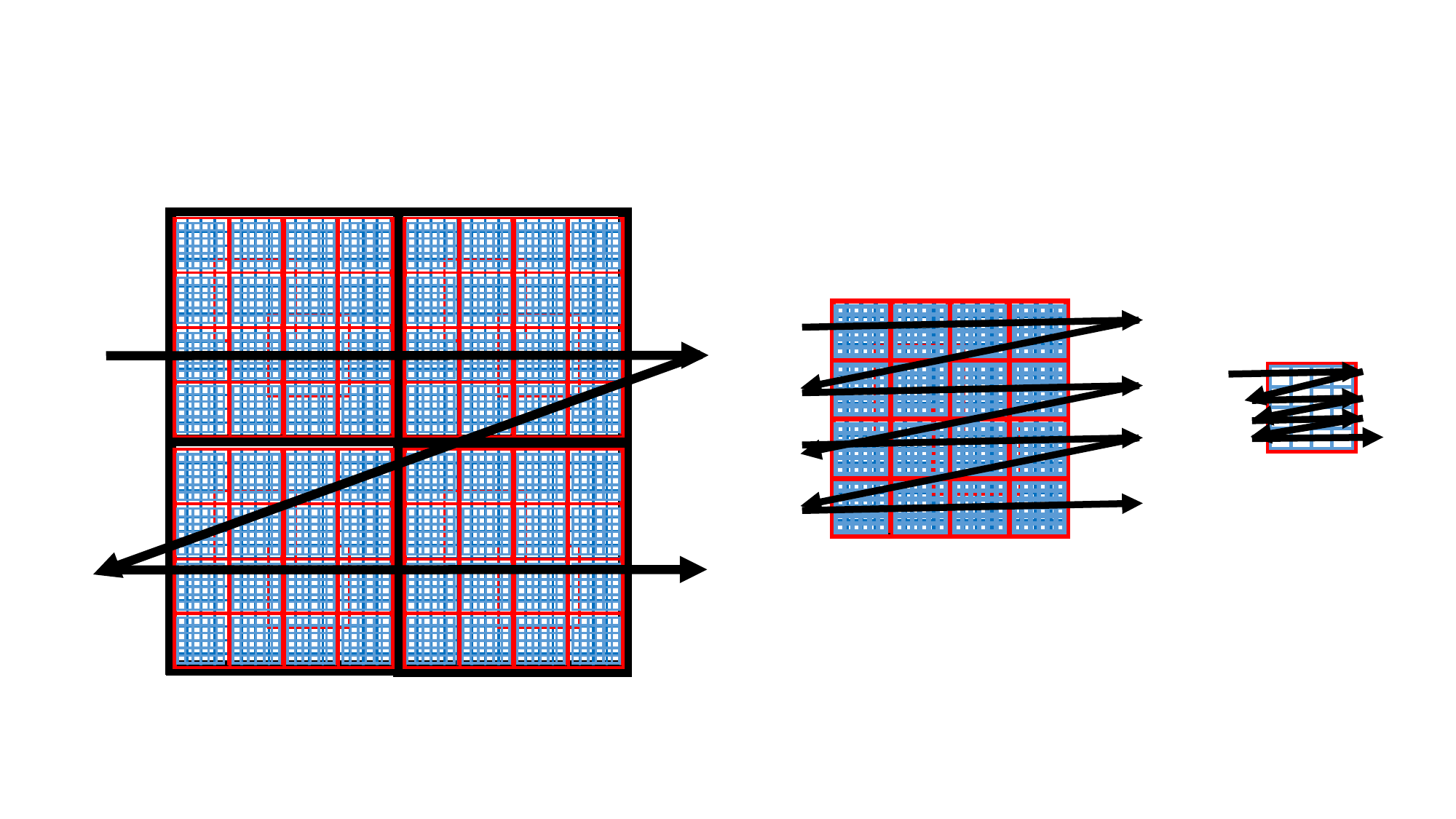}
        \vspace{-0in}
    \caption{The final step is to linearize the redundant grid using a multi-layered (nested) raster scan.}
    \label{coffee4}
\end{figure}

When storing redundant pixels on the tape is allowed, the set of states of the storage system increases, and it is in general not entirely determined by the pixel currently being accessed. The system is now specified by an initial linear arrangement of pixels in the store, and a rule for deciding which state the system moves into given its current state and the next request. For example, with redundancy, we can store many copies of the same pixel on the tape. Then the rule must specify which of the several copies of a requested block is accessed. Nevertheless, we still operate under the assumption that the tape system states form an irreducible Markov chain. 
%The general problem with redundancy involves a mapping between two graphs with the number of vertices for that has the same number of vertices such that a given random walk on the first graph mapping to an induced random walk on the second. The problem is to design the mapping so that the induced random walk on the second graph has as low an expected cost as possible. Specifically, the expected cost is defined as
The expected cost in this case is defined as
\begin{equation*}
    E_C=\sum_{\ell\in L,j\in V}p(\ell)\Pr(\ell,j)c(\ell,j)
\end{equation*}
where $L$ is the redundant state space, and $Pr(\ell,j)$ are the transition probabilities of the Markov chain with redundant state space $L$. As before, $p(\ell)$ stands for the unique invariant probability distribution of the chain, while $c(\ell,j)$ denotes the cost of serving a request for block $j$ when the system is in state $\ell$, given the redundant  mapping of pixels to the tape.
Under this definition, unlike in our setting, it is not clear whether increasing redundancy always reduces the expected cost. For example, when we insert a new copy of a pixel on the tape, the cost of transitions between other states might increase since one needs to potentially cross one additional pixel on the tape. One of the main results in~\cite{coffey2000fundamental} shows
that when the file graph is a 2D $n \times n$ torus (i.e., wrapped grid) and the storage tape is a 1D loop (cycle), for the case of no redundancy ($m=n^{2}$), the optimal value of the objective scales as  $\Theta(n)$. However, with a small amount of redundancy (i.e., for $m=(1+\delta)n^{2}$ with a small constant $\delta$), the optimal value of the objective reduces to  $\Theta(1)$. The latter can be realized by repeatedly duplicating neighboring pixels across different granularities and then linearizing the 2D structure using a multi-layered raster scan, as illustrated in Figures~\ref{coffee1},~\ref{coffee2},~\ref{coffee3}, and~\ref{coffee4}.


\section{Future Work}\label{sec:future}
Many open problems in this new research area remain. Some of them are listed in what follows.
\begin{enumerate}
    \item \emph{Extend the results for the stretch and jump metric to apply to other types of graph-induced file models.} Despite sparse Hamiltonian and tree graphs being a good starting point for file modeling and analysis, it is clear that they represent an abstraction with practical limitations. For example, one can argue that instead of graphs, it is more appropriate to use unions of graphs, each of which models a different data file category. This setting can be analyzed in the same way as a single graph, but when the graphs are sparsely connected to form communities, the problem becomes significantly more challening.
    \item \emph{Extend the  characterization and conduct a more in-depth analysis of the redundancy-fragmentation trade-offs for different type of file graphs.} Since the trade-offs heavily depend on the properties of the file graphs, it would be relevant to first focus on bounds on the redundancy needed to achieve a certain fragmentation level, independent on the file graphs.
    \item \emph{Repeat and extend the results for other fragmentation metrics.} One can envision other metrics that are combinations of the stretch and jump metric (e.g., sum of stretches of individual jump fragments) or that take into account file integrity. In the latter case, not all permutations are allowed for the chunk store, but only those that respect \emph{batches} of unique chunks encountered in a file (i.e., breaking up bursts of unique chunks in files is not allowed).  
    \item \emph{Determine for what type of graphs adding coded redundant chunks outperforms simple chunk repetition.} Our examples show that the performance or repetition and linear coding may also depend on the allowed redundancy, which makes this problem challenging.
    \item \emph{Explore coding methods that involve more than $1$ redundant data chunk}. As deduplication necessitates significant reductions in the number of redundant chunks, to properly control fragmentation one may need to include more than one or two redundant or coded data chunks. We still have no general approaches for identifying which chunks to replicate or encode in this case. 
    \item \emph{Develop corresponding models for online setting.} The challenges associated with the online setting are that unique chunks already present in a file have to be stored in their natural order of appearance in the file. This implies that not all permutations of chunks are allowed for use in the chunk store.
    \item \emph{Simultaneously address fragmentation and robustness issues by using redundant or coded data chunks.} Although our initial analysis of fragmentation phenomena suggests that for file-graph models ``highly connected nodes'' are natural candidates for repetition, it is not clear if other, low-popularity chunks can also help in reducing fragmentation.
    \item \emph{Better analysis for bandwidth problem on Sparse Hamiltonian graph}.
    \item \emph{Characterization the middle part of the trade-off curves $S^{u}(G,m)$ and $J^{u}(G,m)$}.
    %\item repeat analyses for online deduplication (we exclusively worked with offline/postprocessing); in this case, chunks that come from the same file will necessarily have to be placed together, so we cannot arbitrary permute the content.
%    \item nonuniform chunk sizes
%    \item adding caching?
\end{enumerate}

\vspace{0.2in}
\textbf{Acknowledgement.} The authors are grateful to Hsing-Po Wang for suggesting to examine coding strategies other than pure repetition coding.

\section{Appendix}\label{sec:appendix}
\subsection{Proof for Theorem~\ref{lem:generalk}}\label{pf:generalk}
Let $\Earc=\{\e{a_1}{b_1},\ldots,\e{a_k}{b_k}\}$ and rearrange the entries in $\{a_1,b_1,\ldots,a_k,b_k\}$ into $(c_1,\ldots,c_2k)$ so that $c_1<\ldots<c_2k$, i.e., the sequence is increasing. Similar to what we did in the proof of Proposition~\ref{prop:k3}, we construct $h$ using a two-step approach. In the first step, we group the nodes $\dc_{c_1},\ldots,\dc_{c_{2k}}$. In the second step, we find a set of values $r_1,\ldots,r_{L}$ that satisfies Equations~\eqref{eq:diri} and~\eqref{eq:decidingr}, and the constraints on \textcolor{red}{$d_i$,} $i\in[L],$ that is consistent with the grouping of $c_1,\ldots,c_{2k}$ (which is \eqref{eq:group6} in the proof of Proposition \ref{prop:k3}).

Here, the grouping step generalizes that in the proof of Proposition~\ref{prop:k3} as described in Algorithm~\ref{alg:grouping}.
\begin{algorithm}
  \caption{Grouping of nodes $\{c_1,\ldots,c_{2k}\}$\iffalse$(\boldsymbol{a},\boldsymbol{b},\boldsymbol{c})$\fi} 
  \label{alg:grouping}
  \begin{algorithmic}[1]
    \STATE \textbf{Input:} $\Earc$.
    \STATE \textbf{Initialization:} Order the nodes in $\{\dc_{a_1},\ldots,\dc_{a_k},\dc_{b_1},\ldots,\dc_{b_k}\}$ as $\{\dc_{c_1},\ldots,\dc_{c_{2k}}\}$ such that $c_1<\ldots<c_{2k}$. Let
    $E_{adj} = \{\{c_i,c_{i+1}\}:i\in\{1,\ldots,2k--1\}\}$ and $E_{group}=\emptyset$.  
    \STATE \textbf{Step 1:} Find the pair $\{c_{i^*},c_{i^*+1}\}\in E_{adj}$ such that $\dc_{c_{i^*}},\dc_{c_{i^*+1}})\notin \Earc$ and $(c_{i^*+1}-c_{i^*})$ is as small as possible, i.e. $i^*=\arg\min_{\{c_{i},c_{i+1}\}\in E_{adj},(\dc_{c_{i}},\dc_{c_{i+1}})\notin \Earc}(c_{i+1}-c_i)$.
    \STATE \textbf{Step 2:} For each $c_i$ such that $(\dc_{c_{i}},\dc_{c_{i^*}})\in \Earc$ or $(\dc_{c_{i}},\dc_{c_{i^*+1}})\in \Earc$, and each $c_j$ such that $(\dc_{c_j},\dc_{c_{i-1}})\in \Earc$ or $(\dc_{c_j},\dc_{c_{i+1}})\in \Earc$, remove $\{c_{j},c_{j+1}\}$ from $E_{adj}$ if $\{c_{j},c_{j+1}\}\in E_{adj}$. Remove $\{c_{i^*},c_{i^*+1}\}$ from $E_{adj}$ if $\{c_{i^*},c_{i^*+1}\}\in E_{adj}$. Add $\{c_{i^*},c_{i^*+1}\}$ to $E_{group}$, i.e., $E_{group}=E_{group}\cup \{\{c_{i^*},c_{i^*+1}\}\}$.  
    \STATE \textbf{Step 3:} If $E_{adj}\ne \emptyset$, go to \textbf{Step 1}. 
    \STATE \textbf{Return} $E_{group}\cup \{\{c_i\}\}_{\{c_i,c_{i+1}\}\notin E_{group}, \{c_{i-1},c_i\}\notin E_{group}}$.
  \end{algorithmic}
\end{algorithm}
We have the following proposition for the output $E_{group}$ of Algorithm \ref{alg:grouping}.
\begin{proposition}
The output $E_{group}$ of Algorithm~\ref{alg:grouping} satisfies $|E_{group}|\le \lceil \frac{9k+1}{5} \rceil.$   
\end{proposition}
\begin{IEEEproof}
Let 
$$E'_{adj}=\{\{c_{i},c_{i+1}\}:\{c_{i},c_{i+1}\}\in E_{adj}, \{c_{i},c_{i+1}\}\notin \Earc\}$$ 
be the subset of pairs in $E_{adj}$ that are not in $\Earc$. 
Note that in each round of \textbf{Step 2} in Algorithm~\ref{alg:grouping}, at most $5$ adjacent pairs (corresponding to two choices of $i$ for each $i^*$ and two choices of $j$ for each choice of $i$, plus the pair $\{c_{i^*},c_{i^*+1}\}$) are removed from $E_{adj}$, and thus from $E'_{adj}$ as well. In addition, one pair of nodes in $E'_{adj}$ is added to $E_{group}$ in each round. Since at the \textbf{Initialization} round we have $|E'_{adj}|\ge 2k-1-k=k-1$, there are at least $\frac{k-1}{5}$ rounds. Hence, $E_{group}$ contains at least $\frac{k-1}{5}$ pairs of nodes, and thus $|E_{group}|\le 2k-\frac{k-1}{5}=\frac{9k+1}{5}$. The claimed result follows by observing that the bound has to be integer-valued.  
\end{IEEEproof} 
After obtaining the grouping $E_{group}$, we compute the values of $r_1,\ldots,r_{|E_{group}|}$ in a similar manner to~\eqref{eq:defri1} and~\eqref{eq:defri2}. For $i\in \{1,\ldots, |E_{group}|\}$, let $j_i$ be the $i$th smallest integer $j$ such that $\{c_{j},c_{j+1}\}\in E_{group}$ or $\{c_j\}\in E_{group}$. Let $C=\{(c_j,c_\ell):\exists\, \ell\in \{2,\ldots,2k\} \text{ s.t. } \
\{c_{\ell-1},c_\ell\}\in E_{group},\, \{c_j,c_\ell\}\in \Earc\}.$
Note that for any $j\in \{1,\ldots,2k\}$, there is at most one $\ell\in\{1,\ldots,2k\}$ such that $(c_j,c_\ell)\in C$.
For $i\in \{1,\ldots,|E_{group}|\}$, let 
%\begin{align}\label{eq:defrigeneralk}
%r_i=c_i,\text{ if }c_i\in\
%\end{align}
%and 
\begin{align}\label{eq:defrigeneralk}
&r_i=\nonumber\\
&\begin{cases}c_{j_i},~&\text{if }(c_{j_i},c_\ell)\notin C \text{ for any $\ell\in\{1,\ldots,2k\}$}\\
c_{j_i}-(c_{\ell}-c_{\ell-1}),~& \text{if } i\equiv i^*\bmod 2\text{ and }(c_{j_i},c_\ell)\in C\\
    c_{j_i}+(c_{\ell}-c_{\ell-1}),~& \text{if } i\not\equiv i^*\bmod 2\text{ and }(c_{j_i},c_\ell)\in C
    \end{cases}.
\end{align}
The values of $r_i$, $i\in\{1,\ldots,|E_{group}|\},$ are based on~\eqref{eq:decidingr}. Next, we show that the  $r_i$s, $i\in \{1,\ldots,|E_{group}|\},$ are consistent with the grouping $E_{group}$. First,
\begin{align}\label{eq:groupingk}
c_{j_{i+1}-1}<d_{i}\le c_{j_{i+1}}     
\end{align}
Similar to the proof of Proposition~\ref{prop:k3}, we start by showing that either at least one of 
$r_i=c_{j_i}$ and $r_{i+1}=c_{j_{i+1}}$ holds, or that $\{c_{j_i},c_{j_{i+1}}\}\in \Earc$. If $r_i\ne c_{j_i}$ and $r_{i+1}\ne c_{j_{i+1}}$, then either there exist a $\{i_1,i_1+1\}\in E_{group}$ and a $\{i_2,i_2+1\}\in E_{group}$ such that $\{i_1+1,j_i\}\in \Earc$ (or $i_1+1=j_i$) and $\{i_2+1,j_{i+1}\}\in \Earc$, or $\{c_{j_i-1},c_{j_i}\}\in E_{group}$ $\{c_{j_i},c_{j_{i+1}}\}\in \Earc$. The former case contradicts the \textbf{Step 2} in Algorithm~\ref{alg:grouping} where at least one of the sets $\{i_1,i_1+1\}$ and $\{i_2,i_2+1\}$ is removed from $E_{adj}$ and is not in $E_{group}$. In the latter case, we have that $$d_i= \frac{r_i+r_{i+1}}{2}=\frac{c_{j_{i+1}-1}+c_{j_{i+1}}}{2}\in (c_{j_{i+1}-1},c_{j_{i+1}}).$$
On the other hand, if one of the two equalities $r_i=c_{j_i}$ and $r_{i+1}=c_{j_{i+1}}$ holds, we have that
\begin{align*}
    d_i=\frac{r_i+r_{i+1}}{2}=\frac{c_{j_{i}}+c_{j_{i+1}}\pm (c_{\ell}-c_{\ell-1})}{2}
\end{align*}
for some $(c_{j_i},c_\ell)\in C$ or $(c_{j_{i+1}},c_\ell)\in C$, which implies that $\{c_{\ell-1},c_{\ell}\}\in E_{group}$. Then, since $|c_{\ell-1}-c_{\ell}|$ is the minimum among all pairs in $E_{adj}$ we have that 
$$d_i=\frac{c_{j_{i}}+c_{j_{i+1}}\pm (c_{\ell}-c_{\ell-1})}{2}\in 
(c_{j_{i+1}-1},c_{j_{i+1}}).$$
Otherwise, $\{c_{\ell-1},c_{\ell}\}$ should have been removed in the previous rounds. Finally, if $r_i=c_{j_i}$ and $r_{i+1}=c_{j_{i+1}}$ both hold, we have~\eqref{eq:groupingk}. Thus, we conclude that \eqref{eq:groupingk} holds. The rest of the proof is similar to that of Proposition~\ref{prop:k3}.

%\section{Main Results}\label{sec:results}
%\input{\srcpath sec_results.tex}

%\section{Algorithms}\label{sec:algo}
%\input{\srcpath sec_results.tex}
%\input{\srcpath sec_algo.tex}

%\section{Concluding Remarks}\label{sec:conclusion}
%\input{\srcpath sec_conclusion.tex}

%\section*{Acknowledgments}
%\input{\srcpath acknowledgment.tex}

%This work was supported by the National Science Foundation (NSF) under grants CCF 2107344 and CCF 2046991.

\bibliographystyle{IEEEtran}
\bibliography{Ref.bib}

%\newpage
%\onecolumn

%\appendix
%\input{\srcpath sec_proofs.tex}

%%%%%%
%% To balance the columns at the last page of the paper use this
%% command:
%%
%\enlargethispage{-1.2cm} 
%%
%% If the balancing should occur in the middle of the references, use
%% the following trigger:
%%
%\IEEEtriggeratref{3}
%%
%% which triggers a \newpage (i.e., new column) just before the given
%% reference number. Note that you need to adapt this if you modify
%% the paper.  The "triggered" command can be changed if desired:
%%
%\IEEEtriggercmd{\enlargethispage{-20cm}}
%%
%%%%%%

%%%%%%
%% References:
%% We recommend the usage of BibTeX:
%%

\end{document}